\documentclass[rpreprint,3p,10pt,sort&compress, colorlinks,linkcolor=blue, citecolor=blue,fleqn]{elsarticlex}

\usepackage[utf8]{inputenc}

\usepackage{amssymb}
\usepackage{amsbsy}
\usepackage{amsmath,mathrsfs}
\usepackage{fixmath}
\usepackage{color}
\usepackage[dvipsnames]{xcolor}
\usepackage{placeins}
\usepackage{upgreek}

\font\LARGEbsf=cmssdc10 scaled 2100      

\usepackage{mathrsfs}

\usepackage{tikz}
\usetikzlibrary{spy}
\usetikzlibrary{pgfplots.groupplots}  
\usepackage{pgfplots}

\usepackage{siunitx}

\usepackage{stfloats}
\usepackage{siunitx}

\usepackage{bm}

\usepackage{graphicx}

\usepackage{caption}
\usepackage{subcaption}

\usepackage{multicol}

\usepackage{multirow}

\usepackage{mhchem}

\usepackage[toc,page]{appendix}

\usepackage{bm}

\usepackage[displaymath, mathlines, pagewise]{lineno}

\usepackage{float}

\usepackage{amsthm}
\newtheorem*{remark}{Remark}
\usepackage[makeroom]{cancel}
\usepackage{pgfplots}
\pgfplotsset{compat=1.18, width=7cm, height=5.5cm}
\usepackage{gensymb}
\usepackage{siunitx}
\usepgfplotslibrary{groupplots}
\usetikzlibrary{matrix}

\usepackage{accsupp}

\DeclareMathAlphabet\mathbfcal{OMS}{cmsy}{b}{n}

\usepackage[colorlinks,linkcolor=blue,anchorcolor=green,citecolor=blue]{hyperref}



\begin{document}
	
	\begin{frontmatter} 
		\title{A unified multi-perspective quadratic manifold for mitigating the Kolmogorov barrier in multiphysics damage} 
		\author[els]{Qinghua Zhang\corref{cor1}}
		\ead{qinghua.zhang@rwth-aachen.de}
		\author[els]{Stephan Ritzert}
		\author[els]{Jian Zhang}
		\author[els]{Jannick Kehls}
		\author[els,rvt]{Stefanie Reese}
		\author[els]{Tim Brepols}
		\cortext[cor1]{Corresponding author.}
		
		\address[els]{Institute of Applied Mechanics, RWTH Aachen University, Mies-van-der-Rohe-Str. 1, 52074 Aachen, Germany}
		\address[rvt]{University of Siegen, Adolf-Reichwein-Straße 2a, 57076 Siegen, Germany}
		
		\begin{abstract}
			In multiphysics damage problems, material degradation is often modeled using local and/or global damage variables, whose evolution introduces strong nonlinearities and significant computational costs. Linear projection-based reduced-order models (ROMs) are widely used to accelerate these simulations but often fail to capture complex nonlinear damage evolution effectively. This limitation arises from the slow decay of the Kolmogorov $n$-width, which leads to a phenomenon known as the Kolmogorov barrier in linear approximation. To overcome this challenge, this study proposes a novel unified multi-perspective (multi-field and multi-state) quadratic manifold-based ROM framework for thermo-mechanically coupled damage-plasticity problems. A key feature lies in a multi-field and multi-state decomposition strategy that is grounded in the material’s physical response to guide the selection of mode numbers for each coupled field. Moreover, the framework decouples both material states and physical fields, providing clearer insights into the contributions and interactions of each field within the overall multiphysics simulation. Benchmark tests demonstrate that the proposed approach mitigates the Kolmogorov barrier of linear projection-based ROMs by ensuring a smooth and monotonic decrease in error as the number of modes increases. The proposed multi-perspective quadratic manifold framework offers a robust and flexible approach for efficiently reducing complex damage-involved multiphysics problems and shows strong potential for industrial applications.
		\end{abstract}
		
		
	\end{frontmatter}
	
	\section{Introduction}
	\label{Introduction}
	Dimensional reduction, in physical problems often achieved by suitably projecting governing equations from the full-order space onto lower-dimensional subspaces, plays a critical role in accelerating computational simulations. However, many existing studies \cite{radermacher2016pod,ghavamian2017pod,rocha2020accelerating,kastian2023discrete} focus exclusively on linear projections and often neglect the fact that a high compressibility of the reduced order model relies heavily on a fast decay of the Kolmogorov $n$-width \cite{pinkus2012n,cohen2023nonlinear,barnett2022quadratic,greif2019decay}. The Kolmogorov $n$-width limitation in model order reduction (MOR) arises when approximating complex high-dimensional systems using low-dimensional linear projections and is also called the Kolmogorov barrier \cite{peherstorfer2022breaking}. Simply speaking, certain nonlinear systems may not be well approximated using linear projection methods. This results in significant errors in reduced order modeling, for instance, problems involving shocks, discontinuities, or highly localized phenomena (e.g., damage \cite{felder2022thermo,zhang2025multi} or transport-dominated problems \cite{peherstorfer2020model,geelen2023operator}), where it is particularly difficult to precisely capture the nonlinear effects using a low-dimensional linear subspace.
	
	In light of the approximation barrier posed by the slow decay of the Kolmogorov $n$-width \cite{greif2019decay,bachmayr2017kolmogorov,barnett2022quadratic,cohen2023nonlinear}, recent research has increasingly turned to nonlinear approximation techniques.
	These approaches aim to construct reduced models through nonlinear mappings that can better capture the system's complex behavior. As a result, noticeable interest has emerged in the development of nonlinear reduction techniques for finite element (FE)-based simulations due to the widespread application of FE in industry, see \cite{rutzmoser2017generalization,jain2017quadratic,jain2019hyper,ghavamian2017pod}. In the FE framework, projection-based MOR denotes a semi-discretization, wherein global shape and test functions are employed to construct either a Galerkin or Petrov-Galerkin projection to the governing equations \cite{barnett2022quadratic}. This process is realized by collecting and compressing solution snapshots of a discretized equation system using a \textit{thin} singular value decomposition (SVD) to build up a reduced-order basis (ROB). Once the ROB is obtained, the global shape and test functions can be chosen either to be the same or different. If these functions are identical, the resulting formulation corresponds to a Galerkin projection, in which the governing equations are projected onto the reduced subspace. If they differ, a Petrov-Galerkin projection is applied instead. In the context of nonlinear approximation, not only the classical linear projection quantities but also both the nonlinear projection matrix and the quadratic/higher-order operator are needed \cite{geelen2024learning,geelen2023learning}. Such projection-based MOR approaches are commonly referred to as \textit{manifold learning} or \textit{simulation-driven} computational methods, see \cite{barnett2022quadratic,geelen2023operator,jain2019hyper,lee2007nonlinear}.
	
	Over the past few decades, projection-based MOR has achieved significant success in both solid mechanics \cite{ghavamian2017pod,kaneko2021hyper,ghnatios2024new,michel2016model,farhat2015structure,farhat2014dimensional} and fluid mechanics \cite{grimberg2021mesh,tezaur2022robust,grimberg2020stability,washabaugh2012nonlinear,barnett2022quadratic}. In solid mechanics, many studies have focused on materials for which the Kolmogorov $n$-width decays rapidly, such as hyperelastic \cite{radermacher2013comparison,bhattacharjee2016nonlinear,goury2018fast}, plastic \cite{lange2024monolithic,bugas2024grain}, and viscoelastic \cite{radermacher2016pod,radermacher2013comparison,trainotti2024ecsw} materials. As the governing partial differential equations (PDEs) in these models are typically weakly nonlinear, standard single-field-based proper orthogonal decomposition (POD) methods can efficiently compress solution snapshots of both linear ordinary differential equations (ODEs) and weakly nonlinear PDEs. In contrast, for damage models typically characterized by a slow decay of the Kolmogorov $n$-width, reduced-order modeling remains considerably more challenging. Few studies have successfully developed reduced-order approaches that can accurately capture the global softening behavior induced by damage. For complex multiphysics-coupled damage models, the available literature is rare and remains under-investigated. Within the context of reduced-order modeling in finite element simulations, the overall strategies can be broadly classified into two categories: general model order reduction approaches globally decreasing the number of degrees of freedom (e.g., via projection-based techniques such as POD \cite{radermacher2013comparison,ghavamian2017pod,zhang2025multi}) and hyperreduction (e.g., through element or quadrature point sampling to further reduce computational costs \cite{hernandez2017dimensional,lange2024monolithic,wulfinghoff2025empirically}). 
	A deep investigation of the limitations imposed by approximation theory is essential to understand the mechanisms behind instabilities that may occur when applying such reduction strategies.
	For instance, in damage-induced softening simulations, many studies do not focus enough on the slow decay of the Kolmogorov $n$-width and its underlying mechanisms. The problem usually becomes even more pronounced if multi-physical coupling effects play a role. In such cases,  employing a standard single-field-based linear projection theory can lead to strong fluctuations in the reduced-order solution, particularly in the softening region, see \cite{selvaraj2024adaptive,kastian2023discrete,mishra2024enhanced,rocha2020accelerating,oliver2017reduced}.
	
	To mitigate the Kolmogorov barrier, nonlinear manifolds (quadratic or of higher order) were developed to accurately represent the strongly nonlinear character of the governing equations. To the best of the authors' knowledge, the well-accepted contribution of \citet{qiao2012explicit} pioneered the use of nonlinear mapping for dimensional reduction through manifold learning in computer science. Subsequently, a quadratic manifold MOR approach was developed by \citet{jain2017quadratic} for structural dynamic problems discretized by finite elements. Their approach outperformed linear projections that fail to capture the rapid degradation of the dynamic response. In the same year, \citet{rutzmoser2017generalization} extended the quadratic manifold method by incorporating static derivatives to generalize modal derivatives for the displacement field, which improves the accuracy of the quadratic manifold approach in engineering problems. Nevertheless, despite its efficacy in mitigating the Kolmogorov barrier and providing more precise projections, an achievable reduction in computation time remains limited in nonlinear finite element simulations, since the repeated calculation of the full global stiffness matrix and residual vector is still required. To overcome this, \citet{jain2019hyper} built upon \citet{an2008optimizing}'s initial presentation of the energy-conserving sampling and weighting (ECSW) method in computer graphics as well as \citet{farhat2014dimensional,farhat2015structure}'s extension of ECSW-based hyperreduction for solid mechanics by combining ECSW with the quadratic manifold framework for structural dynamics. \citet{jain2019hyper} further provided a rigorous proof that ECSW conserves energy and preserves the Lagrangian structure dictated by Hamilton’s principle, ensuring intrinsic numerical stability in time integration. Subsequently, \citet{barnett2022quadratic} consistently applied a similar ECSW-based manifold coupling strategy in the context of fluid mechanics with promising results. In their work, the slow decay of the Kolmogorov $n$-width for the simulation of the Ahmed body’s turbulent wake flow \cite{ahmed1984some} was emphasized. However, the distribution of the Kolmogorov $n$-width was, in the authors' opinion, not clearly enough presented. Subsequently, \citet{geelen2023operator,geelen2024learning,geelen2023learning} extended the quadratic manifold approach to address the slow decay of the Kolmogorov $n$-width in transport-dominated dynamic problems, yielding promising accuracy via a purely data-driven methodology. Through this non-intrusive approach, constant, linear, and nonlinear operators are determined by minimizing the difference between high-fidelity and approximated solutions, without direct access to stiffness matrices and discretized residuals. However, it should be noted that this approach relies heavily on nonlocal data inference, and its applicability to damage-plasticity models, characterized by numerous local variables and complex global-local coupling, remains unexplored. Furthermore, many studies have demonstrated the instabilities that can arise while using reduced-order damage models, once damage-induced global softening occurs in the material \cite{oliver2017reduced,kerfriden2011bridging,kerfriden2013partitioned,selvaraj2024adaptive,rocha2020accelerating,kastian2023discrete}. 
	
	To overcome the instability induced by damage in the context of reduced-order modeling in the global softening region, \citet{zhang2025multi} proposed an effective strategy using a multi-field decomposed approach based on linear projection. In their work, based on the vectoral equivalence of the discretized multiphysics solutions, the multi-field decomposed approach is realized by individually projecting the governing equations of each physical field onto distinct subspaces at a global level.
	However, this approach still requires a relatively large number of modes across multiple fields due to the inherent limitations of the linear approximation, which is particularly affected by the slow decay of the Kolmogorov $n$-width in the associated solution spaces. In addition, the POD-based Galerkin method employed there remains constrained by the computational bottleneck associated with the repeated evaluation of all elements during the iterative procedures. To overcome these challenges, a novel unified multi-field and multi-state decomposed quadratic manifold MOR formulation is proposed in this work to effectively mitigate the Kolmogorov barrier observed in multi-physical damage problems in a fast manner. To realize a fast simulation for strongly nonlinear problems, the approach is additionally integrated with the ECSW-based hyperreduction technique. The new method aims to significantly reduce the computational cost of thermo-mechanically coupled damage-plasticity simulations.
	
	In light of the observations above, the paper is structured as follows. \hyperref[sec:Modelling]{{Section 2}} briefly recapitulates the derivation of a thermo-mechanically coupled gradient-extended damage model along with its governing partial differential equations. \hyperref[sec:MOR theory]{{Section 3}} provides a mathematical analysis of the slow decay of the Kolmogorov $n$-width associated with the considered multi-physical damage model. A novel unified quadratic manifold-based MOR framework is then proposed, incorporating a multi-field and multi-state decomposition together with a corresponding ECSW-based hyperreduction technique to mitigate the Kolmogorov barrier in damage. \hyperref[sec:Model validation]{{Section 4}} investigates the performance of the newly proposed MOR approach through two academic benchmark problems, focusing on reduction efficiency, predictive accuracy, and numerical precision. \hyperref[Results and discussion]{{Section 5}} conducts a comparative error analysis between the decomposed and non-decomposed strategies applied to the coefficient matrices, highlighting the critical role of field-wise decomposition. Furthermore, a comparison is carried out between the quadratic manifold-based method and a classical POD-based linear approximation, aiming to explain why the quadratic manifold is preferred in the reduced-order modeling of multi-physical damage simulations. Finally, \hyperref[sec:Conclusion]{{Section 6}} summarizes the key findings, discusses the advantages and limitations of the newly proposed MOR approach, and outlines potential directions for future research.
	
	\section{Material model}
	\label{sec:Modelling}	
	
	\subsection{Kinematics and general form of the Helmholtz free energy density function}
	Following the multiplicative decomposition framework proposed by \citet{lu1975decomposition}, the deformation gradient $\bm{F}$ is multiplicatively decomposed into a mechanical part $\bm{F}_m$ and a thermal part $\bm{F}_{\theta}$ \cite{stojanovic1964finite}. The mechanical part $\bm{F}_m$ is further decomposed into elastic and plastic components, denoted by $\bm{F}_e$ and $\bm{F}_p$, respectively:
	\begin{equation}
		\bm{F} = {\bm{F}_m} \, {\bm{F}_\theta } = {\bm{F}_e} \, {\bm{F}_p} \, {\bm{F}_\theta} \quad \text{with} \quad {\bm{F}_\theta } =  \underbrace{\exp {\left( \alpha \, \left( {\theta  - {\theta _0}} \right) \right)}}_{\vartheta \left( \theta  \right)} \, \bm{I}.
		\label{eq1}
	\end{equation}
	\noindent Here, under the assumption of isotropic thermal expansion, the thermal deformation gradient $\bm{F}_{\theta}$ can be expressed in terms of an exponential function $\vartheta (\theta)$ that depends on the temperature $\theta$, see  \citet{stojanovic1964finite}. \hyperref[eq1]{Eq.\eqref{eq1}} involves the reference temperature $\theta_0$ and the second-order identity tensor $\bm{I}$. The parameter $\alpha$ denotes the thermal expansion coefficient, which is assumed to be temperature-independent. Based on the underlying kinematic relations, the elastic and plastic right Cauchy-Green deformation tensors are expressed as:
	\begin{equation}
		{\bm{C}_e} = \bm{F}_e^{\rm{T}} \, {\bm{F}_e} = \frac{1}{{{\vartheta ^2}}} \, \bm{F}_p^{ - \rm{T}} \, \bm{C} \, \bm{F}_p^{ - 1}, \quad {\bm{C}_p} = \bm{F}_p^{\rm{T}} \, {\bm{F}_p}.
		\label{eq3}
	\end{equation}
	
	Subsequently, the general Helmholtz free energy density function per unit reference volume $ \psi $ is assumed to depend on the elastic and plastic right Cauchy-Green tensors, denoted by $\bm{C}_{e}$ and $\bm{C}_{p}$, respectively, as well as the temperature $\theta$ in the following way:
	\begin{equation}
		\psi  = \underbrace {{f_d}(D)\left( {{\psi _e} \, ({\bm{C}_e}, \, \theta ) + {\psi _p} \, ({\bm{C}_p}, \, {{\xi} _p}, \, \theta )} \right)}_{{\rm{Elastoplastic \, part}}}  \, + \underbrace {{\psi _d} \, ({\xi _d}, \, \theta )}_{{\rm{Damage \, hardening \, part}}} + \underbrace {{\psi _{\bar d}} \, (D, \, \bar D, \, \nabla \bar D, \, \theta )}_{{\rm{Micromorphic \, extension}}} + \underbrace {{\psi _\theta } \, (\theta )}_{{\rm{Caloric \, part}}}.
		\label{eq4}
	\end{equation}
	
	\noindent Based on the hypothesis of energy equivalence proposed by \citet{cordebois1982damage}, a quadratic degradation function $f_{d}(D)=(1-D)^2$ is used to degrade the elastoplastic energy whenever the local damage variable $D$ evolves. $\psi _e$ represents the elastic energy density, which depends on both the elastic right Cauchy-Green tensor $\bm{C}_e$ and temperature $\theta$. $\psi _p$ denotes the plastic energy density, which is formulated in terms of the plastic right Cauchy-Green tensor $\bm{C}_p$, the accumulated plastic strain $\xi _p$, and the temperature $\theta$. In addition, a damage-hardening energy density $\psi _d$ is postulated as a function of the damage-hardening variable $\xi _d$ and the temperature $\theta$. 
	
	Following the general micromorphic framework proposed by \citet{forest2009micromorphic}, the energy density contribution $\psi_{\bar{d}}$ introduces a coupling between the local damage variable $D$ and the nonlocal (micromorphic) damage variable $\bar{D}$. It also incorporates the influence of its gradient $\nabla \bar{D}$, defined with respect to the reference configuration. In the present work, the caloric energy density $\psi_{\theta}$ does not require a further specification for certain temperature-dependent materials \cite{ames2009thermo,zhang2022exploring,ruan2024phase}, as a constant heat capacity ($c$:=const.) is assumed. Further details are omitted here for brevity; interested readers are referred to \citet{felder2022thermo}. The energy density is formulated in terms of the invariants of $\bm{C}_e$, which can alternatively be expressed in terms of $\bm{C}$ and $\bm{C}_p$, see \hyperref[eq3]{Eq.\eqref{eq3}}. Accordingly, the total Helmholtz free energy density function is expressed as $\psi \left( \bm{C}, \,  \bm{\zeta}_{{\rm {int}}},  \, \bar{D}, \,  \nabla {\bar D}, \, \theta \right)$, where the internal variables are defined as ${\bm{\zeta}_{{\rm {int}} }} = \{ {\bm{C}_p}, \, { \xi _p}, \, D, \, {\xi _d} \}$.

	\subsection{Specific choice for the Helmholtz free energy density}
	\begin{itemize}
		\item Elastoplasticity
	\end{itemize}
	
	A compressible Neo-Hookean-type material model is adopted for the elastic energy density function $\psi_e$, which is defined in terms of the elastic right Cauchy-Green tensor $\bm{C}_e$ as:
	\begin{equation}
		{\psi _e} = \frac{\mu }{2} \, \left( {{\rm{tr}}({\bm{C}_e}) - 3 - \ln \left(\det \left({\bm{C}_e} \right) \right)} \right) + \frac{\Lambda }{4} \, \left( {\det \left({\bm{C}_e} \right) - 1 - \ln \left( \det \left ({\bm{C}_e} \right) \right)} \right).
		\label{eq5}
	\end{equation}
	
	\noindent The plastic energy density function $\psi_{p}$ is expressed as follows:
	\begin{equation}
		{\psi _p} = \underbrace {\frac{a}{2} \, \left( {{\rm{tr}}({\bm{C}_p}) - 3 - \ln \left( {\det ({\bm{C}_p})} \right)} \right)}_{\rm{linear \,kinematic \, hardening}} + \underbrace { e_{p} \, \left( {{{\xi} _p} + \frac{{\exp ( - f_{p}\,{{\xi} _p}) - 1}}{f_p}} \right)}_{\text{nonlinear isotropic hardening}} + \underbrace {\frac{1}{2} \, P \,{\xi} _p^2.}_{\text{linear isotropic hardening}}
		\label{eq6}
	\end{equation}
	
	\noindent Here, the Lam\'e constants $\mu$ and $\Lambda$ in \hyperref[eq5]{Eq.\eqref{eq5}} as well as the plastic material parameters $a$, $e_p$, $f_p$, and $P$ in \hyperref[eq6]{Eq.\eqref{eq6}} can be temperature dependent.
	\begin{itemize}
		\item Damage hardening
	\end{itemize}
	
	The energy density function associated with damage hardening is defined in analogy to the nonlinear isotropic hardening in plasticity, with $e_d$ and $f_d$ denoting constant material parameters:
	\begin{equation}
		{\psi _d} = {e_d} \, \left( {{{\xi _d}} + \frac{{\exp ( - {f_d}\,{{\xi _d}}) - 1}}{{{f_d}}}} \right).
		\label{eq7}
	\end{equation}
	
	\begin{itemize}
		\item Micromorphic extension
	\end{itemize}
	\begin{equation}
		{\psi _{\bar d}} = \underbrace {\frac{{H}}{2} \, {{\left( {D - \bar D} \right)}^2}}_{{\text{local - nonlocal coupling}}} + \frac{{A}}{2} \, \nabla \bar D \cdot \nabla \bar D 
		\label{eq8}
	\end{equation}
	
	The micromorphic energy density function $\psi_{\bar{d}}$ introduces a strong coupling between the local damage variable $D$ and the nonlocal damage variable $\bar{D}$ through a constant penalty parameter $H$, as proposed by \citet{forest2009micromorphic}. Furthermore, the constant material parameter $A$ serves as a length scale relevant parameter that governs the contribution of the gradient term $\nabla \bar{D}$, the latter of which provides an appropriate regularization of the formulation. The presented model is thermodynamically consistent. Further details are omitted at this point, but can be found in  \cite{brepols2018micromorphic,brepols2020gradient, felder2022thermo} or \hyperref[AppendixA]{Appendix A}.
	
	\subsection{Balance equations}
	The material model in this study is based on the general micromorphic approach proposed by \citet{forest2009micromorphic}, which basically eliminates the well-known mesh dependency issues of conventional local continuum damage models. Under the assumption of quasi-static finite strain deformations and a transient thermal field, the corresponding partial differential equations governing the formulation are expressed as follows:
	\begin{equation}
		\begin{gathered}
			{\underline{\text{Balance of linear momentum}}} \hfill \\
			\begin{array}{*{20}{c}}
				{{\text{Div}}\left( {{\bm{F}  \, \mathbf{S}}} \right) + {\bm{\bar f}}}&{ = \bm{0}}&{{\text{in }}\Omega } \\ 
				{(\bm{F} \, \mathbf{S}) \cdot \bm{n}}&= {\bm{ \bar t}}&{{\text{on }}\partial {\Omega _t}} \\ 
				\bm{u}&{ = \widetilde{\bm {u}}}&{{\text{on }}\partial {\Omega _u}} 
			\end{array} \hfill \\ 
		\end{gathered}
		\label{eq26}
	\end{equation}
	\begin{equation}
		\begin{gathered}
			{\underline{\text{Micromorphic balance}}} \hfill \\
			\begin{array}{*{20}{c}}
				{{\text{Div}}\left( {\bm{b}_i} - {\bm{b}_e} \right) - {a_i} + {a_e}}&{ = {0}}&{{\text{in }}\Omega } \\ 
				{\left({\bm{b}_{i}-\bm{b}_e}\right) \cdot \bm{n}}&{ = {a_c}}&{{\text{on }}\partial {\Omega _c}} \\ 
				{\bar D}&{ = \widetilde {\bar D}}&{{\text{on }}\partial {\Omega _{\bar{D}}}} 
			\end{array} \hfill \\ 
		\end{gathered}
		\label{eq27}
	\end{equation}
	\begin{equation}
		\begin{gathered}
			{\underline{\text{Energy balance}}} \hfill \\
			\begin{array}{*{20}{c}}
				{ \underbrace {- {\dot{e}_{\text{int}}} + {\bf{S}}:\frac{1}{2} \, \dot{\bm{C}} + {a_i} \, \dot {\bar D} + \bm{b}_i \cdot \nabla \dot {\bar D} }_{r_{\rm{int}}} + {r_{{\text{ext}}}}} - {\text{Div}}\left( \bm{q} \right)&{ = 0}&{{\text{in }}\Omega } \\ 
				{\bm{q} \cdot \bm{n}}&{ =  - \bar{q}}&{{\text{on }}\partial {\Omega _q}} \\ 
				\theta &{ = \widetilde \theta }&{{\text{on }}\partial {\Omega _\theta }} 
			\end{array} \hfill \\ 
		\end{gathered}
		\label{eq28}
	\end{equation}
	
	\noindent In the above balance equations, $\Omega$ denotes the domain of the body in the reference configuration. $\bf{S}$ represents the second Piola-Kirchhoff stress tensor. \hyperref[eq27]{Eq.\eqref{eq27}} formulates the micromorphic balance in terms of the internal body forces  $a_i$ and $\bm{b}_i$,  as well as the external micromorphic forces $a_e$ and $\bm{b}_e$. In the present formulation, $a_e$ and $\bm{b}_e$ are assumed to be zero. The vector ${\bm{\bar f}}$ denotes the general body force per unit reference volume, while $\bm{n}$ represents the unit outward normal vector. In \hyperref[eq28]{Eq.\eqref{eq28}}, the internal energy per unit reference volume $e_{\text{int}}$ and its time derivative $\dot{e}_{\text{int}}$ are given by $e_{\text{int}} = \psi + \theta \, \eta$ and $\dot{e}_{\text{int}} = \dot{\psi} + \dot{\theta} \, \eta + \theta \, \dot{\eta}$, respectively. The entropy and volumetric heat capacity are defined as $\eta = -\frac{\partial \psi}{\partial \theta}$ and $c = -\frac{\partial^2 \psi}{\partial \theta^2}$, respectively. Assuming a constant heat capacity and no external heat source ($r_{\text{ext}} = 0$), this leads to $\dot{e}_{\text{int}} = c \, \dot{\theta}$, see \hyperref[AppendixB]{Appendix B} for details. On the Dirichlet boundaries $\partial \Omega _u$, $\partial \Omega _{\bar{D}}$, and $\partial \Omega _{\theta}$, displacements $\widetilde{\bm{u}}$, nonlocal damage $\widetilde{\bar{D}}$, and temperature $\widetilde{\theta}$ are prescribed, respectively. Similarly, Neumann boundary conditions are imposed on $\partial \Omega _t$, $\partial \Omega _{c}$, and $\partial \Omega _{q}$, associated with the macroscopic traction force $\bm{\bar{t}}$, the micromorphic traction force $a_c$, and the heat flux $\bar{q}$, respectively. The total internal heat generation is subsequently expressed as $r_{\rm{int}}={r_e}+{r_p}+{r_d}$, where $r_e$, $r_p$, and $r_d$ denote the elastic, plastic, and damage-related contributions, respectively. Further details can be found in \citet{felder2022thermo} and \citet{reese2001thermomechanische}. 
	
	\subsection{Weak form of the problem}
	Based on the strong form given in Eqs.\eqref{eq26}, \eqref{eq27}, and \eqref{eq28}, the corresponding weak form is obtained by multiplying the equations with appropriate test functions, namely $\delta \bm{u}$, $\delta \bar D$, and $\delta \theta$. The energy residuals for an element associated with the displacement, damage, and temperature fields\footnote{Note that the field quantities in the following belong to an element $e$ and should probably more precisely be denoted as $\boldsymbol{u}^e$, $\bar{D}^e$, and $\theta^e$. However, this is intentionally avoided to prevent a proliferation of notation, and instead understood implicitly.} are denoted as $\mathit{g} _u^e$, $\mathit{g} _d^e$, and $\mathit{g} _{\theta}^e$, respectively:
	
	\begin{itemize}
		\item Displacement field
	\end{itemize}
	\begin{equation}
		\mathit{g} _u^e \left( {\bm{u}}, \, \bar{D}, \, \theta ,\delta {\bm{u}} \right) = \int\limits_{{\Omega _e}} {{\bf{S}} \, \, {:} \, \delta {\bm{E}} \, dV}  - \int\limits_{{\Omega _e}} {{{\bm{\bar f}}} \cdot \delta {\bm{u}} \, dV}  - \int\limits_{\partial {\Omega _e}} {{{\bm{\bar t}}} \cdot \delta {\bm{u}} \, dA}
		\label{eq29}
	\end{equation}
	
	\begin{itemize}
		\item Damage field
	\end{itemize}
	\begin{equation}
		\mathit{g} _d^e \left( {\bm{u}}, \, \bar{D}, \, \theta , \, \delta {\bar{D}} \right) = \int\limits_{{\Omega _e}} {H \, \left( {D - \bar D} \right) \, \delta \bar D \, dV}  - \int\limits_{{\Omega _e}} {A \, \nabla \bar D \cdot \nabla (\delta \bar {D})  \, dV}
		\label{eq30}
	\end{equation}
	
	\begin{itemize}
		\item Temperature field
	\end{itemize}
	\begin{equation}
		\mathit{g} _\theta ^e ( {\bm{u}}, \, \bar D, \, \theta , \, \dot \theta , \, \delta {\theta} ) = \int\limits_{{\Omega _e}} {c \, \dot \theta \, \delta \theta \, dV}  - \int\limits_{{\Omega _e}} {{\bm{q}} \cdot \nabla  (\delta \theta) \, dV}  - \int\limits_{{\Omega _e}} {{r_{{\text{int}}}} \, \delta \theta \, dV}  - \int\limits_{\partial \Omega _e} {\bar q \, \delta \theta \, dA}
		\label{eq31}
	\end{equation}
	
	\noindent Here, $\partial \Omega _e$ represents the relevant Dirichlet ($\partial \Omega_{e, D} = \partial \Omega_{e, u} \cup \partial \Omega_{e, \bar{D}} \cup \partial \Omega_{e, \theta}$) and Neumann ($\partial \Omega_{e, N} = \partial \Omega_{e, t} \cup \partial \Omega_{e, c} \cup \partial \Omega_{e, q}$) boundaries of an element $e$, with $\partial \Omega_{e}=\partial \Omega_{e, D} \cup \partial \Omega_{e, N}$. For brevity, a unified boundary $\partial \Omega _e$ is employed here. $\delta \bm{E} = \frac{1}{2} \, [\bm{F}^{{\rm{T}}} \, \nabla (\delta \bm{u}) + \nabla ^{\rm{T}} (\delta (\bm{u})) \, \bm{F}]$ denotes the test function of the Green-Lagrange strain tensor $\delta \bm{E}$. The element residual force $\bm{R}_e$ is derived from the virtual energy $\mathit{g}^{e}=\mathit{g} _u^{e}+\mathit{g} _d^{e}+\mathit{g} _{\theta}^{e} $ with respect to all element degrees of freedom ${\bm{U}_e} = \left[ {{\bm{u}}, \, {\bar{D}}, \, {\theta} } \right]$, named the displacement $\bm{u}$, nonlocal damage ${\bar{D}}$, and temperature  ${\theta}$:
	\begin{equation}
		\bm{R}_e = \frac{{\partial {\mathit{g}^{e} \, (\bm{u},\, \bar{D}, \, \theta)}}}{{\partial \bm{U}_e}}, \quad \bm{K}^{e} = \frac{{\partial \bm{R}_e \, (\bm{u}, \, \bar{D}, \, \theta)}}{{\partial \bm{U}_e}}.
		\label{eq32}
	\end{equation}
	
	\noindent Specifically, the residual force vector is obtained by taking the partial derivative of the virtual energy $\mathit{g}^{e}$ with respect to the nodal degrees of freedom. Subsequently, the element stiffness matrix $\bm{K}^e$ is obtained by the partial derivative of the residual vector $\bm{R}_e$ with respect to the nodal degrees of freedom. Accordingly, the components of the element stiffness matrix $\bm{K}^e$ are expressed as follows:
	\begin{equation}
		{\bm{K}}_{uu}^e = \frac{{{\partial ^2}{\mathit{g} ^e}}}{{\partial {{\bm{u}}^2}}}, \, {\bm{K}}_{u\bar D}^e = \frac{{{\partial ^2}{\mathit{g} ^e}}}{{\partial {\bm{u}} \partial {{\bar{D}}} }}, 
		\, {\bm{K}}_{u\theta }^e = \frac{{{\partial ^2}{\mathit{g} ^e}}}{{\partial {\bm{u}}\partial {{\theta}}  }}, \, {{K}}_{\bar D\bar D}^e = \frac{{{\partial ^2}{\mathit{g} ^e}}}{{\partial {{{{\bar{D}}} }^2}}},  \,  {{K}}_{\bar D\theta }^e = \frac{{{\partial ^2}{\mathit{g} ^e}}}{{\partial {{\bar{D}}}  \partial {{\theta}}  }}, \, {{K}}_{\theta \theta }^e = \frac{{{\partial ^2}{\mathit{g} ^e}}}{{\partial {{{\theta}}  ^2}}}.
		\label{eq33}
	\end{equation}
	\noindent The omitted matrices $\bm{K}^{e}_{\bar{D}u}$, $\bm{K}^{e}_{\theta u}$, and ${K}^{e}_{\theta \bar{D}}$ are derived analogously. Derivatives are computed by the automated differentiation toolbox \textit{AceGen} \cite{korelc2002multi,korelc2016automation}. The global residual vector $\bm{R}$, stiffness matrices $\bm{K}$, and increment vectors of unknowns $\Delta \bm{U}$ are then assembled from their corresponding element-level counterparts $\bm{R}_e$, $\bm{K}^e$, and $\Delta \bm{U}_e$, respectively, as follows:
	\begin{equation}
		{{\bm{R} = \raise3pt
				\hbox{$\hbox{\scriptsize $N_e$}\atop{\hbox{\LARGEbsf A}\atop {\scriptstyle e=1}}$} {\bm{R}_e}}, \quad {\bm{K} = \raise3pt
				\hbox{$\hbox{\scriptsize $N_e$}\atop{\hbox{\LARGEbsf A}\atop {\scriptstyle e=1}}$} {\bm{K}^{e}}}
			, \quad {\Delta \bm{U} = \raise3pt
				\hbox{$\hbox{\scriptsize $N_e$}\atop{\hbox{\LARGEbsf A}\atop {\scriptstyle e=1}}$} {\Delta \bm{U}_e}}
		},   
		\label{eq34}
	\end{equation}
	\noindent where the operator $\hbox{$\hbox{\scriptsize $N_e$}\atop{\hbox{\LARGEbsf A}\atop {\scriptstyle e=1}}$} \left( \bullet \right)$ in \hyperref[eq34]{Eq.\eqref{eq34}} denotes the well-known assembly operator applied to all elements $N_e$ in the given domain. Finally, the global stiffness matrix $\bm{K}$, the residual force vector $\bm{R}$, and the increment of the global solution vector $\Delta \bm{U}$ are expressed as:
	\begin{equation}
		\underbrace {\left( {\begin{array}{*{20}{c}}
					{{{\bm{K}}_{uu}}}&{{{\bm{K}}_{u\bar D}}}&{{{\bm{K}}_{u\theta }}} \\ 
					{{{\bm{K}}_{\bar Du}}}&{{{\bm{K}}_{\bar D\bar D}}}&{{{\bm{K}}_{\bar D\theta }}} \\ 
					{{{\bm{K}}_{\theta u}}}&{{{\bm{K}}_{\theta \bar D}}}&{{{\bm{K}}_{\theta \theta}}} 
			\end{array}} \right)}_{\bm{K}}\underbrace {\left( {\begin{array}{*{20}{c}}
					{\Delta {\bm{u}}} \\ 
					{\Delta \bar {\bm{D}} } \\ 
					{\Delta {\bm{\theta }}} 
			\end{array}} \right) }_{\Delta {\bm{U}}} =  - \underbrace {\left( {\begin{array}{*{20}{c}}
					{{{\bm{R}}_u}} \\ 
					{{{\bm{R}}_{\bar D}}} \\ 
					{{{\bm{R}}_\theta }} 
			\end{array}} \right) }_{\bm{R}}.
		\label{eq35}
	\end{equation}
	
	\section{Nonlinear manifold model order reduction}
	\label{sec:MOR theory}
	In a nonlinear finite element computation based on the Newton-Raphson method, \hyperref[eq35]{Eq.\eqref{eq35}} needs to be solved repeatedly for the high-dimensional increment vector $\Delta \boldsymbol{U}$. However, this approach can become computationally expensive, particularly for damage problems where a large number of finite elements are usually required. Therefore, projection-based MOR techniques are employed to enable more efficient simulations.
	
	\subsection{Reducibility of Kolmogorov $n$-width barrier in damage}
	In the case of damage problems, when the local damage initiates, the associated stress tensors are degraded by a quadratic degradation function $f_d(D)=(1-D)^2$, see Eqs. \eqref{eq4}, \eqref{eqA3}, \eqref{eqA4}, and \eqref{eqA5}. In more complex scenarios, incorporating the spectral decomposition for the stress tensor to distinguish the tensile and compressive responses in an asymmetric manner results in a highly nonlinear constitutive relation \cite{fassin2019efficient,zhang2025hierarchical,wu2017unified,wu2020phase,ruan2023thermo,miehe2010phase,narayan2021fracture}.
	The high nonlinearity of the presented themo-mechanically coupled damage-plasticity model typically leads to the phenomenon that the singular values of the snapshot matrix decay very slowly during the entire simulation process, see \hyperref[figC16]{Fig.C.16} in \hyperref[AppendixC]{Appendix C}. This slow decay of singular values introduces a Kolmogorov barrier \cite{greif2019decay,peherstorfer2022breaking,barnett2022quadratic,geelen2023operator,cohen2023nonlinear}, posing a challenge for approximations based on a linear projection method. The Kolmogorov $n$-width with the dimension of the subspace $n$ is defined as:
	\begin{equation}
		d_n(\mathcal{S}) := \inf_{\substack{\mathcal{Y}_n \subseteq \mathcal{Y} \\ \dim(\mathcal{Y}_n) \leq n}} \ \sup_{u \in \mathcal{S}} \ \inf_{u_n \in \mathcal{Y}_n} \| u - u_n \|_{\mathcal{Y}}, 
		\label{35-1}
	\end{equation}
	which formulates the maximum distance between the manifold solution point $u$ and the reconstructed solution point $u_n$ while projecting the solution points of the solution manifold $\mathcal{S}$ onto a Hilbert or Banach space $\mathcal{Y}$ by incorporating the supremum and the second infimum. Here, $\mathcal{S}$ denotes a set of collected solutions of PDEs and is therefore named \textit{solution manifold}. $\mathcal{Y}_n$ represents the linear subspace \cite{greif2019decay,barnett2022quadratic}. For certain linear coercive parameterized problems, the Kolmogorov $n$-width $d_n$ decays drastically and exhibits an exponential decay according to $d_n \leq C \, e^{-\beta \, n}$, where $0<C< \infty$ and $\beta > 0$ \cite{buffa2012priori,greif2019decay,ohlberger2015reduced}. A fast decay of $d_n$ indicates that the equation system is highly reducible using a linear projection-based method. However, in the case of elliptic or parabolic PDEs, $d_n$ typically decays very slowly since $d_n \geq 1/4  (n^{-1/2})$ \cite{greif2019decay,barnett2022quadratic}, where $d_n \sim \sigma_{n+1}$. Here, $\sigma_{n+1}$ denotes the $n+1$-th singular value of the solution manifold. In other words, if the singular values decay rapidly, the PDE system is highly reducible. In this case, an accurate linear approximation is achievable using a low-dimensional reduced basis $\mathcal{Y}_n$. Conversely, if $d_n$ decays slowly, a linear approximation method lacks the accuracy to reduce the equation system, leading to the Kolmogorov barrier in approximation, see the distribution of the normalized singular values in \hyperref[fig4]{Fig.4 (a)} and \hyperref[AppendixC]{Appendix C}.
	
	In this study, the presence of a Kolmogorov $n$-width approximation barrier in the thermo-mechanically coupled damage-plasticity formulation arises naturally from the type and nature of the governing PDEs. The corresponding solution manifold is highly nonlinear and spatially localized in certain regions. More specifically, the mechanical equilibrium equation \eqref{eq26} for displacement and the balance equation \eqref{eq27} for micromorphic damage are elliptic \cite{ostwald2019implementation}, while the transient heat conduction equation \eqref{eq28} is parabolic \cite{mersel2025dynamic,ostwald2019implementation}. Additionally, the micromorphic balance equation leads to a second-order spatial regularization. Despite the smoothing effect of the nonlocal gradient term, the damage field often exhibits strong spatial localization, such as sharp crack-like zones, whose location, orientation, and intensity can vary significantly for different parameters, e.g., the damage threshold $Y_0$, penalty parameter $H$, and gradient parameter $A$. The localized features result from the softening and coupling among the different field quantities, making the solution manifold strongly nonlinear. Even small variations in the input parameters can easily lead to qualitatively different damage patterns (e.g., distinct crack paths). A linear subspace cannot suitably capture these features. This leads to a slow decay of the Kolmogorov $n$-width. To mitigate the Kolmogorov barrier in linear projection, a new multi-perspective nonlinear manifold MOR approach is developed in this study with the intention to significantly accelerating multi-physical damage simulations.
	
	\subsection{Nonlinear manifolds for dimensionality reduction} 
	Dimension reduction using nonlinear representation is realized by enriching linear approximations via low-order polynomial terms, which are dependent on the reduced state $\widetilde{\bm{U}}$ in the subspaces. To the authors' knowledge, the original work for nonlinear manifolds can be traced back to \citet{qiao2012explicit}. Recently, numerous applications of nonlinear manifold MOR based on the single-field projection have been presented in \cite{rutzmoser2017generalization,benner2020operator,barnett2022quadratic,geelen2023operator,zhang2024nonlinear,geelen2024learning}. The objective of the nonlinear manifold MOR is to seek a mapping ${\mathbold{\Gamma}} (\bullet)$ that can approximate full states $\boldsymbol{U} \in \mathbb{R}^{n}$ via reduced states $\widetilde{\boldsymbol{U}} \in \mathbb{R}^{r}$ in a nonlinear manner as follows: 
	\begin{equation}
		\bm{U} \approx {\mathbold{\Gamma}} (\widetilde{\bm{U}}) := \underbrace{\mathbf{\Phi}_{\text{lin}} \, \widetilde{\bm {U}}}_{\text{linear}} + \underbrace{ {\mathbf{\bar \Phi}}_{\text{nonlin}} \, (\widetilde{\bm{U}} \otimes \widetilde{\bm{U}})}_{\text{nonlinear}},
		\label{n1}
	\end{equation}
	where $\mathbf{\Phi}_{\text{lin}} \in {\mathbb{R}}^{n \times r}$ and ${\mathbf{\bar \Phi}}_{\text{nonlin}} \in {\mathbb{R}}^{n \times q}$ denote linear and nonlinear projection matrices, respectively. $n$ is the number of degrees of freedom in a discretized equation system, $r$ denotes the selected column number of the global linear projection matrices. Note that the operator $\otimes$ represents the Kronecker product\footnote{The Kronecker product of a vector $\boldsymbol{U}$ with itself is defined as $\bm{U} \otimes \bm{U} = [u_1^2, u_1u_2, \cdots, u_1u_n,u_2u_1, u_2^2, \cdots, u_2u_n, \cdots u_n^2]^{\text{T}} \in \mathbb{R}^{n^2}$, where $\bm{U} = [u_1, u_2, \cdots u_n]^{\text{T}} \in \mathbb{R}^{n}$. When duplicate elements are removed, the dimension of the product becomes $\bm{U} \otimes \bm{U} \in \mathbb{R}^{n(n+1)/2}$.} between two vectors. $q$\footnote{Here, $q$ can be expressed as $q=r^2$ or $q=r(r+1)/2$, depending on the chosen form of the Kronecker product. In this study, $q$ is independent of $r$ due to the employment of $\bf{\Xi}$.} specifies the number of selected columns in the global nonlinear projection matrices. Here, unlike dynamic cases, the reference state $\bm{U}_{\text{ref}}$ (initial condition)  is not further discussed in the quasi-static case in this study, as shown in \cite{geelen2023learning,benner2023quadratic,jain2019hyper,jain2017quadratic}. Since the initial condition is addressed separately within the Newton iteration, the primary focus is placed on the unknown quantities.
	Hence, following the pioneering contributions by \citet{geelen2024learning,geelen2023learning,geelen2023operator} in nonlinear manifold modeling, \hyperref[n1]{Eq.\eqref{n1}} is further simplified by introducing an additional coefficient matrix $\mathbf{\Xi} \in \mathbb{R}^{q \times (p-1)r}$ together with a polynomial operator $\bm{g} (\bullet)$, thereby avoiding the large computational cost associated with the Kronecker product. $\mathbf{\Xi}$ plays a key role in weighting the nonlinear projection basis and the higher-order polynomials of the reduced states:
	\begin{equation}
		\bm{U} \approx {\mathbold{\Gamma}} (\widetilde{\bm{U}}) :=   \mathbf{\Phi}_{\text{lin}} \, \widetilde{\bm {U}} +  {\mathbf{\bar \Phi}}_{\text{nonlin}} \,  \mathbf{\Xi} \, \mathbold{g} (\widetilde{\bm{U}}).
		\label{n2}
	\end{equation}
	The operator $\bm{g}({\widetilde{\bm{U}}})$ for the reduced state is expressed with the selected polynomial order $p$ as: 
	\begin{equation}
		\bm{g}({\widetilde{\bm{U}}}) = \begin{pmatrix}
			\widetilde{\bm{U}}^2(t) \\
			\vdots \\
			\widetilde{\bm{U}}^p(t) \\
		\end{pmatrix}
			\label{n3}
		\end{equation}
		
		As outlined by \citet{volkwein2013proper}, \citet{ghavamian2017pod}, and \citet{zhang2025multi}, the key idea of reduced-order modeling is to minimize the discrepancy between the full-order states and their reduced representations. This process aims to seek a mapping ${\mathbold{\Gamma}}$, which can be achieved in terms of a linear or nonlinear approach. As $\mathbold{\Gamma}$ is determined, it can be used to perform forward and backward projections between the high-fidelity space and the reduced subspace, enabling the reduced equations to be efficiently solved within the subspace. The minimization process is expressed in \hyperref[n4]{Eq.\eqref{n4}}, where $r_A$ represents the rank of the snapshot matrix and plays a key role in defining the dimension of the objective function:
		\begin{equation}
			\mathbold{\Gamma}= \mathop {{\text{arg min}}}\limits_{\{ \mathbf{\Phi}_{\text{lin}}, \, {\mathbf{\bar \Phi}}_{\text{nonlin}} , \, \mathbf{\Xi}\} } {\sum\limits_{j = 1}^{{r_A}} {\left\|  {{ {\bm{U}}_j - \mathbf{\Phi}_{\text{lin}} \, \widetilde{\bm {U}}_j -  {\mathbf{\bar \Phi}}_{\text{nonlin}} \,  \mathbf{\Xi} \, \mathbold{g} (\widetilde{\bm{U}}_j) }} \right\| ^2 _2} }.
			\label{n4} 
		\end{equation}
		To facilitate the minimization of \hyperref[n4]{Eq.\eqref{n4}}, the error between the full-order (high-fidelity) states and their reduced representations is expressed using the Frobenius norm, denoted as $\mathcal{F}$. As demonstrated by \citet{geelen2023operator} as well as \citet{grussler2018low}, the regularized nuclear Frobenius norm is one of the most widely used for minimization problems, which aims to reduce the error while considering a huge amount of data. In this regard, $\mathcal{F}$ is expressed in terms of several quantities $\mathbf{\Phi}_{\text{lin}}$, ${\mathbf{\bar \Phi}}_{\text{nonlin}}$,  $\mathbf{\Xi}$, and $\widetilde{\bm{U}}_{\text{snap}}$ as:
		\begin{equation}
			\mathcal{F} \, ( \mathbf{\Phi}_{\text{lin}}, \, {\mathbf{\bar \Phi}}_{\text{nonlin}} , \, \mathbf{\Xi}, \, \widetilde{\bm{U}}_{\text{snap}})= \frac{1}{2} \, {\left\|  \bm{U}_{\text{snap}}  -  
				\begin{bmatrix}
					\mathbf{\Phi}_{\text{lin}} &  {\mathbf{\bar \Phi}}_{\text{nonlin}} 
				\end{bmatrix}
				\begin{bmatrix}
					\widetilde{\bm{U}}_{\text{snap}} \\
					\mathbf{\Xi} \, \bm {g} (\widetilde{\bm{U}}_{\text{snap}})
				\end{bmatrix}
				\right\| ^2 _F}.
			\label{n5}
		\end{equation}
		
		\noindent Here, ${{\bm{U}}_{\text{snap}}} = \left[ {{{\bm{U}}_1}, {{\bm{U}}_2}, \ldots {{\bm{U}}_{n_{\text{step}}} }} \right] \in {\mathbb{R}^{n \times n_{\text{step}}} }$ is a snapshot matrix consisting of all converged nodal solutions of the precomputed full-order simulations. $n_{\text{step}}$ denotes the step number of performed full-order simulations. In \hyperref[n5]{Eq.\eqref{n5}}, the projection matrices need to be orthonormal as $\left( \mathbf{\Phi}_{\text{lin}}^{\text{T}}, \, {\mathbf{\bar \Phi}}_{\text{nonlin}}^{\text{T}} \right)^ {\text{T}} \,\left( \mathbf{\Phi}_{\text{lin}}, \, {\mathbf{\bar \Phi}}_{\text{nonlin}}\right)  = \bm{I}_{(r+q)}$, where $\bm{I}_{(r+q)}$ is the identity matrix with the dimension of $\mathbb{R}^{(r+q) \times (r+q)}$. This constraint forms a smooth submanifold of the subspace ${\mathbb{R}}^{n\times (r+q)}$, namely the Stiefel manifold \cite{geelen2024learning}. However, solely employing the Frobenius norm in \hyperref[n5]{Eq.\eqref{n5}} will lead to an overfitting problem in the minimization process \cite{geelen2023operator,geelen2023learning,geelen2024learning}, especially when a huge amount of data is involved. To address this issue, an additional regularization in terms of the coefficient matrix $\mathbf{\Xi}$ is employed with a regularization parameter $\gamma \geq 0$ as: 
		\begin{equation}
			\left\lbrace \mathbf{\Phi}_{\text{lin}}, \, {\mathbf{\bar \Phi}}_{\text{nonlin}} , \, \mathbf{\Xi}, \, \widetilde{\bm{U}}_{\text{snap}} \right\rbrace  = \mathop {{\text{arg min}}} \limits_{\{ \mathbf{\Phi}_{\text{lin}}, \, {\mathbf{\bar \Phi}}_{\text{nonlin}} , \, \mathbf{\Xi}, \, \widetilde{\bm{U}}_{\text{snap}}\} } 
			\Big[  {\mathcal{F}  \, ( \mathbf{\Phi}_{\text{lin}}, \, {\mathbf{\bar \Phi}}_{\text{nonlin}} , \, \mathbf{\Xi}, \, \widetilde{\bm{U}}_{\text{snap}} )}
			+ \frac{\gamma}{2} \left\| \mathbf{\Xi} \right\|^2_F \Big].
			\label{n6}
		\end{equation}
		
		\subsection{Proper orthogonal decomposition-based nonlinear manifolds}
		There are many different ways to realize a minimization for \hyperref[n6]{Eq.\eqref{n6}} in order to obtain $\mathbf{\Phi}_{\text{lin}}$, ${\mathbf{\bar \Phi}}_{\text{nonlin}}$, and $\mathbf{\Xi}$, such as, by means of POD \cite{geelen2024learning,barnett2022quadratic,geelen2023learning} or data-driven \cite{geelen2023operator,geelen2023learning,prume2025direct}, and operator inference \cite{geelen2023operator,gruber2023canonical,gosea2023learning} based methods. 
		Here, a classical POD-based nonlinear manifold method is employed for two reasons. On the one hand, this is due to its widespread usage and its stable characteristics, see \cite{ghavamian2017pod,radermacher2016pod,zhang2025multi}. On the other hand, the employment of the POD method leads to a comprehensible expression for the coefficient matrices, avoiding a black-box approximation.
		In this context, the nodal snapshot matrix $\bm{U}_{\text{snap}}$ is decomposed via a \textit{thin} singular value decomposition. 
		\begin{equation}
			{\bm{U} _{{\text{snap}}}} = {\bm{S} \, \bm{\Sigma}} \,{{\bm{V}}^{\text{T}}} \quad \text{with} \quad
			{{\bm{S}}} = \Big[ { \underbrace{{{\bm{S}}_1}, \ldots {{\bm{S}}_r}}_{\mathbf{\Phi}_{\text{lin}}}, \underbrace{\ldots \bm{S}_{r+q}}_{ {\mathbf{\bar \Phi}}_{\text{nonlin}}}, \ldots {{\bm{S}}_{n_{\text{step}}} }} \Big] \in {\mathbb{R}^{n \times {n_{\text{step}}} }}
			\label{n7}
		\end{equation}
		\noindent Here, $\bm{S}$ denotes a left orthonormal basis which contains the left singular vectors. Subsequently, the linear and nonlinear projection matrices are obtained by truncating $\bm{S}$ with a user-defined accuracy. Having $\mathbf{\Phi}_{\text{lin}}$ and ${{\mathbf{\bar \Phi}}_{\text{nonlin}}}$ at hand, the next step is to determine the coefficient matrix $\mathbf{\Xi}$ through the minimization of \hyperref[n6]{Eq.\eqref{n6}}. In addition, the reduced snapshot matrix is expressed as $\widetilde{\bm{U}}_{\text{snap}} = \mathbf{\Phi}_{\text{lin}}^{\rm T} \, {\bm {U}}_{\text{snap}}$. For brevity, the objective function of the regularized Frobenius norm is defined as $\mathcal{M}$ and it follows:
		\begin{subequations}
			\begin{align}
				\mathcal{M} & = {\mathcal{F}  \, ( \mathbf{\Phi}_{\text{lin}}, \, {\mathbf{\bar \Phi}}_{\text{nonlin}} , \, \mathbf{\Xi}, \, \widetilde{\bm{U}}_{\text{snap}} )}
				+ \frac{\gamma}{2} \, \left\| \mathbf{\Xi} \right\|^2_F  \\
				\label{n8-b}
				& = \frac{1}{2} \, \| {{{\bm{U}}_{{\rm{snap}}}} - {{\mathbf{\Phi }}_{{\rm{lin}}}} \, {{{\bm{\widetilde U}}}_{{\rm{snap}}}} - {{{{\mathbf{\bar \Phi}  }}}_{{\rm{nonlin}}}} \, {\mathbf{\Xi }} \, \underbrace {{\bm{g}}({{{\bm{\widetilde U}}}_{{\rm{snap}}}})}_{:=\bf{W}}} \|_F^2 + \frac{\gamma }{2}\left\| {\mathbf{\Xi }} \right\|_F^2 \\
				& = \frac{1}{2} \, \left\| {\left( {{{\bm{I}}_n} - {{\mathbf{\Phi }}_{{\rm{lin}}}} \, {\mathbf{\Phi }}_{{\rm{lin}}}^{\rm{T}}} \right){{\bm{U}}_{{\rm{snap}}}} - {{{\mathbf{\bar \Phi }}}_{{\rm{nonlin}}}} \, {\mathbf{\Xi} \, \bf W}} \right\|_F^2 + \frac{\gamma }{2}\left\| {\mathbf{\Xi }} \right\|_F^2 . \label{n80}
			\end{align}
			\label{n8}
		\end{subequations}
		The Frobenius norm of an arbitrary matrix $\bm{X}$ can be calculated as $\left\| \bm{X} \right\|_F^2 := \text{tr} (\bm{X}^{\text{T}} \, \bm{X}) =  \sqrt {{\sum_{i=1}^{r_{A}}} \sigma_i^2 }$, where $r_A$ is the rank of the matrix $\boldsymbol{X}$ and $\sigma_i$ is $i$-th singular value. Hence, using the trace operator $\text{tr}(\bullet)$, the regularized Frobenius norm $\mathcal{M}$ in \hyperref[n80]{Eq.\eqref{n80}} can be expressed as:
		\begin{equation}
			\mathcal{M} = \frac{1}{2} \, {\rm{tr}}\Big[ {{{\left( {\left( {{{\bm{I}}_n} - {{\bf{\Phi }}_{{\rm{lin}}}} \, {\bf{\Phi }}_{{\rm{lin}}}^{\rm{T}}} \right){{\bm{U}}_{{\rm{snap}}}} - {{{\bf{\bar \Phi }}}_{{\rm{nonlin}}}} \, {\bf{\Xi \, W}}} \right)}^{\rm{T}}}\left( {\left( {{{\bm{I}}_n} - {{\bf{\Phi }}_{{\rm{lin}}}} \, {\bf{\Phi }}_{{\rm{lin}}}^{\rm{T}}} \right){{\bm{U}}_{{\rm{snap}}}} - {{{\bf{\bar \Phi }}}_{{\rm{nonlin}}}} \, {\bf{\Xi \, W}}} \right)} \Big] + \frac{\gamma }{2} \, {\rm{tr}}({{\bf{\Xi }}^{\rm{T}}} \, {\bf{\Xi }}).
			\label{n9}
		\end{equation}
		To minimize $\mathcal{M}$, its partial derivative with respect to the coefficient matrix $\bf \Xi$ needs to be computed. With 
		\begin{equation}
			{\left( {{{\bm{I}}_n} - {{\bf{\Phi }}_{{\rm{lin}}}} \, {\bf{\Phi }}_{{\rm{lin}}}^{\rm{T}}} \right)^{\rm{T}}} 
			= {{\bm{I}}_n} - {{\bf{\Phi }}_{{\rm{lin}}}} \, {\bf{\Phi }}_{{\rm{lin}}}^{\rm{T}},
			\label{n10-1}
		\end{equation}
		where $\bm{I}_n \in {\mathbb{R}^{n \times n} }$ denotes the identity matrix, this leads to
		\begin{equation}
			\frac{{\partial \mathcal{M}}}{{\partial {\bf{\Xi }}}} =  - {\bm{U}}_{{\rm{snap}}}^{\rm{T}}\left( {{{\bm{I}}_n} - {{\bf{\Phi }}_{{\rm{lin}}}} \, {\bf{\Phi }}_{{\rm{lin}}}^{\rm{T}}} \right){{{\bf{\bar \Phi }}}_{{\rm{nonlin}}}} \, {\bf{W}} + {{\bf{W}}^{\rm{T}}} \, {{\bf{\Xi }}^{\rm{T}}}\underbrace {{\bf{\bar \Phi }}_{{\rm{nonlin}}}^{\rm{T}} \, {{{\bf{\bar \Phi }}}_{{\rm{nonlin}}}}}_{\bm{I}_{(p-1)q}} {\bf{W}} + \gamma \, {\bf{\Xi }}. 
			\label{eqM1}
		\end{equation}
		By setting Eq.\eqref{eqM1} to $\bm 0$, the following equation is obtained:
		\begin{subequations}
			\begin{gather}
				{{{\bf{\bar \Phi }}}_{{\rm{nonlin}}}}\left( {{{\bm{I}}_n} - {{\bf{\Phi }}_{{\rm{lin}}}} \, {\bf{\Phi }}_{{\rm{lin}}}^{\rm{T}}} \right) {\bm{U}}_{{\rm{snap}}}^{\rm{T}} \, {\bf{W}} = ({{\bf{W}}^{\rm{T}}} \, {\bf{W}} + \gamma \, {{\bm{I}}_{(p - 1) r}}) \, {\bf{\Xi }}. 
			\end{gather}
			\label{n10-3}
		\end{subequations}
		\noindent Due to the convexity of the Frobenius norm \cite{powell2004least,grussler2018low}, the global minimum is guaranteed using \hyperref[eqM1]{Eq.\eqref{eqM1}}. Finally, the coefficient matrix ${\bf{\Xi}}$ is obtained as:
		\begin{equation}
			{\bf{\Xi }} = {\bf{\bar \Phi }}_{{\rm{nonlin}}}^{\rm{T}} \, {\left( {{{\bm{I}}_n} - {{\bf{\Phi }}_{{\rm{lin}}}} \, {\bf{\Phi }}_{{\rm{lin}}}^{\rm{T}}} \right){{\bm{U}}_{{\rm{snap}}}}} \,  {\bf{W}}^{\text{T}} \left( {{\bf{W}} \, {{\bf{W}}^{\rm{T}}} + \gamma \, {{\bm{I}}_{(p - 1) r}}} \right)^ {-1}.
			\label{n10}
		\end{equation} 
		
		\subsection{Multi-field decomposed nonlinear manifold reduction}
		The key point of the multi-field decomposed MOR approach is to decompose the nodal solution vector into three individual fields, see \hyperref[eq40]{Eq.\eqref{eq40}}, namely the displacement field $\bm{{u}}_{u}$, damage field $\bm{{u}}_{\bar{D}}$, and temperature field $\bm{{u}}_{\theta}$. The motivation for additively decomposing the full field into multiple individual fields is the strong nonlinearity of the formulation and its impact on developing a suitable reduced-order model, see Eqs.\eqref{eq3}, \eqref{eqA3}, and \eqref{eqA7}. For reduced-order modeling, it is challenging to capture the strongly nonlinear softening behavior in low-dimensional subspaces using a single-field projection-based method. This problem was also observed by \citet{kerfriden2011bridging,kerfriden2013partitioned}, \citet{selvaraj2024adaptive}, \citet{mishra2024enhanced}, and \citet{zhang2025multi}. To overcome this problem, the multi-field decomposed MOR approach is employed (see \cite{zhang2025multi}) by splitting the solution vector associated with a single node as follows: 
		\begin{equation}
			{{\bm{u}}} = \left( {\begin{array}{*{20}{c}}
					u \\ 
					v \\ 
					w \\ 
					{\bar D} \\ 
					\theta  
			\end{array}} \right) = \underbrace {\left( {\begin{array}{*{20}{c}}
						u \\ 
						v \\ 
						w \\ 
						0 \\ 
						0 
				\end{array}} \right)}_{{{{\boldsymbol{u}}}_u}} + \underbrace {\left( {\begin{array}{*{20}{c}}
						0 \\ 
						0 \\ 
						0 \\ 
						{\bar D} \\ 
						0 
				\end{array}} \right)}_{{{\boldsymbol{u}}_{\bar D}}} + \underbrace {\left( {\begin{array}{*{20}{c}}
						0 \\ 
						0 \\ 
						0 \\ 
						0 \\ 
						\theta  
				\end{array}} \right)}_{{{\boldsymbol{u}}_\theta }}.
			\label{eq40}
		\end{equation}
		
\begin{figure}[!ht]
			\centering
			\includegraphics[width=18cm]{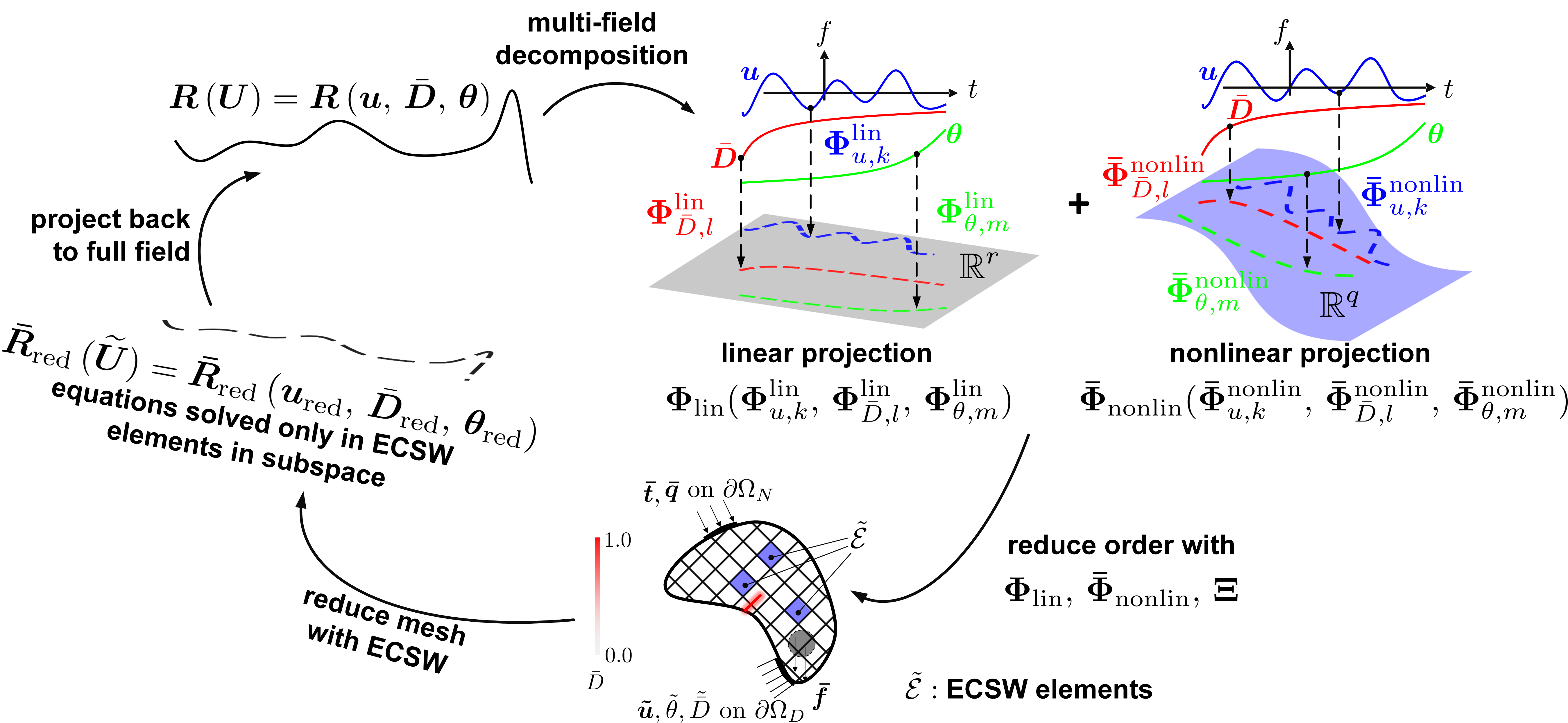}
			\caption{Schematic illustration of the multi-field decomposed nonlinear manifold MOR process and its corresponding projections. The solid and dashed lines represent the functions of the full-order model (full space) and the reduced-order model (reduced subspace), respectively. The full-order residual vector $\bm{R}$ is decomposed and projected onto both linear and nonlinear subspaces. $\bf{\Phi}_{\text{lin}}$ denotes the linear projection matrix which comprises individual projection matrices $\mathbf{\Phi}_{u,k}^{\text{lin}}$, $\mathbf{\Phi}_{\bar{D},l}^{\text{lin}}$, and $\mathbf{\Phi}_{\theta,m}^{\text{lin}}$ for the displacement, nonlocal damage, and temperature fields, respectively. $\bf{\bar \Phi}_{\text{nonlin}}$ represents the nonlinear projection that corresponds to the nonlinear terms. $\bar{\bm{R}}_{\text{red}}$ denotes the reduced residual vector for the ECSW selected elements $\mathcal{\widetilde{E}}$ (see \autoref{sec:inf_hyper}) in terms of the reduced quantities $\bm{u}_{\text{red}}$, $\bar {\bm{D}}_{\text{red}}$, and $\bm{\theta}_{\text{red}}$.}
			\label{fig1}
		\end{figure}

		The objective of the multi-field decomposed nonlinear manifold MOR is to seek the low-dimensional orthonormal linear projection matrix $\mathbf{\Phi}_{\text{lin}} \in \mathbb{R}^{n \times r}$ and nonlinear projection matrix ${\mathbf{\bar \Phi}}_{\text{nonlin}} \in \mathbb{R}^{n \times q}$, which project the global residual vector $\bm{R}$ and stiffness matrix $\bm{K}$ onto reduced subspaces, as defined in \hyperref[eq35]{Eq.\eqref{eq35}} and schematically illustrated in \hyperref[fig1]{Fig.1}. These projections are realized not only in a linear and nonlinear manner but also individually for the displacement, nonlocal damage, and temperature fields.
		
		Hence, the global solution vector is decomposed into a displacement field ${{\bm{U}}_u} = {\left[ {{\bm{ u}}_u^1, \, {\bm{ u}}_u^2, \, \ldots  {\bm{ u}}_u^{\lambda}} \right]^{\text{T}}}$, a nonlocal damage field ${{\bm{U}}_{\bar D}} = {\left[ {{\bm{ u}}_{\bar D}^1, \, {\bm{ u}}_{\bar D}^2, \, \ldots   \, {\bm{ u}}_{\bar D}^{\lambda}} \right]^{\text{T}}}$, and a temperature field ${{\bm{U}}_\theta } = {\left[ {{\bm{ u}}_\theta ^1, \, {\bm{ u}}_\theta ^2, \,  \ldots  \, {\bm{ u}}_\theta ^{\lambda}} \right]^{\text{T}}}$, respectively. Here, the symbol $\lambda$ denotes the total number of nodes in the unreduced discretization $\mathcal{E}$. The decomposition of the solution vector of a single node $\bm{u}$, as presented in \hyperref[eq40]{Eq.\eqref{eq40}}, enables the decoupling of distinct field quantities, thereby allowing separate projection operations for each component. From an implementation perspective, filling the corresponding vectors with zeros ensures consistency in the dimensions of vectors and matrices associated with different field quantities within the reduced-order modeling framework \cite{zhang2025multi}. As a result, decomposed snapshots are defined as ${\bm{U}}_{{\text{snap}}}^u$, ${\bm{U}}_{{\text{snap}}}^{\bar D}$, and ${\bm{U}}_{{\text{snap}}}^\theta$ corresponding to the displacement, nonlocal damage, and temperature fields as:
		\begin{equation}
			{\bm{U}}_{{\text{snap}}}^u = \left[ {{\bm{U}}_u^1, \ldots {\bm{U}}_u^k, \ldots {\bm{U}}_u^{n_{\text{step}}}} \right], \quad {\bm{U}}_{{\text{snap}}}^{\bar D} = \left[ {{\bm{U}}_{\bar D}^1, \ldots {\bm{U}}_{\bar D}^l, \ldots {\bm{U}}_{\bar D}^{n_{\text{step}}}} \right], \quad {\bm{U}}_{{\text{snap}}}^\theta  = \left[ {{\bm{U}}_\theta ^1, \ldots {\bm{U}}_\theta ^m, \ldots {\bm{U}}_\theta ^{n_{\text{step}}}} \right].
			\label{eq41}
		\end{equation}
		\noindent The indices $k$, $l$, and $m$ denote independently selected truncation numbers corresponding to the number of displacement, nonlocal damage, and temperature modes, respectively, thereby defining the dimensions of the projection matrices for each field. By employing a \textit{thin} singular value decomposition $\mathbf{SVD} (\bullet)$ for the different snapshot matrices, one obtains:
		\begin{equation}
			\left\{ {{{\bm{S}}_u}, \, {{\bm{\Sigma }}_u}, \, {\bm{V}}_u^{\text{T}}} \right\} = {\mathbf{SVD}} ( {{\bm{U}}_{{\text{snap}}}^u} ), \quad
			\left\{ {{{\bm{S}}_{\bar D}}, \, {{\bm{\Sigma }}_{\bar D}}, \, {\bm{V}}_{\bar D}^{\text{T}}} \right\} = {\mathbf{SVD}}  ( {{\bm{U}}_{{\text{snap}}}^{\bar D}} ), \quad
			\left\{ {{{\bm{S}}_\theta }, \, {{\bm{\Sigma }}_\theta }, \, {\bm{V}}_\theta ^{\text{T}}} \right\} = {\mathbf{SVD}} ( {{\bm{U}}_{{\text{snap}}}^\theta } ).
			\label{eq42}
		\end{equation}
		Subsequently, the projection matrices of the displacement field $\mathbf{\Phi}_{u, \, k}^{\text{lin}}$, nonlocal damage field $\mathbf{\Phi}_{\bar{D}, \, l}^{\text{lin}}$, and temperature field $\mathbf{\Phi}_{\theta, \, m}^{\text{lin}}$ are computed by truncating $\bm{S}_u$, $\bm{S}_{\bar {D}}$, and $\bm{S}_{\theta}$ at $k$, $l$, and $m$, respectively:
		\begin{equation}
			\begin{split}
				{{\bm{S}}_u} &= \Big[\underbrace {{\bm{S}}_u^1, \cdots {\bm{S}}_u^k}_{{\bf{\Phi }}_{u,k}^{{\rm{lin}}}},\underbrace { \cdots {\bm{S}}_u^{k + q}}_{{\bf{\bar \Phi }}_{u,k}^{{\rm{nonlin}}}}, \cdots {\bm{S}}_u^{n_{\text{step}}} \Big], \quad
				{{\bm{S}}_{\bar D}} = \Big[\underbrace {{\bm{S}}_{\bar D}^1, \cdots {\bm{S}}_{\bar D}^l}_{{\bf{\Phi }}_{\bar D,l}^{{\rm{lin}}}},\underbrace { \cdots {\bm{S}}_{\bar D}^{l + q}}_{{\bf{\bar \Phi }}_{\bar D,l}^{{\rm{nonlin}}}}, \cdots {\bm{S}}_{\bar D}^{n_{\text{step}}} \Big], \\
				{{\bm{S}}_\theta } &= \Big[\underbrace {{\bm{S}}_\theta ^1, \cdots {\bm{S}}_\theta ^m}_{{\bf{\Phi }}_{\theta ,m}^{{\rm{lin}}}},\underbrace { \cdots {\bm{S}}_\theta ^{m + q}}_{{\bf{\bar \Phi }}_{\theta ,m}^{{\rm{nonlin}}}}, \cdots {\bm{S}}_\theta ^{n_{\text{step}}} \Big].
			\end{split}
			\label{eq43}
		\end{equation}
		The nonlinear components ${{{\bf{\bar \Phi}}} }_{u,k}^{\text{nonlin}}$, ${\bf{\bar{{\Phi}}} }_{\bar{D},l}^{\text{nonlin}}$, and ${\bf{\bar{{\Phi}}} }_{\theta,m}^{\text{nonlin}}$ are obtained as expressed in \hyperref[eq43]{Eq.\eqref{eq43}}. For simplicity and consistency, the number of nonlinear modes is uniformly selected as $q$ across the different physical fields. A similar strategy for constructing the nonlinear projection basis has been adopted in prior studies \cite{cohen2023nonlinear,geelen2024learning,geelen2023learning}. Consequently, the global linear projection matrix ${\mathbf{\Phi }}_{\text{lin}} \in {\mathbb{R}^{n \times r}}$ and the global nonlinear projection matrix ${{\mathbf{\bar \Phi }}}_{\text{nonlin}} \in {\mathbb{R}^{n \times 3q}}$ are obtained using the mode numbers $k$, $l$, and $m$ associated with the three respective fields as:
		\begin{equation}
			{{\bf{\Phi }}_{{\rm{lin}}}} = \left[ {{\bf{\Phi }}_{u,k}^{{\rm{lin}}}, \, {\bf{\Phi }}_{\bar D,l}^{{\rm{lin}}}, \, {\bf{\Phi }}_{\theta ,m}^{{\rm{lin}}}} \right], \quad 
			{{\bf{\bar \Phi }}_{{\rm{nonlin}}}} = \left[ {{\bf{\bar \Phi }}_{u,k}^{{\rm{nonlin}}}, \, {\bf{\bar \Phi }}_{\bar D,l}^{{\rm{nonlin}}}, \, {\bf{\bar \Phi }}_{\theta ,m}^{{\rm{nonlin}}}} \right].
			\label{eq44}
		\end{equation}
		
		To ensure a sufficient reduction, the total number of linear $r$ and nonlinear $q$ modes should satisfy $r \ll n$ and $q \ll n$, where $r=k + l + m$. Corresponding to the multi-field decomposed linear and nonlinear projection matrices, the decomposed coefficient matrices for the displacement, nonlocal damage, and temperature fields are denoted as $\mathbf{{\Xi}}_u$, $\mathbf{{\Xi}}_{\bar{D}}$, and $\mathbf{{\Xi}}_{\theta}$, respectively. Accordingly, the objective function is expressed in terms of additional penalty parameters ${\gamma_u}$, ${\gamma_{\bar{D}}}$, and ${\gamma_{\theta}}$ applied to the corresponding fields as follows:
		\begin{equation}
			{\bf{\Xi}} = \mathop{\text{arg min}} \limits_{\{{{\bf{\Xi}}_u}, \, {{\bf{\Xi}}_{\bar{D}}}, \, {{\bf{\Xi}}_{\theta}}\}} \left[ 
			\mathcal{F} \, ( \mathbf{\Phi}_{\text{lin}}, \, {\mathbf{\bar\Phi}}_{\text{nonlin}} , \, \mathbf{\Xi}, \, \widetilde{\bm{U}}_{\text{snap}} ) 
			+ \frac{\gamma_u}{2} \left\|   {\bf \Xi}_u  \right\| _F^2 
			+ \frac{\gamma_{\bar{D}}}{2} \left\|  {\bf \Xi}_{\bar D}  \right\| _F^2
			+ \frac{\gamma_{\theta}}{2} \left\|  {\bf \Xi}_{\theta}  \right\| _F^2
			\right].
			\label{n11}
		\end{equation}
		
		\noindent Here, the global coefficient matrix ${\bf{\Xi}} \in \mathbb{R}^{3q \times (p-1)r}$ is expressed as a block diagonal matrix composed of the contributions from different fields as:
		\begin{equation}
			{\bf{\Xi }} = \begin{pmatrix}
				{{{\bf{\Xi }}_u}}&{\bf{0}}&{\bf{0}}\\
				{\bf{0}}&{{{\bf{\Xi }}_{\bar D}}}&{\bf{0}}\\
				{\bf{0}}&{\bf{0}}&{{{\bf{\Xi }}_\theta }}
			\end{pmatrix}.
			\label{n14}
		\end{equation}
		
		By decomposing \hyperref[n11]{Eq.\eqref{n11}} into three individual fields corresponding to displacement, nonlocal damage, and temperature, the associated objective function for minimization is expressed as:
		\begin{equation}
			\begin{split}
				{\bf{\Xi}} =  \mathop{\text{arg min}} \limits_{\{{{\bf{\Xi}}_u}, \, {{\bf{\Xi}}_{\bar{D}}}, \, {{\bf{\Xi}}_{\theta}}\}} \Big[ & \underbrace{\mathcal{F}_u \, (\mathbf{\Phi}_{u,k}^{\text{lin}}, \, \mathbf{\bar \Phi}_{u,k}^{\text{nonlin}}, \, \mathbf{\Xi}_u, \, \bm{\widetilde U}_{\text{snap}}^u) + \frac{\gamma_u}{2} \, \|\mathbf{\Xi}_u\|_F^2}_{\text{displacement field}} 
				+ \underbrace{\mathcal{F}_{\bar D} \, (\mathbf{\Phi}_{\bar D,l}^{\text{lin}}, \, \mathbf{\bar \Phi}_{\bar D,k}^{\text{nonlin}}, \, \mathbf{\Xi}_{\bar D}, \, \bm{\widetilde U}_{\text{snap}}^{\bar D}) + \frac{\gamma_{\bar D}}{2} \, \|\mathbf{\Xi}_{\bar D}\|_F^2}_{\text{nonlocal damage field}} \\
				&+ \underbrace{\mathcal{F}_\theta \, (\mathbf{\Phi}_{\theta,k}^{\text{lin}}, \, \mathbf{\bar \Phi}_{\theta,k}^{\text{nonlin}}, \, \mathbf{\Xi}_\theta, \, \bm{\widetilde U}_{\text{snap}}^\theta) + \frac{\gamma_\theta}{2} \, \|\mathbf{\Xi}_\theta\|_F^2}_{\text{temperature field}} \Big].
			\end{split}
			\label{n12}
		\end{equation}
		
		\noindent Compared with the merged minimization approach presented in \cite{geelen2023learning,geelen2023operator,geelen2024learning}, a scaling issue can arise particularly in the case of multiphysics problems, as noted in \cite{washabaugh2016faster,lindsay2022preconditioned,parish2023impact}. To address this issue and build upon prior work on multi-field decomposed reduced-order modeling \cite{zhang2025multi}, the objective function in \hyperref[n6]{Eq.\eqref{n6}} is newly decomposed into three separate fields, as shown in Eqs.\eqref{n11} and \eqref{n12}. This strategy enhances the reduction performance since the coefficient matrices can be individually tuned for the displacement, nonlocal damage, and temperature fields. Accordingly, these matrices can be similarly derived as in \hyperref[n10]{Eq.\eqref{n10}}:
		\begin{subequations}
			\begin{align}
				{{\bf{\Xi }}_u} &= {( {{\bf{\bar \Phi }}_{u,k}^{{\rm{nonlin}}}} )^{\rm{T}}} \, \left( {{{\bm{I}}_k} - {\bf{\Phi }}_{u,k}^{{\rm{lin}}} \, {{\left( {{\bf{\Phi }}_{u,k}^{{\rm{lin}}}} \right)}^{\rm{T}}}} \right) \, {\bm{U}}_{{\rm{snap}}}^u \, {\bf{W}}_u^{\rm{T}} \, {\left( {{{\bf{W}}_u} \, {\bf{W}}_u^{\rm{T}} + {\gamma _u } \, {{\bm{I}}_{(p - 1)  k}}} \right)^{ - 1}}, \\
				{{\bf{\Xi }}_{\bar D}} &= {( {{\bf{\bar \Phi }}_{\bar D,l}^{{\rm{nonlin}}}} )^{\rm{T}}} \, \left( {{{\bm{I}}_l} - {\bf{\Phi }}_{\bar D,l}^{{\rm{lin}}} \, {{( {{\bf{\Phi }}_{\bar D,l}^{{\rm{lin}}}} )}^{\rm{T}}}} \right) \, {\bm{U}}_{{\rm{snap}}}^{\bar D} \, {\bf{W}}_{\bar D}^{\rm{T}} \, {\left( {{{\bf{W}}_{\bar D}} \, {\bf{W}}_{\bar D}^{\rm{T}} + {\gamma _{\bar D}} \, {{\bm{I}}_{(p - 1) l}}} \right)^{ - 1}}, \\
				{{\bf{\Xi }}_\theta } &= {( {{\bf{\bar \Phi }}_{\theta,m}^{{\rm{nonlin}}}} )^{\rm{T}}} \, \left( {{{\bm{I}}_m} - {\bf{\Phi }}_{\theta,m}^{{\rm{lin}}} \, {{( {{\bf{\Phi }}_{\theta,m}^{{\rm{lin}}}} )}^{\rm{T}}}} \right) \, {\bm{U}}_{{\rm{snap}}}^\theta \, {\bf{W}}_\theta ^{\rm{T}} \, {\left( {{{\bf{W}}_\theta } \, {\bf{W}}_\theta ^{\rm{T}} + {\gamma _\theta } \, {{\bm{I}}_{(p - 1) m}}} \right)^{ - 1}}.
			\end{align}
			\label{n13}
		\end{subequations}
		
		\subsection{Nonlinear manifold inferred hyperreduction}
		\label{sec:inf_hyper}
		In FE simulations, the residual equation involving the nonlinear residual vector $\bm{R}$ is typically linearized and solved via the Newton-Raphson method, as shown in \hyperref[eq35]{Eq.\eqref{eq35}}. In the context of reduced-order modeling, the primary computational cost can be attributed to the following cause:
		\begin{itemize}
			\item Evaluating the reduced equations in each global iteration scales with the dimension of the high-fidelity model rather than that of the reduced subspace. 
			This is particularly computationally expensive when assembling $\bm{K}$ in large-scale simulations, as it requires information from every integration point of the unreduced system. Consequently, this leads to a computational bottleneck, even when advanced MOR techniques are employed.
		\end{itemize}
		To overcome this computational bottleneck, hyperreduction techniques \cite{ryckelynck2005priori,farhat2014dimensional,farhat2015structure,an2008optimizing} offer an effective solution. The general idea is to evaluate only a few selected elements that are most relevant in the simulations, and compensate for the neglected \textit{virtual work} through the non-negative weighting coefficients. However, capturing the strong nonlinearity induced by local/nonlocal damage remains challenging for classical single-field projection methods, as was shown in prior studies \cite{oliver2017reduced,selvaraj2024adaptive,mishra2024enhanced,zhang2025multi}. Therefore, the ECSW-based hyperreduction method proposed by \citet{farhat2014dimensional,farhat2015structure} is integrated with the newly developed multi-perspective decomposed quadratic manifold approach proposed in this study. The ECSW method is selected among the various hyperreduction techniques (e.g., ECM\footnote{Empirical Cubature Method (ECM).} \cite{hernandez2017dimensional,lange2024monolithic,wulfinghoff2025empirically} or DEIM\footnote{Discrete Empirical Interpolation Method (DEIM).} \cite{radermacher2016pod,ghavamian2017pod,rutzmoser2018model}) in this study due to its already proven capability in large-scale simulations, ranging from solid to fluid mechanics \cite{farhat2014dimensional,tezaur2022robust,grimberg2020stability,barnett2022quadratic}. Using ECSW, the reduced residual vector $\bm{\bar R}_{\rm{red}}$, the reduced stiffness matrix $\bm{K}_{\rm{red}}$, as well as the stiffness matrix ${{{\bm{\bar K}}}_{{\rm{red}}}}$ coupled between linear and nonlinear projections are approximated as:
		\begin{equation}
			{{\bm{\bar R}}_{{\text{red}}}} \approx \sum_{e\in \mathcal{\widetilde E}} {\xi_e\, (\bm{L}_e  {\bf{\Phi}_{{\text{lin}}})^{\text{T}} }  \bm{R}_e}, \quad \bm{K}_{\text{red}}  \approx \sum_{e\in \mathcal{\widetilde E}} {\xi_e\, (\bm{L}_e  {\bf{\Phi}_{{\text{lin}}})^{\text{T}} }   \bm{K}^e} (\bm{L}_e   {\bf{\Phi}_{{\text{lin}}}}), \quad \bm{\bar K}_{\text{red}} \approx \sum_{e\in \mathcal{\widetilde E}} {\xi_e\, (\bm{L}_e   {\bf{\Phi}_{{\text{lin}}})^{\text{T}} }   \bm{K}^e  (\bm{L}_e   {\bf{\bar \Phi}_{{\text{nonlin}}}}) }.
			\label{eq45}
		\end{equation}
		Here, the number of selected elements $n_s$ in the hyperreduced mesh $\mathcal{\widetilde E}$ is much smaller than the number of elements $N_e$ in the original (unreduced) mesh $\mathcal{E}$, denoted as $n_s = | \mathcal{\widetilde E} |$ and $N_e = | \mathcal{E} |$ with $\mathcal{\widetilde E} \subseteq \mathcal{E}$ and $n_s \ll N_e$. As expressed in \hyperref[eq45]{Eq.\eqref{eq45}}, the stiffness matrix $\bm{\bar{K}}_{\text{red}}$ composed of linear and nonlinear projections is formulated as a Petrov-Galerkin projection of $\bm{K}$, due to the fact $\bf{\Phi}_{\text{lin}} \neq \bf{\bar \Phi}_{\text{nonlin}}$. In \hyperref[eq45]{Eq.\eqref{eq45}}, $\xi_e$ denotes a non-negative weighting coefficient associated with the element $e$, which compensates for the neglected elements to fulfill the equivalence of \textit{virtual work} in the discretized system. $\bm{L}_e \in \{0, 1\}^{{n_s} \times n}$ represents a Boolean matrix that localizes/selects the corresponding DOFs of an element $e$ from the global high-dimensional vector of unknowns. $\bm{K}^e$ and $\bm{R}_e$ are the stiffness matrix and residual vector of the selected element $e$ defined in \hyperref[eq32]{Eq.\eqref{eq32}}, respectively. Here, the weighting coefficients associated with the selected elements $\bm{\xi} = \{ \xi_1, \, \xi_2, \cdots, \, \xi_e \}$ are determined in an offline training stage at the element level. For a given element $e$, the reduced residual $\bm{c}_{je}$ at a certain time step $j$, along with its accumulated counterpart over time $\bm{d}_e$, is computed using the reconstructed nodal solution $\bm{U}_j'$, which is derived from the reduced state $\bm{\widetilde{U}}_j \in \mathbb{R}^r$. The reduced state $\bm{\widetilde{U}}_j$ is obtained by solving a nonlinear least-squares problem $\bm{\widetilde U}_j = \mathop {{\text{arg min}}}  \| \bm{\Gamma}(\bm{\widetilde U}_j) - \bm{U}_j \|^2_2$ \cite{barnett2022quadratic,jain2019hyper}. All reconstructed snapshots are thus given by $\bm{U}_j' = \bm{\Gamma}(\bm{\widetilde{U}}_j)$. Accordingly, the reduced residual $\bm{c}_{je}$ and the accumulated residual $\bm{d}_e$ can be expressed as follows:
		\begin{equation}
			\bm{c}_{je} = (\bm{L}_e {\bf{\Phi}}_{\text{lin}} ) ^{\text{T}} \bm{R}_e (\bm{U}'_{j}) , \quad \bm{d}_e = \sum_{e\in \mathcal{\widetilde E}} {\bm{c}_{je}} \quad \Rightarrow \quad \bm{C} = \begin{pmatrix}
				{\bm{c}}_{11} &  \cdots & {\bm{c}}_{1e} \\
				\vdots &  \ddots & \vdots \\
				{\bm{c}}_{j 1} & \cdots & {\bm{c}}_{j e}
			\end{pmatrix}, \quad 
			\bm{d} = \begin{pmatrix}
				\bm{d}_1\\
				\vdots \\
				\bm{d}_e 
			\end{pmatrix}.
			\label{eq45-1}
		\end{equation}
		Next, the constructed matrix $\bm{C} \in \mathbb{R}^{n \, n_{\text{step}} \times n_s}$, which collects the reduced residuals of each element over all time steps, should fulfill the \textit{virtual work} equivalence condition given by:
		\begin{equation}
			\bm{C} \, \bm{\xi} - \bm{d} = \bm{0} \quad \text{with} \quad \lbrace  \mathcal{\widetilde E}, \, \bm{\xi}  \rbrace = \mathop {{\text{arg min}}} \limits_{ \{ \mathcal{\widetilde E},  \, \bm{\xi} \}}  \lbrace \left\|  \bm{C} \, \bm{\xi} - \bm{d}  \right\|^2 _2 \leq \uptau \left\| \bm{d} \right\| ^2 _2, \, \bm{\xi} \ge \bm{0} \rbrace . 
			\label{eq45-2}
		\end{equation}
		Hence, the hyperreduced mesh $\mathcal{\widetilde{E}}$ and its associated weight coefficients $\bm{\xi}$ are finally being determined by solving \hyperref[eq45-2]{Eq.\eqref{eq45-2}} via a sparse non-negative least squares (sNNLS) algorithm \cite{lawson1995solving,chapman2017accelerated,rutzmoser2018model} using a threshold-based early termination criterion. When $\bm{\xi}=\bm{1} \in \mathbb{R}^{N_e}$, it means that all elements of the original mesh are selected ($\mathcal{\widetilde{E}}=\mathcal{E}$) with a weight value $\xi_e = 1$ for all elements $e$. The threshold parameter $\uptau$ is typically chosen to be small, satisfying $0<\uptau \ll 1$, with its specific value determined by the required accuracy of the user and the degree of nonlinearity of the underlying PDEs. As $\uptau$ decreases, more elements are sampled by the ECSW method to achieve the desired accuracy in reduced-order simulations. Conversely, if $\uptau$ is set too large, the number of selected elements becomes insufficient to accurately approximate the full-order solution.
		\begin{remark}
			$\bm{c}_{je}$ in \hyperref[eq45-1]{Eq.\eqref{eq45-1}} can be constructed using the reduced stiffness matrix $\bm{c}_{je}=\bm{T} \, (\bm{L}_e {\bf{\Phi_{\rm{lin}}}})^{\rm{T}} \bm{K}^e (\bm{L}_e {\bf{\Phi_{\rm{lin}}}}) \in \mathbb{R}^{r^2}$, where $\bm{T}$ denotes the row-wise vectorization operator that converts a $r \times r$ matrix into a column vector in $\mathbb{R}^{r^2}$, as proposed by \citet{tezaur2022robust}. This construction is particularly suitable for weakly nonlinear systems where the reduced dimension $r$ remains small (e.g., $r<20$). However, in the present setting, the number of retained linear modes is defined as $r=k+l+m$, which can lead to a rapid growth in $r^2$ depending on the choice of parameters $k$, $l$, and $m$. As a consequence, the dimension of $\bm{c}_{je}$ increases quadratically with $r$, which can easily lead to a significant computational burden.
		\end{remark}
		
		Next, the primary task of the Newton-Raphson method is to iteratively compute a suitable global solution vector $\bm{U}=[\bm{u}, \, \bm{\bar{D}}, \, \bm{\theta}]$ by solving the reduced equation ${\bm{\bar R}}_{\text{red}} \, ({{\bm{U}}_{i + 1}^j}) \approx \bm{0}$, see \hyperref[fig1]{Fig.1}. Here, $j$ and $i$ denote the time step and iteration number, respectively. Considering a nonlinear projection within the Newton-Raphson scheme \cite{barnett2022quadratic,jain2019hyper,rutzmoser2017generalization}, the first-order Taylor expansion of the residual in the reduced-order space can be expressed as:
		\begin{equation}
			\begin{gathered}
				{\bf{\Phi }}_{{\rm{lin}}}^{\rm{T}} \, {\bm{R}} \, ({{\bm{U}}_{i + 1}^j}) =  {\bf{\Phi }}_{{\rm{lin}}}^{\rm{T}} \,{\bm{R}} ( {{\bm{U}}_i^j}) +   {\bf{\Phi }}_{{\rm{lin}}}^{\rm{T}} \,
				\underbrace{
					\frac{\partial {{\bm{R}} ( {{\bm{U}}_{i}^j})} }{\partial {\bm{U}}^j_{i}}
				}_{\bm{K}} \,
				\underbrace{
					\frac{\partial \bm{U}^j_{i}}{\partial \Delta \widetilde{\bm{U}}^j_i}
				}_{\bm{\mathcal{H}}}
				\,\Delta \widetilde {\bm{U}}_{i + 1}^j \approx \bm{0},
			\end{gathered}
			\label{eq47-1}
		\end{equation}´
		where the global solution vector $\boldsymbol{U}^j_i$ and $\boldsymbol{\mathcal{H}}$ can be expressed as:
		\begin{subequations}
			\begin{gather}
				{\bm{U}}_{i}^j = {\bm{U}}_{i-1}^{j} + {{\bf{\Phi }}_{{\text{lin}}}} \, \Delta {{\bm{\widetilde U}}}_{i}^j + {{{\bf{\bar \Phi }}}_{{\rm{nonlin}}}} \, {\bf{\Xi}} \, {\bm{g}} ( {\Delta {{\bm{\widetilde U}}}_{i}^j}), \\ 
				\bm{\mathcal{H}} = {\bf{\Phi}}_{\text{lin}} + {{\bf{\bar \Phi}}}_{\text{nonlin}} \, {\bf{\Xi}} \, \underbrace{\text{diag}(2 \, {\Delta {{\bm{\widetilde U}}}_{i}^j}) }_{ \bm{\mathcal{W}} }.
			\end{gather}
			\label{eq47-2}
		\end{subequations} 
		\noindent By employing the linear and nonlinear projections as well as the ECSW method, the reduced equations are expressed as: 
		\begin{subequations}
			\begin{gather}
				{{\bf{\Phi}}^{\text{T}}_{\text{lin}}} \,{\bm{R}} \, ({{\bm{U}}_{i + 1}^j}) = {{\bf{\Phi}}^{\text{T}}_{\text{lin}}} \, {\bm{R}} ( {{\bm{U}}_i^j}) + {{\bf{\Phi}}^{\text{T}}_{\text{lin}}} \, {\bm{K}} \, ( 
				{\bf{\Phi}}_{\text{lin}} + {{\bf{\bar \Phi}}}_{\text{nonlin}} \, {\bf{\Xi}} \, \bm{\mathcal{W}} )\, {\Delta {{\bm{\widetilde U}}}_{i+1}^j}, \label{eq47a} \\
				\underbrace{
					{{\bf{\Phi}}^{\text{T}}_{\text{lin}}} \, {\bm{R}} ( {{\bm{U}}_i^j})}_{:\approx {\bm{\bar R}}^j_{i,\text{red}} } +
				\underbrace{{{\bf{\Phi}}^{\text{T}}_{\text{lin}}} \, {\bm{K}} \, \bf{\Phi}_{\text{lin}} }_{:\approx \bm{K}^j_{i, \text{red}}} \, \Delta {{\bm{\widetilde U}}}_{i + 1}^j + 
				\underbrace{{{\bf{\Phi}}^{\text{T}}_{\text{lin}}} \, {\bm{K}} \,{\bf{\bar \Phi}}_{\text{nonlin}} }_{:\approx {\bm{\bar K}}^j_{i, \text{red}}} \, {\bf{\Xi}} \, \bm{\mathcal{W}} \, {\Delta {{\bm{\widetilde U}}}_{i+1}^j} \approx \bm{0}, \label{eq47b-1} \\
				{\Delta {{\bm{\widetilde U}}}_{i + 1}^j} =  - {(\bm{K}^j_{i, \text{red}} + {\bm{\bar K}}^j_{i, \text{red}} \, {\bf{\Xi}} \, \bm{\mathcal{W}})^{-1}} \, {\bm{\bar R}}^j_{i, \text{red}} , \label{eq47c}\\
				{\bm{U}}_{i + 1}^j = {\bm{U}}_i^j +  {{\bf{\Phi }}_{{\text{lin}}}} \, \Delta {{\bm{\widetilde U}}}_{i + 1}^j + {{{\bf{\bar \Phi }}}_{{\text{nonlin}}}} \, {\bf{\Xi}} \, \bm{g} (\Delta {{\bm{\widetilde U}}}_{i + 1}^j), \label{eq47d}\\
				\left\| {{{\bm{\bar R}}_{\text{red}}} ( {{\bm{U}}_{i + 1}^j})} \right\| \leqslant tol, \label{eq47e}\\
				i \leftarrow i + 1. \label{eq47f}
			\end{gather}
			\label{eq47}
		\end{subequations}
		\noindent As observed in \hyperref[eq47]{Eq.\eqref{eq47}}, utilizing the inverse of the reduced stiffness matrices in \hyperref[eq47c]{Eq.\eqref{eq47c}} leads to a noticeable reduction in computational complexity due to the lower dimension of both $\bm{K}_{\text{red}}$ and $\bm{K}$ compared to those of full-order simulations. Note that \hyperref[eq47]{Eq.\eqref{eq47}} incorporates not only the traditional linear term $\bm{K}_{\text{red}}$ but also the additional enrichment from the quadratic approximation terms: $\bm{\bar{K}}_{\text{red}}$, ${\bf{\Xi}}$, and $\bm{\mathcal{W} }$. These quantities are crucial to achieve a higher accuracy compared to the classical POD throughout the iteration process. When the norm of the reduced residual ${\bm{\bar R}}_{\text{red}}$ falls below a predefined tolerance ($tol$), the converged increment in the reduced-order solution $\Delta \bm{\widetilde{U}}$ is deemed found. Simultaneously, the reduced solution is projected back into the full-order equation system to update the solution vector $\bm{U}$ throughout the iteration process.
		
		\subsection{Multi-state and multi-field decomposed nonlinear manifold MOR for prediction}
		\label{sec:multi_pers}	
\begin{figure}[!ht]
			\centering
			\includegraphics[width=18cm]{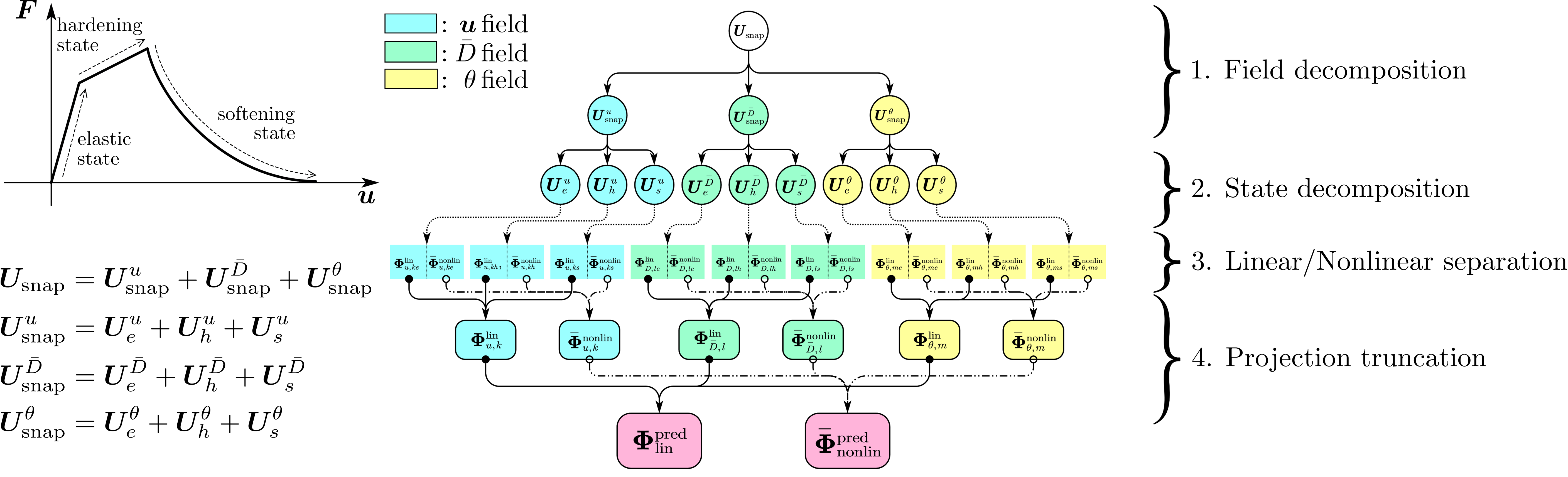}
			\caption{Schematic illustration of the multi-field and multi-state decomposed nonlinear manifold MOR process proposed in this work and the corresponding projections.}
			\label{fig2}
		\end{figure}

		To assess the predictive performance of the proposed MOR approach in the numerical examples later on, the \textit{predictive} linear and nonlinear projection matrices $\mathbf{\Phi}^{\text{pred}}_{\text{lin}}$ and ${\mathbf{\bar \Phi} }^{\text{pred}}_{\text{nonlin}}$ are constructed using the multi-field and multi-state decomposition method, see \hyperref[fig2]{Fig.2}. 
		Concretely speaking, these matrices are built by taking into account additional snapshots from full-order simulations. These simulations are conducted with deviations in the elastic parameters $\mu$ and $\Lambda$ of  $\pm d\%$, where $d$ is a chosen value. Importantly, only half of the snapshots are collected to maintain dimensional consistency with \hyperref[eq41]{Eq.\eqref{eq41}} in terms of the number of snapshots. The resulting decomposed snapshots for a positive deviation ($+d\%$) are separated into displacement, nonlocal damage, and temperature fields according to:
		\begin{equation}
			{\bm{U}}_{{\text{snap,}+d}}^u = \Big[ {{\bm{U}}_u^1, \, {\bm{U}}_u^3, \ldots {\bm{U}}_u^{{n_{\text{step}}}-1}} \Big], \quad {\bm{U}}_{{\text{snap,}+d}}^{\bar D} = \Big[ {{\bm{U}}_{\bar D}^1, \, {\bm{U}}_{\bar D}^3, \ldots {\bm{U}}_{\bar D}^{{n_{\text{step}}}-1}} \Big], \quad
			{\bm{U}}_{{\text{snap,}+d}}^\theta  = \Big[ {{\bm{U}}_\theta ^1, \, {\bm{U}}_\theta ^3, \ldots {\bm{U}}_\theta ^{{n_{\text{step}}}-1}} \Big].
			\label{eq44-1}
		\end{equation}
		The snapshots for the negative deviation ($-d\%$) are obtained analogously to \hyperref[eq44-1]{Eq.\eqref{eq44-1}} as ${\bm{U}}_{{\text{snap,}-d}}^u$, ${\bm{U}}_{{\text{snap,}-d}}^{\bar D}$, and ${\bm{U}}_{{\text{snap,}-d}}^\theta$ respectively. The combined snapshots are then expressed as:
		\begin{equation}
			{\bm{U}}_{{\text{snap, pred}}}^u = \Big[ {\bm{U}}_{{\text{snap,}-d}}^u; \, {\bm{U}}_{{\text{snap,}+d}}^u \Big], \quad {\bm{U}}_{{\text{snap, pred}}}^{\bar D} = \Big[ {\bm{U}}_{{\text{snap,}-d}}^{\bar D}; \, {\bm{U}}_{{\text{snap,}+d}}^{\bar D} \Big], \quad {\bm{U}}_{{\text{snap, pred}}}^\theta  = \Big[ {\bm{U}}_{{\text{snap,}-d}}^\theta; \, {\bm{U}}_{{\text{snap,}+d}}^\theta \Big].
			\label{eq44-2}
		\end{equation}
		By employing Eqs.\eqref{eq42}, \eqref{eq43}, and \eqref{eq44}, the \textit{predictive} projection matrices $\mathbf{\Phi}^{\text{pred}}_{\text{lin}}$ and ${\mathbf{\bar \Phi} }^{\text{pred}}_{\text{nonlin}}$ are evaluated and later used to predict the thermo-mechanical response within the material parameter intervals $[(1-d\%) \, \mu, (1+d\%) \, \mu]$ and $[(1-d\%) \, \Lambda,(1+d\%) \, \Lambda]$.
		\noindent As a result, the final global snapshot matrix $\bm{U}_{\text{snap}}$, which consists of the decomposed snapshots of three fields, is given as follows:
		\begin{subequations}
			\begin{gather}
				\bm{U}_{\text{snap}} = \bm{U}^u_{\text{snap}} + \bm{U}^{\bar{D}}_{\text{snap}}  + \bm{U}^\theta_{\text{snap}},\\
				\bm{U}^u_{\text{snap}} = \bm{U}_e^u + \bm{U}_h^u + \bm{U}_s^u , \quad \bm{U}^{\bar{D}}_{\text{snap}} = \bm{U}_e^{\bar{D}} + \bm{U}_h^{\bar{D}} + \bm{U}_s^{\bar{D}} , \quad \bm{U}^\theta_{\text{snap}} = \bm{U}_e^\theta + \bm{U}_h^\theta + \bm{U}_s^\theta.
			\end{gather}
		\end{subequations}
		
		To accurately select the number of modes, a multi-state decomposed approach inspired by \citet{oliver2017reduced} is newly proposed in this study, in which the displacement field snapshots are additionally decomposed into three distinct mechanical states: state in the elastic regime $\bm{U}_e^u$, state where plastic hardening takes place $\bm{U}_h^u$, and state where softening occurs $\bm{U}_s^u$, see  \hyperref[fig2]{Fig.2}. Based on this multi-state decomposition, the snapshots of the displacement field are separated as follows:	
		\begin{equation}
			\underbrace{ \left(\begin{array}{c} {\bm{U}}_u^{1} \\\vdots \\ {{{{\bm{U}}}}_u^{n_e}} \\ \vdots \\ {{ {{\bm{U}}}}_u^{n_h}}  \\  \vdots \\ {{ {{\bm{U}}}}_u^{n_{\text{step}}}} \end{array}\right) }_{ \bm{U}^u_{\text{snap}}}=
			\underbrace{ \left(\begin{array}{c} {\bm{U}}_u^{1} \\\vdots \\ {{{{\bm{U}}}}_u^{n_e}}  \\ \vdots \\ \bm{0}  \\  \vdots \\ {\bm{0}} \end{array}\right) }_{{\bm{U}}_e^u}+
			\underbrace{\left(\begin{array}{c} \bm{0} \\\vdots \\ {{{{\bm{U}}}}_u^{n_{e}+1 }}  \\ \vdots \\ {{ {{\bm{U}}}}_u^{n_h}}  \\  \vdots \\ \bm{0} \end{array}\right) }_{{\bm{U}}_h^u}+
			\underbrace{ \left(\begin{array}{c} \bm{0} \\\vdots \\ \bm{0}  \\ \vdots \\ {{ {{\bm{U}}}}_u^{n_h+1}}  \\  \vdots \\ {{ {{\bm{U}}}}_u^{n_{\text{step}}}} \end{array}\right) }_{{\bm{U}}_s^u}.
		\end{equation}
		Here, $n_e$ and $n_h$ denote the step number at the end of the elastic state and the hardening state, respectively. Subsequently, individual \textit{thin} singular value decompositions are applied to different displacement field snapshot states: 
		\begin{equation}
			{\bm{U}}_e^u = {\bm{S}}_e^u \, \bm{\Sigma} _e^u \, {\bm{V}_e^u}^{\text{T}}, \quad
			{\bm{U}}_h^u = {\bm{S}}_h^u \, \bm{\Sigma} _h^u \, {\bm{V}_h^u}^{\text{T}}, \quad
			{\bm{U}}_s^u = {\bm{S}}_s^u \, \bm{\Sigma} _s^u \, {\bm{V}_s^u}^{\text{T}}, \quad \text{and} \quad \bm{U}^u_{\text{snap}} = {\bm{S}}^u \, \bm{\Sigma} ^u \, {\bm{V}^u}^{\text{T}}.
		\end{equation}
		For snapshots in the displacement field, the following relation can hold
		\begin{equation}
			\bm{U}_e^u + \bm{U}_h^u + \bm{U}_s^u 
			\quad \Rightarrow \quad \bm{S}^u_{ehs} = \begin{bmatrix} {\bm{S}}_e^u & {\bm{S}}_h^u & {\bm{S}}_s^u \end{bmatrix} \in {\mathbb{R}}^{n \times  {n_{\text{step}}} }.
			\label{k51}
		\end{equation}
		
		\hyperref[k51]{Eq.\eqref{k51}}  defines the merged projection matrix $\bm{S}^u_{ehs}$ for the displacement field, constructed from the multi-state decomposed snapshots corresponding to the elastic, hardening, and softening regions. However, this merged basis does not satisfy $\bm{S}^u_{ehs} =  \bm{S}_u$ and cannot directly replace $\bm{S}_u$ in \hyperref[eq42]{Eq.\eqref{eq42}} for approximating the full-order displacement field. This is because simply concatenating the projection matrices ${\bm{S}}_e^u$, ${\bm{S}}_h^u$, and ${\bm{S}}_s^u$, derived from different mechanical states, does not ensure mutual orthogonality among the resulting basis vectors. Thus, a randomized singular value decomposition (rSVD) \cite{woolfe2008fast,liberty2007randomized,halko2011finding,rokhlin2010randomized} is applied to the combined bases. This approach does not only preserve orthogonality across different physical states in reduced-order simulations but also enables precise control over the number of modes. More concretely, the merged subspaces are projected onto a lower-dimensional space via the multiplication with a Gaussian random matrix $\bm{\Omega} \in {\mathbb{R}}^{(r_e + r_h + r_s) \times \kappa}$, where $\kappa$ denotes the target rank of the reduced subspace. Specifically, the random projection is expressed as $\bm{\Omega} \sim \mathcal{N}(0, 1), \, \bm{Y}_u = \bm{S}^u_{ehs} \, \bm{\Omega} \in {\mathbb{R}}^{n \times \kappa}$. This intermediate matrix $\bm{Y}_u$ is essential in capturing the dominant features of the combined bases ${\bm{S}}_e^u$, ${\bm{S}}_h^u$, and ${\bm{S}}_s^u$. An orthonormal approximation of the left singular vectors is then obtained by applying a QR decomposition \cite{gander1980algorithms,colin1993computation} to $\bm{Y}_u$, yielding
		\begin{equation}
			\left\lbrace  \bm{Q}, \, \mathbfcal{R} \right\rbrace  = \mathcal{Q} \mathcal{R} \, (\bm{Y}_u),
		\end{equation}
		where $\bm{Q} \in {\mathbb{R}}^{n \times \kappa} $ is orthonormal and captures the main directions of the combined subspace, and $\mathbfcal{R} \in {\mathbb{R}}^{\kappa \times \kappa}$ denotes an upper triangular matrix ($\mathcal{R}_{ij}=0, \, i>j$). Here, $\mathcal{QR}(\bullet)$ denotes the operator for the QR decomposition. To combine the contributions of ${\bm{S}}_e^u$, ${\bm{S}}_h^u$, and ${\bm{S}}_s^u$, they are projected onto the subspace spanned by $\bm{Q}$. This process ensures that the combined information from ${\bm{S}}_e^u$, ${\bm{S}}_h^u$, and ${\bm{S}}_s^u$ is accurately reflected in $\bm{S}_u$. This is achieved by projecting each matrix as follows:
		\begin{equation}
			\widetilde{\bm{S}}_e^u = \bm{Q}^{\text{T}} {\bm{S}}_e^u, \quad \widetilde{\bm{S}}_h^u = \bm{Q}^{\text{T}}  {\bm{S}}_h^u, \quad \widetilde{\bm{S}}_s^u = \bm{Q}^{\text{T}}  {\bm{S}}_s^u.
			\label{k2}
		\end{equation}
		
		The projections $\widetilde{\bm{S}}_e^u $,  $\widetilde{\bm{S}}_h^u$, and $\widetilde{\bm{S}}_s^u$ capture the contributions of the left singular matrices of $\bm{U}_e^u$, $\bm{U}_h^u$,  and $\bm{U}_s^u$, respectively, to the subspace defined by $\bm{Q}$. These projected components are subsequently combined to construct a unified multi-state projection $\bm{S}_{ehs}^u$ in the displacement field. As the projections in \hyperref[k2]{Eq.\eqref{k2}} are expressed within the subspace spanned by $\bm{Q}$, they are concatenated as follows:
		\begin{equation}
			{\bm{S}}^u_{ehs} = \bm{Q} \begin{bmatrix} \bm{Q}^{\text{T}}  {\bm{S}}_e^u & \bm{Q}^{\text{T}}  {\bm{S}}_h^u & \bm{Q}^{\text{T}}  {\bm{S}}_s^u \end{bmatrix} \quad \Rightarrow \quad  \bm{S}^u_{ehs} \approx \bm{S}_u.
			\label{k3}
		\end{equation}
		Having those consistent left singular matrices at hand, the global projection matrix in the displacement field over multiple states is obtained by truncating the left singular matrices $\bm{H}_e^u = \bm{Q}\,(\bm{Q}^{\text{T}} {\bm{S}}_e^u)$, $\bm{H}_h^u = \bm{Q}\,(\bm{Q}^{\text{T}} {\bm{S}}_h^u)$, and $\bm{H}_s^u = \bm{Q}\,(\bm{Q}^{\text{T}} {\bm{S}}_s^u)$ with the number of selected modes $k_e$, $k_h$, and $k_s$, respectively:
		\begin{equation}
			\begin{aligned}
				{{\bm{H}}_e^u} = & \Big[ 
				{\underbrace {{\bm{H}}_e^{u,1}, \cdots {\bm{H}}_e^{u,i}}_{{{\mathbf{\Phi }}_{u, \, k_e}^{\text{lin}}}}}, \, \underbrace{\cdots \, {{\bm{H}}_e^{u,j}}}_{{{\mathbf{\bar \Phi }}_{u, \, k_e}^{\text{nonlin}}}}, \, \cdots \,{{\bm{H}}_e^{u,r}}
				\Big], \quad
				{{\bm{H}}_h^u} = \Big[
				{\underbrace {{\bm{H}}_h^{u,1}, \cdots {\bm{H}}_h^{u,i}}_{{{\mathbf{\Phi }}_{u, \, k_h}^{\text{lin}}}}}, \, \underbrace{\cdots \, {{\bm{H}}_h^{u,j}}}_{{{\mathbf{\bar \Phi }}_{u, \, k_h}^{\text{nonlin}}}}, \, \cdots \, {{\bm{H}}_h^{u,r}} 
				\Big], \\
				{{\bm{H}}_s^u} = & \Big[
				{\underbrace {{\bm{H}}_s^{u,1}, \cdots {\bm{H}}_s^{u,i}}_{{{\mathbf{\Phi }}_{u, \, k_s}^{\text{lin}}}}}, \, \underbrace{\cdots \, {{\bm{H}}_s^{u,j}}}_{{{\mathbf{\bar \Phi }}_{u, \, k_s}^{\text{nonlin}}}}, \, \cdots \,{{\bm{H}}_s^{u,r}} 
				\Big].
				\label{k4}
			\end{aligned}
		\end{equation}
		Analogously, following Eqs. \eqref{k2}, \eqref{k3}, and \eqref{k4}, one can obtain the predictive linear and nonlinear global projection matrices over different fields as $\mathbf{\Phi}_{\text{lin}} = \mathbf{\Phi}^{\text{pred}}_{\text{lin}}$ and ${\mathbf{\bar \Phi} }_{\text{nonlin}} = {\mathbf{\bar \Phi} }^{\text{pred}}_{\text{nonlin}}$:
		\begin{subequations}
			\begin{gather}
				\mathbf{\Phi}^{\text{pred}}_{\text{lin}} = \Big[ \underbrace{{{\mathbf{\Phi }}_{u, \, k_e}^{\text{lin}}}, \, {{{\mathbf{\Phi }}_{u, \, k_h}^{\text{lin}}}}, \, {{{\mathbf{\Phi }}_{u, \, k_s}^{\text{lin}}}}}_{\mathbf{\Phi}_u^{\text{lin}}}, \, 
				\underbrace{{{\mathbf{\Phi }}_{\bar D, \, l_e}^{\text{lin}}}, \, {{{\mathbf{\Phi }}_{\bar D, \, l_h}^{\text{lin}}}}, \, {{{\mathbf{\Phi }}_{\bar D, \, l_s}^{\text{lin}}}}}_{\mathbf{\Phi}_{\bar D}^{\text{lin}}}, \, 
				\underbrace{{{\mathbf{\Phi }}_{\theta, \, m_e}^{\text{lin}}}, \, {{{\mathbf{\Phi }}_{\theta, \, m_h}^{\text{lin}}}}, \, {{{\mathbf{\Phi }}_{\theta, \, m_s}^{\text{lin}}}}}_{\mathbf{\Phi}_{\theta}^{\text{lin}}}
				\Big], \\
				{\mathbf{\bar \Phi} }^{\text{pred}}_{\text{nonlin}} = \Big[ \underbrace{{{\mathbf{\bar \Phi }}_{u, \, k_e}^{\text{nonlin}}}, \, {{{\mathbf{\bar \Phi }}_{u, \, k_h}^{\text{nonlin}}}}, \, {{{\mathbf{\bar \Phi }}_{u, \, k_s}^{\text{nonlin}}}}}_{ {\mathbf{\bar \Phi}}_u^{\text{nonlin}}}, \, 
				\underbrace{{{\mathbf{\bar \Phi }}_{\bar D, \, l_e}^{\text{nonlin}}}, \, {{{\mathbf{\bar \Phi }}_{\bar D, \, l_h}^{\text{nonlin}}}}, \, {{{\mathbf{\bar \Phi }}_{\bar D, \, l_s}^{\text{nonlin}}}}}_{{\mathbf{\bar \Phi}}_{\bar D}^{\text{nonlin}}}, \, 
				\underbrace{{{\mathbf{\bar \Phi }}_{\theta, \, m_e}^{\text{nonlin}}}, \, {{{\mathbf{\bar \Phi }}_{\theta, \, m_h}^{\text{nonlin}}}}, \, {{{\mathbf{\bar \Phi }}_{\theta, \, m_s}^{\text{nonlin}}}}}_{ {\mathbf{\bar \Phi}}_{\theta}^{\text{nonlin}}}
				\Big].
			\end{gather}
		\end{subequations}
		
		\subsection{Measurement of accuracy and computational cost}
		To enable a clearer comparison between reduced-order and full-order simulations in terms of both computational accuracy and efficiency, two measurement quantities are defined: the relative error $\varepsilon _r$ and two CPU time ratios $\eta_s$ and $\eta_w$. The latter are the time required to solve the system a) without including the assembly time ($\eta_s$) and b) including the assembly time ($\eta_w$):
		\begin{equation}
			\varepsilon _{r}  = \frac{1}{N}\sum\limits_{j = 1}^N {\frac{{| {u_{p, \, \rm{red}}^j - u_{p, \, \rm{full}}^j} |}}{{| {u_{p, \, \rm{full}}^j} |}}}, \quad 
			{\eta_s}  = \sum\limits_{j = 1}^N {\frac{{T_{\text{red}, s}^j}}{{{T^{j}_{\text{full}, s}}}}}, \quad  \text{and} \quad {\eta_w}  = \sum\limits_{j = 1}^N {\frac{{T_{\text{red}, w}^j}}{{{T^{j}_{\text{full}, w}}}}}
			\label{eq49}
		\end{equation}
		
		\noindent Here, $N$ represents the actual number of time steps. $u^{j}_{p, \, \rm{red}}$ and $u^{j}_{p,\, \rm{full}}$ denote the considered displacement components obtained using the reduced and full-order model at a selected nodal point $p$ at time $j$. $T^{j}_{\text{red},s}$ and $T^{j}_{\text{full},s}$ are the solution time costs of the reduced and full order models at time $j$, without counting the assembly time, respectively. $T^{j}_{\text{red},w}$ and $T^{j}_{\text{full},w}$ are the wall-time costs of the reduced and full-order models at time $j$, respectively.
		The smaller the CPU time ratios $\eta_s$ and $\eta_w$, the better the performance of the reduced-order model becomes. In the context of nonlocal damage, a relative error measure is unsuitable, since a division by zero occurs in undamaged regions of the structure. On account of this issue, an absolute error in damage $\varepsilon_{a}$ is adopted in this study to quantify the accuracy of the nonlocal damage field:
		\begin{equation}
			\varepsilon _{a}  = \frac{1}{N}\sum\limits_{j = 1}^N {{| {\bar{D}_{p,\, \rm{red}}^j - \bar{D}_{p, \, \rm{full}}^j} |}}.
			\label{eq50}
		\end{equation}
		Here, $\bar{D}^{j}_{p, \, \rm{red}}$ and $\bar{D}^{j}_{p, \, \rm{full}}$ represent the nonlocal damage variables at a selected nodal point $p$ at time $j$, computed using the reduced-order and full-order models, respectively.
		
		\section{Numerical examples}
		\label{sec:Model validation}
		In the following, several three-dimensional numerical examples are employed to investigate the performance of the newly developed multi-field and multi-state decomposed nonlinear manifold MOR approach. The material model in this study and its solver are implemented in \textit{FEAP} \cite{taylor2014feap} and \textit{Python}, respectively. All contour plots are shown using the open-source software \textit{Paraview} \cite{ahrens200536}. 
		\subsection{Test 1: Rectangular specimen} 
		\label{Test1}
		\begin{figure}[!ht]
			\centering
			\includegraphics[width=18cm]{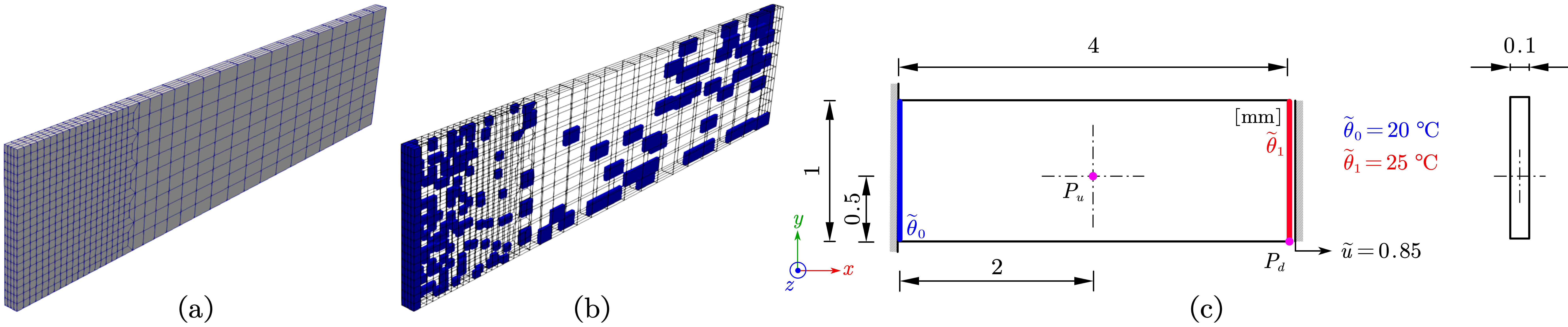} 
			\caption{(a) Schematic illustration of the meshed rectangular specimen. The prescribed displacement on the right surface is  $\widetilde{u}=0.85$ mm. Different constant temperatures are prescribed on the left and right surfaces with $\widetilde{\theta}_0 = \SI{20}{\celsius}$ and $\widetilde{\theta}_1 = \SI{25}{\celsius}$, respectively. Points $P_u$ and $P_d$ are selected to compare the error of both the displacement component $u$ and temperature $\theta$, as well as nonlocal damage variable $\bar{D}$, respectively; (b) selected ECSW elements (shown in blue) with $n_s = 245$ elements over a meshed discretization of 2240 elements; (c) geometry and considered boundary value problem.}
			\label{fig3}
		\end{figure}
		
		In this section, Test 1 is designed to investigate the performance of the developed multi-field decomposed nonlinear manifold MOR approach for fully thermo-mechanically coupled damage-plasticity simulations without using the multi-state decomposition method, see \hyperref[fig3]{Fig.3}. The specimen is discretized by linear 8-node hexahedral elements\footnote{An 8-node hexahedral element possesses a linear shape function for all considered corresponding fields in this study.}, resulting in a total number of 2240 elements. The thickness direction of the specimen is meshed with four elements. Each node possesses five degrees of freedom: three displacement components ($u$, $v$, and $w$), nonlocal damage $\bar D$, and temperature $\theta$. The errors between the reduced- and full-order simulations associated with displacement and nonlocal damage fields are evaluated at the selected points ${P_u} \, (2, \, 0.5, \, 0)$ and ${P_d} \, (4, \, 0, \, 0)$, respectively. The temperatures on the left and right surfaces of the specimen are prescribed by constant values of $\widetilde{\theta}_0 = \SI{20}{\celsius}$ and $\widetilde{\theta}_1 = \SI{25}{\celsius}$, respectively. More details on the boundary conditions are given in \hyperref[fig3]{Fig.3}. The simulation is carried out in 100 timesteps. Mechanical and thermal parameters are taken from \citet{brepols2020gradient} and \citet{dittmann2020phase}, respectively, see \hyperref[T1]{Table 1}.
		
		\begin{table}[!h]
			\centering
			\begin{tabular}{lllll}
				\hline
				\textbf{Symbol}                      & \textbf{Parameter description}                          & \textbf{Value}  & \textbf{Unit}                    & \textbf{Eq.}				\\ \hline
				\textbf{elastic parameters} &                               &                             & \multicolumn{1}{c}{}    \\  \hline
				$\Lambda$                   & $1^{\rm{st}}$ Lam\'e constant                 & 25000                       & MPa		& \eqref{eq5} \\
				$\mu$                       & $2^{\rm{nd}}$ Lam\'e constant (shear modulus)                 & 55000                       & MPa 		& \eqref{eq5} \\ \hline
				\textbf{plastic parameters} &                               &                             &                   &           \\ \hline
				$\sigma _{y0}$              & initial yield stress          & 100                         & MPa           &  \eqref{eqA9}         \\
				$a$                       & kinematic hardening parameter    & 62.5                        & MPa          &  \eqref{eq6}           \\
				$P$                       & linear isotropic hardening parameter   & 0                           & MPa           &  \eqref{eq6}         \\
				$e_{p}$                   & $1^{\rm{st}}$ nonlinear isotropic hardening parameter          & 125                         & MPa           &  \eqref{eq6}         \\
				$f_{p}$                   & $2^{\rm{nd}}$ nonlinear isotropic hardening parameter         & 5                           & -             &  \eqref{eq6}       \\ \hline
				\textbf{damage parameters}  &                               &                             &               &          \\ \hline
				$Y_0$                       & initial damage threshold              & 2.5                         & MPa           & \eqref{eqA9}        \\
				$e_{d}$                     & $1^{\rm{st}}$ damage hardening parameter      & 5.0                         & MPa           & \eqref{eq7}          \\
				$f_{d}$                     & $2^{\rm{nd}}$ damage hardening parameter       & 100                         & -             & \eqref{eq7}        \\
				$A$                         & gradient damage parameter         & 75                          & MPa $\rm{mm^2}$    & \eqref{eq8}          \\
				$H$                         & micromorphic penalty parameter             & $10^6$                      & MPa           & \eqref{eq8}          \\ \hline
				\textbf{thermal parameters} &                               &                             &               &           \\ \hline
				$c$                         & volumetric heat capacity      & 3.59                         & $\rm{mJ/(mm^3 \, K}$)     & \eqref{eq31}      \\
				$\alpha$                    & thermal expansion coefficient & 1.1                         & $\rm{10^{-5}/K}$        & \eqref{eq1}    \\
				$K_0$                       & initial heat conductivity             & 50.2                        & $\rm{mW/(mm \, K}$)     &  \eqref{eqA7}        \\
				$\theta _0$                 & reference temperature         & 273.15                      & $\rm{K}$                 & \eqref{eq1}      \\
				$\omega$                    & thermal softening parameter   & 0.002                       & $\rm{1/K}$               & \eqref{eqA8}     \\ \hline
				\textbf{regularization parameters} &                               &                             &               &           \\ \hline
				$\gamma _u$                 & regularization parameter for displacement        & 3.0                      & -                & \eqref{n12}      \\
				$\gamma _{\bar D}$                 & regularization parameter for damage         & 1.0                      & -                & \eqref{n12}      \\
				$\gamma _{\theta}$                 & regularization parameter for temperature         & 3.0                      & -                & \eqref{n12}      \\ \hline
			\end{tabular}
			\label{T1}
			\caption{Material and regularization parameters as well as their corresponding equations.}
		\end{table}
		
		\subsubsection{Comparison of singular values for different fields}
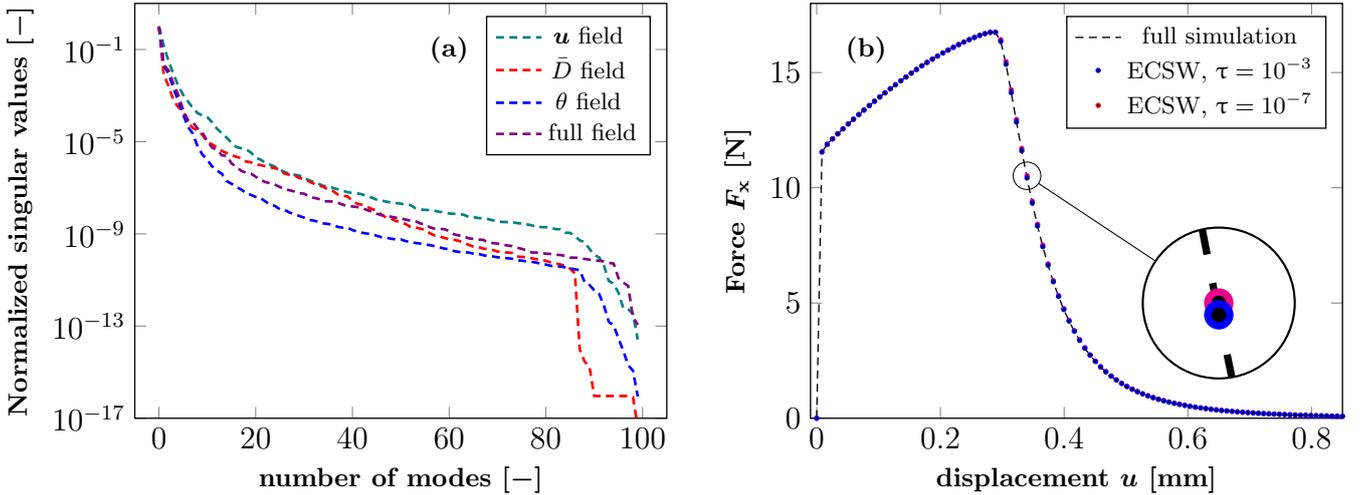
\begin{figure*}[!h]
			\begin{minipage}[b]{0.5\linewidth}
				\begin{tikzpicture}
					\begin{axis}[name=plotA,
						xmin=-5,xmax=105,
						ymin=1e-17,ymax=10,
						scale only axis,
						ticklabel style={font=\large},
						xlabel={number of modes $\bm{[-]}$}, 
						ylabel={Normalized singular values $\bm{[-]}$},
						line width=0.5pt,
						ymode=log,
						log basis y={10},
						axis line style={line width=0.5pt},
						font=\bfseries,
						legend pos=north east,
						legend style={fill=none, font=\small}
						]
						
						\addplot[
						line width=1.0pt,
						color=teal, densely dashed,
						] table[x=modes,y=U]{data/RT_modes_SVD.txt};
						
						\addplot[
						line width=1.0pt,
						color=red, densely dashed,
						] table[x=modes,y=D]{data/RT_modes_SVD.txt};
						
						\addplot[
						line width=1.0pt,
						color=blue, densely dashed,
						] table[x=modes,y=T]{data/RT_modes_SVD.txt};
						
						\addplot[
						line width=1.0pt,
						color=violet, densely dashed,
						] table[x=modes,y=full]{data/RT_modes_SVD.txt};
						
						\legend{
							{\text{$\bm{u}$ field   }},
							{\text{$\bar{D}$ field   }},
							{\text{$\theta$ field   }},
							{\rm{full field}}
						}
						\node[align=center] at (60,0.1) {(a)};
					\end{axis}
				\end{tikzpicture}
			\end{minipage}
			\begin{minipage}[b]{0.5\linewidth}
				\centering
				\begin{tikzpicture}[spy using outlines={circle, magnification=5.5, connect spies}]
					\begin{axis}[
						name=plotA,
						clip=true,
						xmin=-0.01,xmax=0.85,
						ymin=0,ymax=18,
						scale only axis,
						log ticks with fixed point,
						ticklabel style={font=\large},
						xlabel={displacement $\bm{u}$ $\textbf{[mm]}$}, 
						ylabel={Force $\bm{F_{\rm{x}}}$ \textbf{[N]}},
						line width=0.5pt,
						font=\bfseries,
						legend pos=north east,
						legend style={fill=none, font=\small},
						scatter/classes={%
							a={mark=*,draw=blue},
							b={mark=*,draw=red},
							c={mark=*,draw=magenta},
							d={mark=*,draw=blue}
						}
						]
						
						\addplot[
						color=black, densely dashed,
						] table[x=u2240full,y=f2240full]{data/RT_FU_full.txt};
						
						\addplot[
						scatter, only marks,
						mark=*,mark options={solid},
						scatter src=explicit symbolic,
						mark size=0.75pt,
						] table[x=u2240mor,y=f2240mor,meta=label]{data/RT_FU_ECSW_1E3.txt};
						
						\addplot[
						scatter, only marks,
						mark=*,mark options={solid},
						scatter src=explicit symbolic,
						mark size=0.75pt,
						] table[x=u2240mor,y=f2240mor,meta=label]{data/RT_FU_ECSW_1E7.txt};
						
						\legend{\text{full simulation},\text{ECSW, $\uptau=10^{-3}$}, \text{ECSW, $\uptau=10^{-7}$}}
						\node[align=center] at (0.08,16) {(b)};
						\coordinate (spypoint) at (axis cs:0.34, 10.52);
						\coordinate (magnifyglass) at (axis cs:0.65, 5);
					\end{axis}
					\spy [black, size=2cm] on (spypoint) in node[fill=white] at (magnifyglass);
				\end{tikzpicture}
			\end{minipage}
			\caption{(a) Normalized singular values in the case of the multi-field decomposed MOR approach and the classical non-decomposed MOR approach (full field); (b) force-displacement curve comparison between the full-order simulation and the reduced-order simulation using nonlinear manifold-based ECSW with different tolerances $\uptau$.}
			\label{fig4}
		\end{figure*}

		As shown in \hyperref[fig4]{Fig.4}, the normalized singular values of the displacement, temperature, and damage fields exhibit a quite slow decay, indicating that the solution manifold of this fully thermo-mechanically coupled damage-plasticity problem resists an efficient reduction via linear projection, due to the Kolmogorov barrier. Notably, the normalized singular values of the damage field stabilize as the number of modes exceeds approximately 90. This stabilization results from the natural value interval of the nonlocal damage, which lies between 0 and 1 to distinguish undamaged and fractured phases. When the specimen is fractured, the material's stiffness is almost fully degraded and the nodal damage variables remain nearly constant. From this point on, increasing the number of modes yields negligible improvement, and the accuracy of the reduced-order model related to the damage field saturates.
		
		In \hyperref[fig4]{Fig.4 (b)}, a comparison of the force-displacement curves between the full-order simulation and the reduced-order simulation is shown. The numbers of linear and nonlinear modes used for the reduced-order simulations are selected as $90$ and $5$ for the three fields, respectively.  As can be seen, the newly developed multi-field decomposed nonlinear manifold approach agrees very well with the full-order simulation in terms of the force-displacement curve, even though a pretty small number of selected elements ($n_s=245$ for $ \uptau=10^ {-3}$; $n_s=365$ for $ \uptau=10^ {-7}$) is used.
		
		\subsubsection{Influence of ECSW's tolerance $\uptau$ on the accuracy}
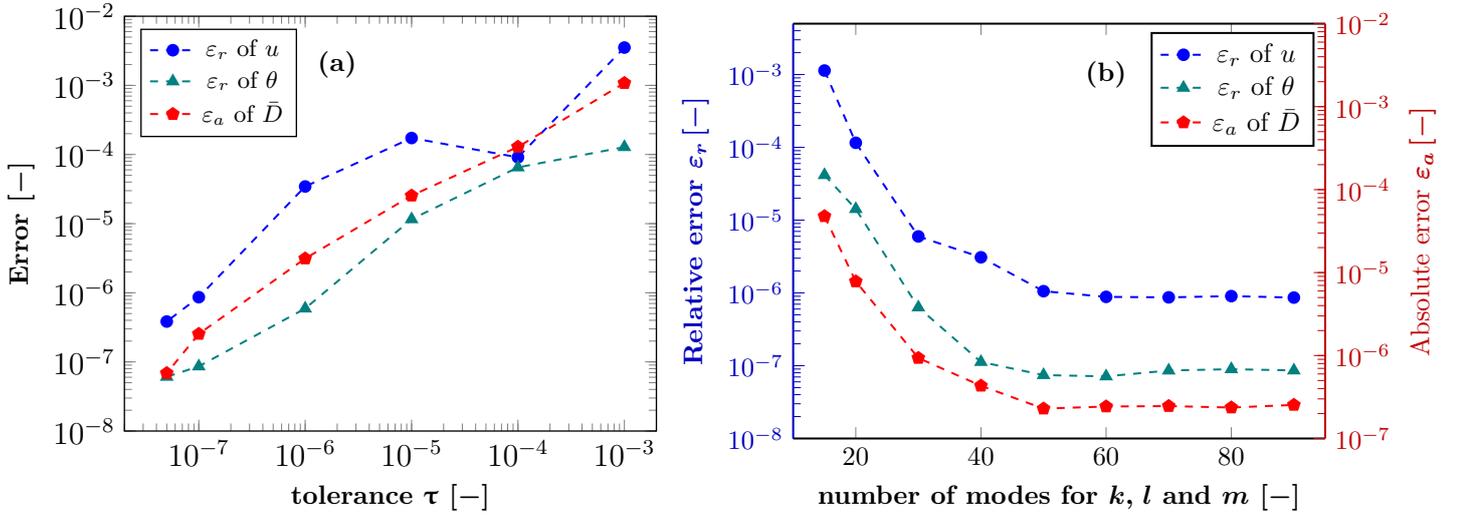
\begin{figure*}[!ht]
			\begin{minipage}[b]{0.4\linewidth}
				\begin{tikzpicture}
					\begin{axis}[
						xmin=2e-8,xmax=2e-3,
						ymin=1e-8,ymax=1e-2,
						scale only axis,
						ticklabel style={font=\large},
						xlabel={tolerance $\bm{\uptau}$ $\bm{[-]}$}, 
						ylabel={Error $\bm{[-]}$},
						line width=0.5pt,
						ymode=log,
						log basis y={10},
						xmode=log,
						log basis x={10},
						font=\bfseries,
						legend pos=north west,
						legend style={fill=none, font=\small}
						]
						\addplot[
						line width=0.75pt,
						color=blue, dashed,
						mark=*,
						mark size=2pt,
						mark options={solid},
						] table[x=Tau,y=U]{data/Error_2240_Tau.txt};
						
						\addplot[
						line width=0.75pt,
						color=teal, dashed,
						mark=triangle*,mark size=2.25pt,
						mark options={solid},
						] table[x=Tau,y=T]{data/Error_2240_Tau.txt};
						
						\addplot[
						line width=0.75pt,
						color=red, dashed,
						mark=pentagon*,
						mark size=2.25pt,
						mark options={solid},
						] table[x=Tau,y=D]{data/Error_2240_Tau.txt};
						
						\legend{
							{\text{$\varepsilon_{r}$ of $u$}},
							{\text{$\varepsilon_{r}$ of $\theta$}},
							{\text{$\varepsilon_{a}$ of $\bar{D}$}}
						}
						\node[align=center] at (2e-6,2e-3) {(a)};
					\end{axis}
				\end{tikzpicture}
			\end{minipage}
			\hspace{0.08\linewidth}
			\begin{minipage}[b]{0.4\linewidth}
				\begin{tikzpicture}
					\pgfplotsset{
						scale only axis,
						xmin=10, xmax=95,
						y axis style/.style={
							yticklabel style=#1,
							ylabel style=#1,
							y axis line style=#1,
							ytick style=#1
						}
					}
					
					\begin{axis}[
						line width=0.75pt,
						axis y line*=left,
						y axis style=blue!75!black,
						ymin=1e-8, ymax=5e-3,
						font=\bfseries,
						xlabel={number of modes for $\bm{k, \, l}$ and $\bm{m}$ $\bm{[-]}$},
						ylabel={Relative error $\bm{\varepsilon _r} \, \bm{[-]}$},
						ymode=log,
						log basis y={10}
						]
						
						\addplot[
						color=blue, dashed,
						mark=*,
						mark size=2pt,
						mark options={solid},
						line width=0.75pt,
						] table[x=modes,y=U]{data/Error_2240_UDT.txt}; \label{plot1},
						
						\addplot[
						color=teal, dashed,
						mark=triangle*,
						mark size=2.25pt,
						mark options={solid},
						line width=0.75pt,
						] table[x=modes,y=T]{data/Error_2240_UDT.txt}; \label{plot2},
						\node[align=center] at (60,1e-3) {(b)};
					\end{axis}
					
					\begin{axis}[
						line width=0.75pt,
						axis y line*=right,
						axis x line=none,
						ymin=1e-7, ymax=1e-2,
						ylabel={Absolute error $\bm{\varepsilon _a} \, \bm{[-]}$},
						ymode=log,
						log basis y={10},
						y axis style=red!75!black
						]
						\addlegendimage{/pgfplots/refstyle=plot1}\addlegendentry{\text{$\varepsilon_{r}$ of $u$}}
						\addlegendimage{/pgfplots/refstyle=plot2}\addlegendentry{\text{$\varepsilon_{r}$ of $\theta$}}
						\addplot[
						color=red, dashed,
						mark=pentagon*,
						mark size=2.25pt,
						mark options={solid},
						line width=0.75pt,
						] table[x=modes,y=D]{data/Error_2240_UDT.txt};\addlegendentry{\text{$\varepsilon_{a}$ of $\bar D$}}
					\end{axis}
				\end{tikzpicture}
			\end{minipage}
			\caption{(a) Error depending on ECSW's training tolerance $\uptau$ varied from $5 \times 10^{-8}$ to $10^{-3}$; (b) error depending on the number of linear modes per field from 15 to 90 per field. The employed tolerance and the number of nonlinear modes per field are set to $\uptau=10^{-7}$ and $q=5$, respectively.} 
			\label{fig5}
		\end{figure*}

		In this section, the influence of the ECSW's training tolerance $\uptau$ on the accuracy of the reduced-order simulation is systematically investigated. For this purpose, $\uptau$ is varied from $5 \times 10^{-8}$ to $10^{-3}$. As shown in \hyperref[fig5]{Fig.5 (a)}, the displacement component $u$, the nonlocal damage variable $\bar{D}$, and the temperature variable $\theta$ obtained from the reduced-order simulations are compared against their corresponding full-order results. Contour plots comparing the results of the full-order and reduced-order simulations are presented in \hyperref[fig7]{Fig.7}. Since the contour plots of the nonlocal damage field are similar to those of the temperature field, they are not shown here.
		
		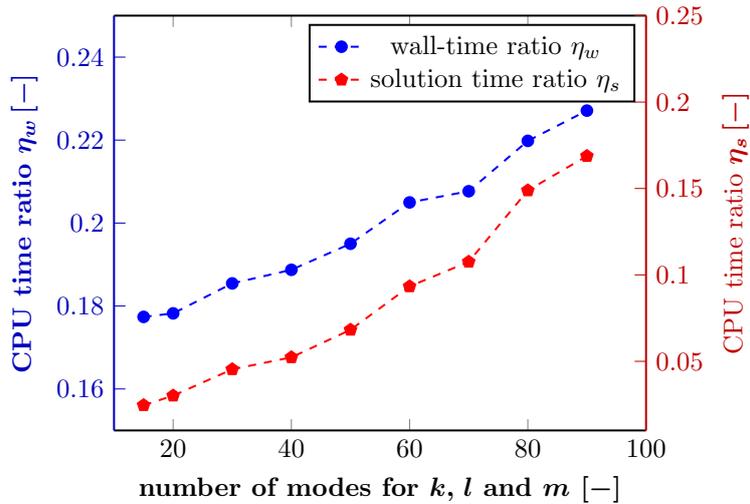
\begin{figure*}[!ht]
				\centering
				\begin{tikzpicture}
					\pgfplotsset{
						scale only axis,
						xmin=10, xmax=100,
						y axis style/.style={
							yticklabel style=#1,
							ylabel style=#1,
							y axis line style=#1,
							ytick style=#1
						}
					}
					
					\begin{axis}[
						line width=0.75pt,
						axis y line*=left,
						y axis style=blue!75!black,
						ymin=0.15, ymax=0.25,
						font=\bfseries,
						xlabel={number of modes for $\bm{k, \, l}$ and $\bm{m}$ $\bm{[-]}$},
						ylabel={CPU time ratio $\bm{\eta_w} \, \bm{[-]}$},
						]
						
						\addplot[
						color=blue, dashed,
						mark=*,
						mark size=2pt,
						mark options={solid},
						line width=0.75pt,
						] table[x=modes,y=Tstep]{data/TCPU_ECSW_4420.txt}; \label{plot_one},
					\end{axis}
					
					\begin{axis}[
						line width=0.75pt,
						axis y line*=right,
						axis x line=none,
						ymin=0.01, ymax=0.25,
						y tick label style={/pgf/number format/fixed, /pgf/number format/precision=2},
						ylabel={CPU time ratio $\bm{\eta_s} \, \bm{[-]}$},
						y axis style=red!75!black
						]
						\addlegendimage{/pgfplots/refstyle=plot_one}\addlegendentry{{wall-time ratio ${\eta_w}$}}
						\addplot[
						color=red, dashed,
						mark=pentagon*,
						mark size=2.25pt,
						mark options={solid},
						line width=0.75pt,
						] table[x=modes,y=Tsolve]{data/TCPU_ECSW_4420.txt};\addlegendentry{solution time ratio ${\eta_s}$}
					\end{axis}
				\end{tikzpicture}
			\caption{CPU time ratio comparison for different numbers of linear modes for the displacement field, nonlocal damage field, and temperature field using the ECSW and nonlinear projection-equipped reduced-order simulations.}
			\label{fig6}
		\end{figure*}
		As shown in \hyperref[fig5]{Fig.5 (a)}, the errors decrease for the different fields while decreasing the ECSW tolerance $\uptau$. This phenomenon is attributed to an increase in the number of selected evaluation elements, which enhances the approximation accuracy of the reduced-order model by providing a richer sampling. However, incorporating more elements inevitably increases the computational cost during the assembly and evaluation phases. Furthermore, setting a higher value for the tolerance sacrifices accuracy to obtain a higher computational efficiency. Therefore, selecting an appropriate ECSW's tolerance for multiphysics damage simulations requires careful consideration of practical engineering requirements as well as a balance between computational accuracy and efficiency. \hyperref[fig5]{Fig.5 (b)} reveals an expected key trend: increasing the number of linear modes brings a better precision for the different fields. However, beyond roughly 50 modes in each field, the error reaches a plateau, meaning that further increasing the number of modes does not lead to a higher accuracy anymore while incurring unnecessary computational cost. Therefore, 50 linear modes and 5 nonlinear modes per field are sufficient to ensure both stability and accuracy for the reduced-order simulation in \hyperref[Test1]{Test 1}.
		
		\begin{figure}[!ht]
			\centering
			\includegraphics[width=16cm]{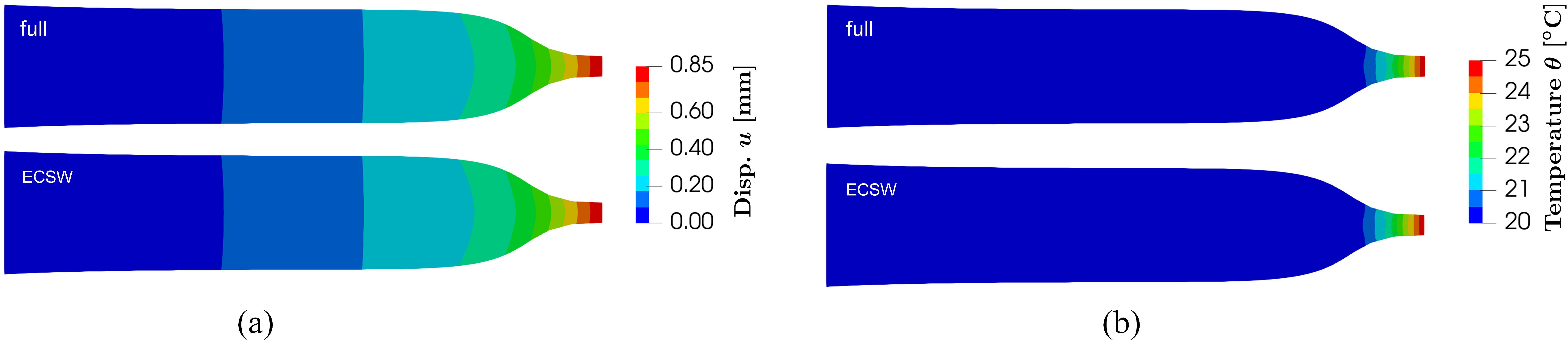}
			\caption{(a) Comparison of contour plots for the displacement component $u$ between full-order and reduced-order simulations, where the employed numbers of linear modes per field are set to 90. The number of ECSW elements and the corresponding tolerance $\uptau$ are selected as $n_s = 245$ and $10^{-7}$, respectively. (b) Comparison of contour plots for the temperature $\theta$ between full-order and reduced-order simulations.} 
			\label{fig7}
		\end{figure}
		The CPU time ratios are employed to evaluate the performance of the proposed reduced-order model while varying the number of linear modes from 15 to 90. The number of linear modes $r$ is varied instead of $q$ because $r$ predominantly determines the dimension of the reduced equation system, which significantly influences the computational efficiency in reduced-order simulation, see \hyperref[eq45]{Eq.\eqref{eq45}}. In addition, both the wall-time ($\eta_w$) and the solution time ($\eta_s$) ratios are considered. As shown in \hyperref[fig6]{Fig.6}, the reduction is more pronounced for the solution time due to the quadratic manifold's capacity to accurately capture nonlinear behavior using a few numbers of modes. The wall-time is more sensitive to the number of selected elements in the ECSW procedure. For Test 1, a relatively larger number of elements is required in comparison to simulations involving weakly nonlinear materials (e.g., hyperelastic models \cite{de2023hyper,rutzmoser2017lean}), owing to the higher complexity and nonlinearity of the present damage model, which leads to global softening in the structure and thus demands a denser sampling to fulfill the accuracy.

		\subsection{Test 2: HC specimen}
		\begin{figure*}[!h]
			\centering
			\includegraphics[width=18cm]{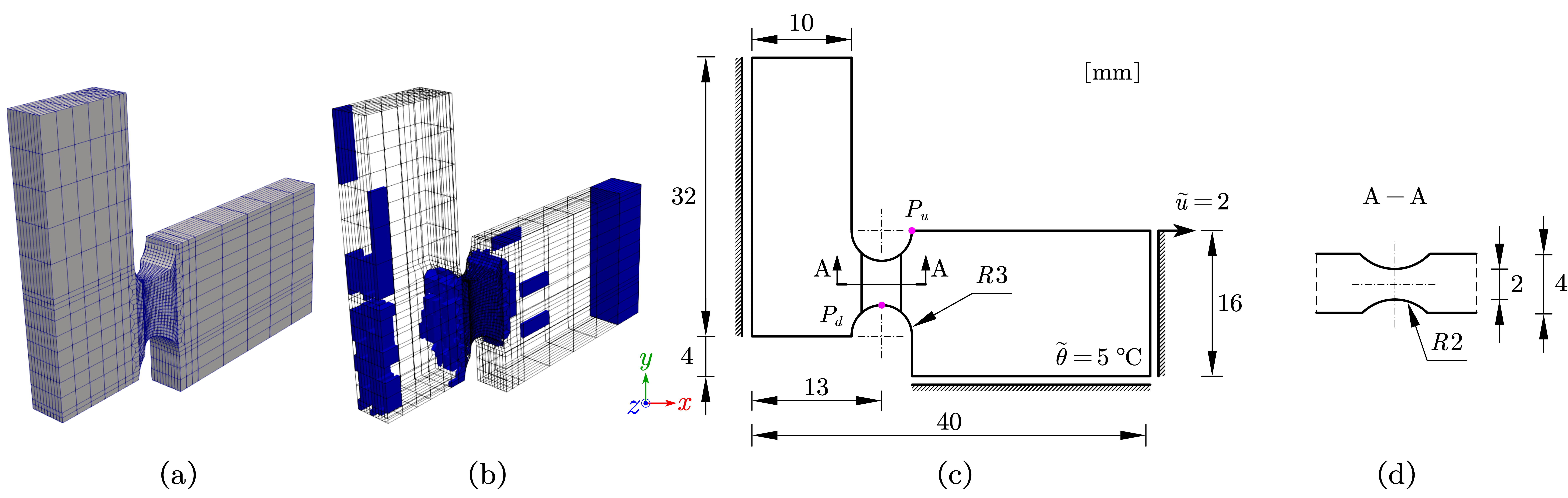}
			\caption{Schematic illustration of the HC specimen and the considered boundary value problem; (a) meshed HC specimen; (b) selected ECSW elements (shown in blue with $n_s = 2106$ elements) for a mesh discretization of 10730 elements; (c) dimension and boundary conditions of HC specimen taken from \cite{wei2023damage,wei2024numerical}; (d) cross-section of HC specimen at section A-A.}
			\label{fig8}
		\end{figure*}

		Test 2 is conducted on an HC specimen in accordance with the experimental setup outlined by \citet{wei2023damage,wei2024numerical}, under a constant temperature of $\widetilde{\theta} = \SI{5}{\celsius}$, see \hyperref[fig8]{Fig.8}. On the right surface of the specimen, a displacement is prescribed in $x$ direction with a value of $\widetilde{u}=2$ mm. Note that, due to symmetry, only a quarter of the structure is used and shown in \hyperref[fig8]{Fig.8}. To investigate the precision of the reduced-order simulation, points ${P_u} \, (16, \, 16, \, 4)$ and ${P_d} \, (13, \, 13, \, 2)$ are selected to measure the nodal values of displacement component $u$ and nonlocal damage $\bar{D}$, respectively.
		
		\newpage
		\subsubsection{Multi-state and multi-field decomposed MOR for prediction}
		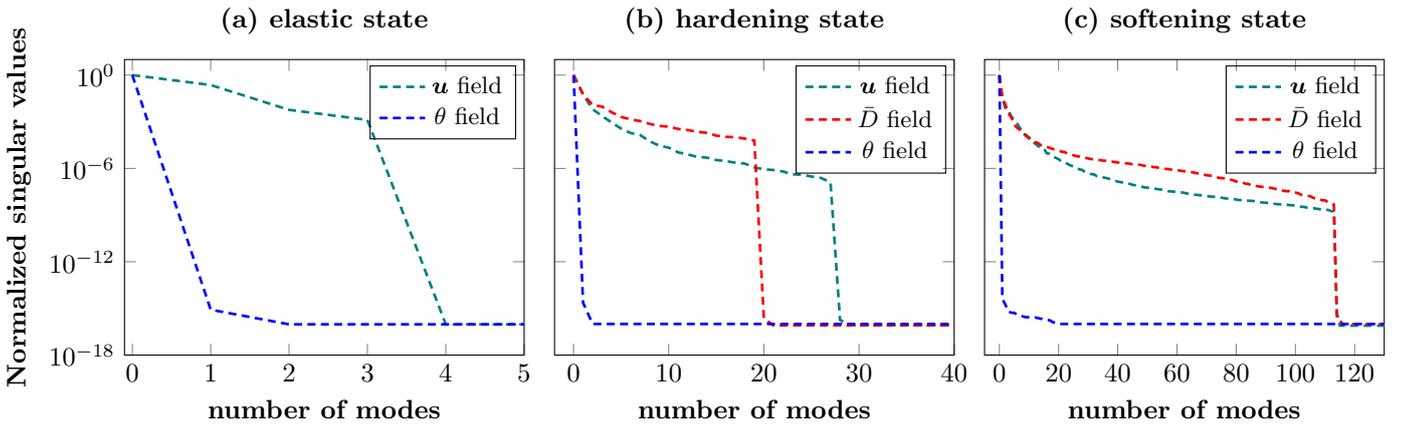
\begin{figure}[!h]
			\begin{tikzpicture}
				\begin{groupplot}[
					group style={
						group size=3 by 1,  
						horizontal sep=0.4cm,  
						vertical sep=0.5cm,    
					},
					width=0.38\textwidth,  
					height=5.5cm,  
					xlabel={\bfseries number of modes},
					every axis/.append style={enlargelimits=false},  
					yticklabel style={anchor=east},  
					]
					
					\nextgroupplot[
					title={\textbf{(a) elastic state}},
					ylabel={\bfseries Normalized singular values},  
					xmin=-0.1, xmax=5,
					ymin=1e-18, ymax=1e1,
					ymode= log,
					legend style={fill=none, font=\small},
					samples=100
					]
					\addplot[
					line width=1.0pt,
					color=teal, densely dashed,
					] table[x=modes,y=U]{data/linearState_sigma.txt};
					
					\addplot[
					line width=1.0pt,
					color=blue, densely dashed,
					] table[x=modes,y=T]{data/linearState_sigma.txt};
					
					\legend{
						{\text{$\bm{u}$ field}},
						{\text{$\theta$ field}}
					}
					
					\nextgroupplot[
					title={\textbf{(b) hardening state}},
					xmin=-2, xmax=40,
					ymin=1e-18, ymax=1e1,
					ymode= log,
					legend style={fill=none, font=\small},
					yticklabels={}  
					]
					\addplot[
					line width=1.0pt,
					color=teal, densely dashed,
					] table[x=modes,y=U]{data/hardeningState_sigma.txt};
					
					\addplot[
					line width=1.0pt,
					color=red, densely dashed,
					] table[x=modes,y=D]{data/hardeningState_sigma.txt};
					
					\addplot[
					line width=1.0pt,
					color=blue, densely dashed,
					] table[x=modes,y=T]{data/hardeningState_sigma.txt};
					
					\legend{
						{\text{$\bm{u}$ field}},
						{\text{$\bar{D}$ field}},
						{\text{$\theta$ field}}
					}
					
					\nextgroupplot[
					title={\textbf{(c) softening state}},
					xmin=-5, xmax=130,
					ymin=1e-18, ymax=1e1,
					ymode= log,
					legend style={fill=none, font=\small},
					yticklabels={}  
					]
					\addplot[
					line width=1.0pt,
					color=teal, densely dashed,
					] table[x=modes,y=U]{data/softeningState_sigma.txt};
					
					\addplot[
					line width=1.0pt,
					color=red, densely dashed,
					] table[x=modes,y=D]{data/softeningState_sigma.txt};
					
					\addplot[
					line width=1.0pt,
					color=blue, densely dashed,
					] table[x=modes,y=T]{data/softeningState_sigma.txt};
					
					\legend{
						{\text{$\bm{u}$ field}},
						{\text{$\bar{D}$ field}},
						{\text{$\theta$ field}}
					}
				\end{groupplot}
			\end{tikzpicture}
			\caption{Normalized singular values while using the multi-field and multi-state decomposition approach proposed in this study for a mesh discretization of 10730 elements in (a) elastic, (b) plastic hardening, and (c) softening states.}
			\label{fig9}
		\end{figure}
		
		This part employs the newly developed multi-field and multi-state decomposed nonlinear manifold MOR approach and assesses its predictive performance for multiphysics damage simulations. Two full-order simulations are conducted with a deviation of $\pm 50\%$ in the Lam\'e constants, as shown in \hyperref[T1]{Table 1}. Subsequently, the linear and nonlinear predictive projection matrices are derived along with the coefficient matrix. As detailed in \autoref{sec:multi_pers}, the simulation process for the tensile test of the HC specimen is decomposed into the elastic, hardening, and softening phases. Numbers of snapshots selected for the elastic, hardening, and softening states are specified as 4, 28, and 114, respectively, based on the overall mechanical performance in the simulations. As illustrated in \hyperref[fig9]{Fig.9 (a)}, the dominant contribution in the elastic region is primarily governed by the displacement field, without an influence from the damage field. A relatively small influence of the temperature field can be expected, since it is set constant. This behavior is attributed to the slower decay of normalized singular values in the displacement field compared to those in the temperature field. The fast decay of singular values in the temperature field stems from the constant temperature throughout the specimen, which can be effectively represented using only a few numbers of temperature modes. A similar trend is also observed in Figs.9 \hyperref[fig9]{(b)} and \hyperref[fig9]{(c)}.
		
		Next, the focus turns to the distribution of the normalized singular values associated with the nonlocal damage field. These values are not shown in \hyperref[fig9]{Fig.9 (a)} because the nonlocal damage variable remains inactive during the elastic phase. At this stage, the evolution of the local damage variable is governed by the damage threshold at the integration point level and remains zero, unless the damage-driving force $Y$ exceeds the critical threshold $Y_0$, see \hyperref[eqA9]{Eq.\eqref{eqA9}}. Due to the penalization strategy coupling the local and nonlocal damage variables, the nonlocal damage field is also identically zero. Under these conditions, the corresponding damage projection matrices are not needed to approximate the nonlocal damage variable (see \hyperref[eq7]{Eq.\eqref{eq7}}).
		
		One interesting observation in \hyperref[fig9]{Fig.9 (b)} is that the influence of the nonlocal damage field on the material's hardening response does not persist throughout the process. This is reflected in the rapid decay of its normalized singular values that drop to near-zero at 20 modes, even though 28 snapshots are collected to formulate the entire hardening phase.
		This phenomenon highlights two aspects in this example when selecting the number of modes for reduced-order modeling. From a computational perspective, only a relatively small number of nonlocal damage modes is needed to accurately capture the deformation-induced hardening behavior of the HC specimen. This enables a more efficient selection of modes across different physical fields to improve the overall computational performance in reduced-order simulations. From a physical perspective, the displacement field primarily governs the thermo-mechanically coupled hardening response in Test 2, whereas the nonlocal damage and temperature fields play a comparatively minor role.
		 
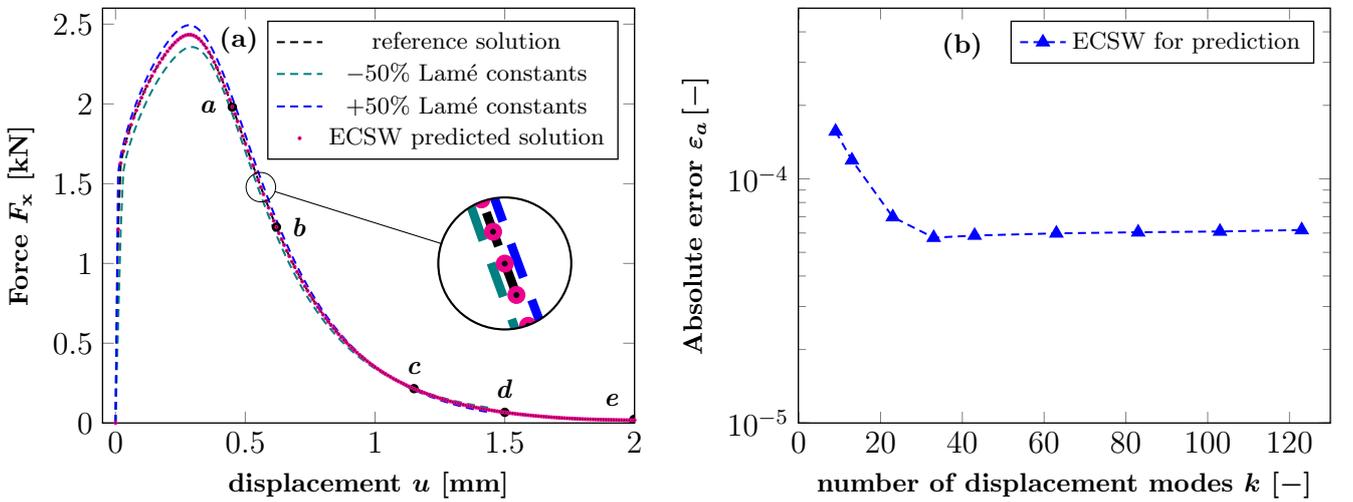
\begin{figure*}[!h]
			\begin{minipage}[b]{0.5\linewidth}
				\centering
				\begin{tikzpicture}[spy using outlines={circle, magnification=4.5, connect spies}]
					\begin{axis}[
						name=plotA,
						clip=true,
						xmin=-0.05,xmax=2,
						ymin=0,ymax=2.6,
						scale only axis,
						log ticks with fixed point,
						ticklabel style={font=\large},
						xlabel={displacement $\bm{u}$ \textbf{[mm]}}, 
						ylabel={Force $\bm{F_{\rm{x}}}$ \textbf{[kN]}},
						line width=0.5pt,
						font=\bfseries,
						legend pos=north east,
						legend style={fill=none, font=\small}, 
						scatter/classes={%
							a={mark=*,draw=magenta},
							b={mark=*,draw=magenta}
						}
						]
						\addplot[
						line width=0.75pt,
						color=black, densely dashed,
						] table[x=u10730,y=f10730]{data/FU10730HCfull.txt};
						
						\addplot[
						line width=0.75pt,
						color=teal, densely dashed,
						] table[x=u10730N50,y=f10730N50]{data/FU10730HCN50.txt};
						
						\addplot[
						line width=0.75pt,
						color=blue, densely dashed,
						] table[x=u10730P50,y=f10730P50]{data/FU10730HCP50.txt};
						
						\addplot[
						scatter, only marks,
						mark=*,mark options={solid},
						scatter src = explicit symbolic,
						mark size=0.5pt,
						] table[x=u10730pred,y=f10730pred,meta=label]{data/FU10730HCpred_U33D120T2.txt};
						
						\node[align=center] at (0.475,2.4) {(a)};
						\node[circle, fill=black, inner sep=1.25pt, label=left:\textit{a}] at (0.45, 1.98) {};
						\node[circle, fill=black, inner sep=1.25pt, label=right:\textit{b}] at (0.62, 1.228) {};
						\node[circle, fill=black, inner sep=1.25pt, label=above:\textit{c}] at (1.15, 0.2153) {};
						\node[circle, fill=black, inner sep=1.25pt, label=above:\textit{d}] at (1.5, 0.0668) {};
						\node[circle, fill=black, inner sep=1.5pt, label=above left:\textit{e}] at (2.0, 0.0176) {};
						
						\legend{\text{reference solution},\text{$- 50\%$ Lam\'e constants},\text{$+ 50\%$ Lam\'e constants},\text{ECSW predicted solution}}
						
						\coordinate (spypoint) at (axis cs:0.56, 1.478); 
						\coordinate (magnifyglass) at (axis cs:1.5, 1);
					\end{axis}
					\spy [black, size=1.75cm] on (spypoint) in node[fill=white] at (magnifyglass);
				\end{tikzpicture}
			\end{minipage}
			\begin{minipage}[b]{0.5\linewidth}
				\begin{tikzpicture}
					\begin{axis}[
						xmin=-0.1,xmax=130,
						ymin=1e-5,ymax=5e-4,
						scale only axis,
						ticklabel style={font=\large},
						xlabel={number of displacement modes $\bm{k}$ $\bm{[-]}$}, 
						ylabel={Absolute error $\bm{\varepsilon _{a}} \, \bm{[-]}$},
						line width=0.5pt,
						ymode=log,
						log basis y={10},
						font=\bfseries,
						legend pos=north east,
						legend style={fill=none, font=\small}
						]
						
						\addplot[
						line width=0.75pt,
						color=blue, densely dashed,
						mark=triangle*,
						mark size=2.5pt,
						mark options={solid},
						] table[x=modes,y=predicted]{data/HC_predict_errors.txt};
						
						\legend{
							{\text{ECSW for prediction}}
						}
						\node[align=center] at (40,3.5e-4) {(b)};
					\end{axis}
				\end{tikzpicture}
			\end{minipage}
			\caption{(a) Force-displacement curves, comparing the reference solution to the predicted result using ECSW for the HC specimen. The number of displacement modes in the predictive simulation for the displacement field is selected as $k=33$, whereas the  corresponding mode numbers for nonlocal damage and temperature fields are defined as $l=120$ and $m=2$, respectively; (b) the predictive error in displacement component $u$ for different numbers of displacement modes $k$.}
			\label{fig10}
		\end{figure*}

		By contrast, \hyperref[fig9]{Fig.9 (c)} shows that the distribution of the normalized singular values for both the damage and displacement fields persists longer during the softening phase, where the corresponding singular values decay more slowly. These fields exhibit dominant modal contributions, highlighting their critical role in the evolution of damage. This observation is in agreement with physical expectations: the initialization and evolution of damage are strongly related to strain localization, where the local damage variable governs the degradation of stress-like quantities, such as the second Piola-Kirchhoff stress $\mathbf{S}$, Mandel stress $\bm{M}$, and back-stress tensors $\bm{\chi}$, while the nonlocal damage field enforces spatial regularity. Therefore, a larger number of modes is required in these fields to accurately capture the complex interactions associated with damage-induced softening. To evaluate the predictive performance of the reduced-order model, \hyperref[fig10]{Fig.10 (a)} compares the force-displacement responses obtained from both reduced- and full-order simulations. The reference solution is additionally computed without any deviation in the Lam\'e constants in order to provide a consistent baseline for assessing the accuracy of the reduced model.
		
\begin{figure*}[!ht]
			\centering
			\includegraphics[width=18cm]{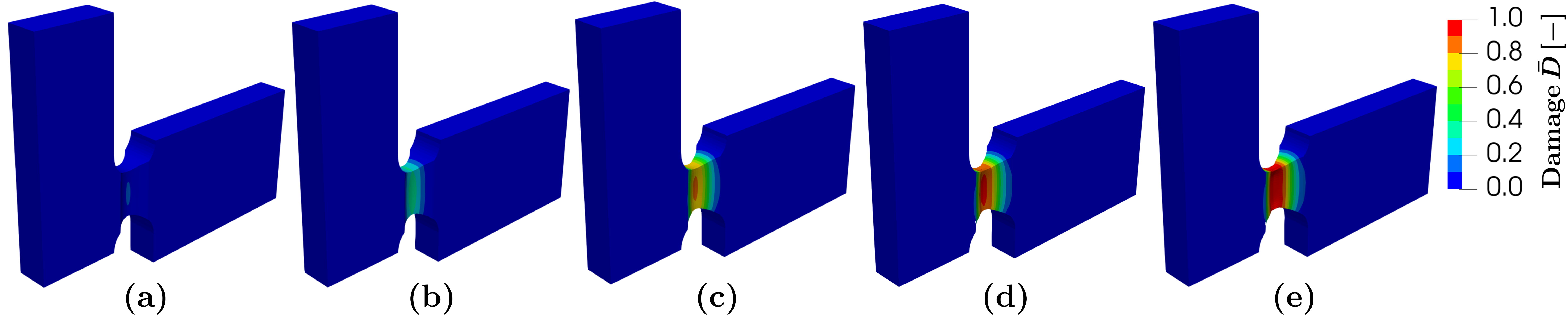}
			\caption{A schematic illustration of the nonlocal damage evolution in a reduced-order simulation for the HC specimen. The corresponding numbers of modes are $k=33$, $l=120$, and $m=2$. The associated figures depict the specimen's state at loading displacements of $u=0.45$, $0.62$, $1.15$, $1.50$, and $2.00$ mm, as indicated by points $a$, $b$, $c$, $d$, and $e$ in \hyperref[fig10]{Fig.10 (a)}.}
			\label{fig11}
		\end{figure*}

		As shown in \hyperref[fig10]{Fig.10 (a)}, the proposed multi-field and multi-state decomposition technique leads to a good predictive performance across the entire simulation process. The evolution of the nonlocal damage in the reduced-order simulation is illustrated in \hyperref[fig11]{Fig.11}, while \hyperref[fig10]{Fig.10 (b)} presents the corresponding absolute error in the predictive scenario. In addition, as the number of displacement modes $k$ exceeds 33, the predictive error stabilizes at approximately $5.74 \times 10^{-5}$. At this level of accuracy, discrepancies in the force-displacement curves and contour plots become visually indistinguishable, see \hyperref[fig10]{Fig.10 (a)}. Importantly, the employment of the multi-state decomposed nonlinear manifold enables a stabilized error with a certain number of modes. These results collectively demonstrate the robustness and reliability of the proposed multi-field and multi-state decomposed nonlinear manifold MOR approach for mitigating the Kolmogorov barrier in damage.
		
		\section{Results and discussion}
		\label{Results and discussion}
		\subsection{Decomposed versus non-decomposed nonlinear manifold MOR}
		\label{sec:RAD}
		
		\begin{figure}[!h]
			\centering
			\includegraphics[width=17cm]{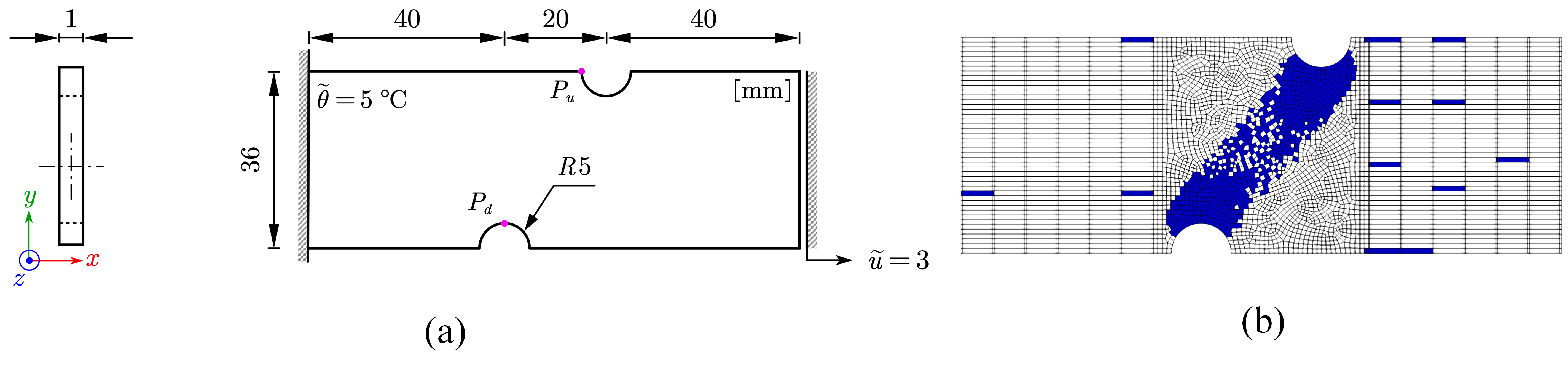}
			\caption{(a) Schematic illustration of an asymmetrically notched specimen and the considered boundary value problem. The displacement in the horizontal direction on the right surface is prescribed by a value of $\widetilde{u}=3$ mm. A constant temperature is prescribed on the specimen $\widetilde{\theta}=\SI{5}{\celsius}$. Points $P_u$ and $P_d$ are selected to compare the errors of the displacement and nonlocal damage variables, respectively. The thickness direction is constrained. (b) The selected ECSW elements ($ n_s = 708$) are shown in blue in a discretized specimen of 2912 elements.}
			\label{fig12}
		\end{figure}
		
		As shown in Eqs.\eqref{n12} and \eqref{n14}, a novel decomposed strategy is proposed in the optimization process in order to obtain a more accurate coefficient matrix $\bf{\Xi}$ in the context of the quadratic manifold approach. Here, a comparative analysis is performed for an asymmetrically notched specimen (\hyperref[fig12]{Fig.12}) to assess the performance of the newly developed method. The distribution of normalized singular values for the asymmetrically notched specimen is shown in \hyperref[figC16]{Fig.C.16}. The objective of \autoref{sec:RAD} is to investigate the influence of the field decomposition technique on the coefficient matrix, when the objective function is optimized in terms of the Frobenius norm in \hyperref[n12]{Eq.\eqref{n12}}. For the sake of simplicity, the specimen thickness is discretized using a single element.
		
		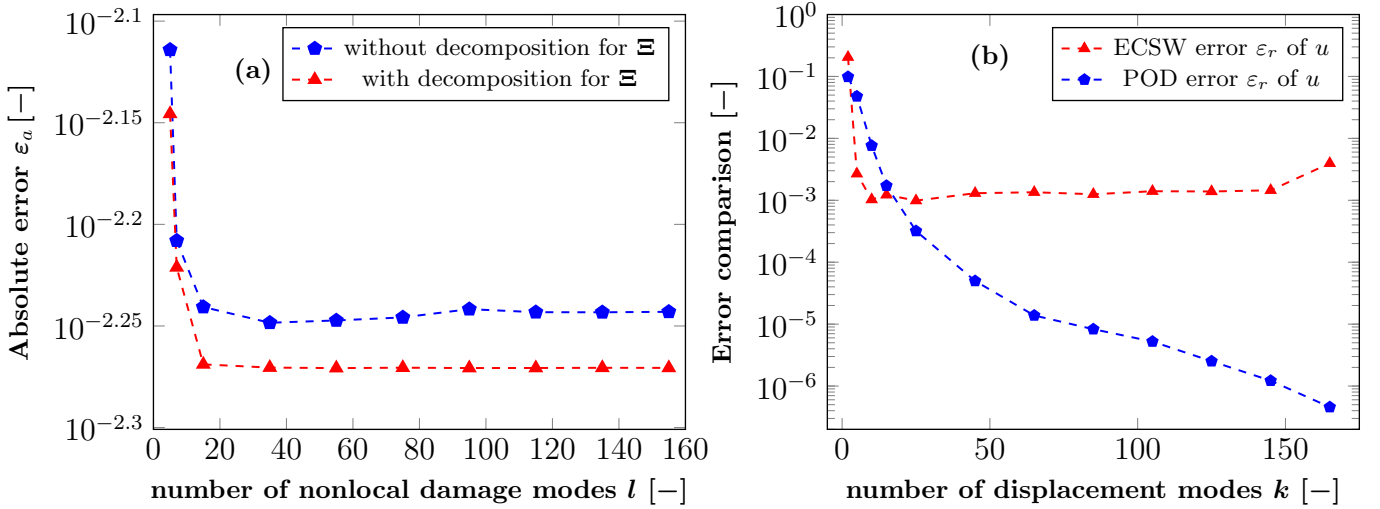
\begin{figure*}[!h]
			\begin{minipage}[b]{0.5\linewidth}
				\begin{tikzpicture}
					\begin{axis}[
						xmin=0,xmax=160,
						ymin=5e-3,ymax=8e-3,
						scale only axis,
						ticklabel style={font=\large},
						xlabel={number of nonlocal damage modes $\bm{l}$ $\bm{[-]}$}, 
						ylabel={Absolute error $\bm{\varepsilon}_{a} \, \bm{[-]}$},
						line width=0.5pt,
						ymode=log,
						log basis y={10},
						font=\bfseries,
						legend pos=north east,
						legend style={fill=none, font=\small}
						]
						\addplot[
						line width=0.75pt,
						color=blue, dashed,
						mark=pentagon*,
						mark size=2.5pt,
						mark options={solid},
						] table[x=modes,y=nSPL]{data/Asy-splvsnspl.txt};
						
						\addplot[
						line width=0.75pt,
						color=red, dashed,
						mark=triangle*,
						mark size=2.5pt,
						mark options={solid},
						] table[x=modes,y=SPL]{data/Asy-splvsnspl.txt};
						
						\legend{
							{\text{without decomposition for $\bf{\Xi}$}},
							{\text{with decomposition for $\bf{\Xi}$}}
						}
						\node[align=center] at (30, 7.5e-3) {(a)};
					\end{axis}
				\end{tikzpicture}
			\end{minipage}
			\begin{minipage}[b]{0.5\linewidth}
				\centering
				\begin{tikzpicture}
					\begin{axis}[
						clip=true,
						xmin=-5,xmax=175,
						ymin=2e-7,ymax=1,
						scale only axis,
						ticklabel style={font=\large},
						xlabel={number of displacement modes $\bm{k}$ $\bm{[-]}$}, 
						ylabel={Error comparison $\bm{[-]}$},
						line width=0.5pt,
						font=\bfseries,
						legend pos=north east,
						legend style={fill=none, font=\small},
						ymode=log,
						log basis y={10},
						]
						
						\addplot[
						color=red, dashed,
						mark=triangle*,
						mark size=2pt,
						mark options={solid},
						line width=0.75pt,
						] table[x=modes,y=UError]{data/Asy8979_ECSW_UDerrors.txt};
						
						\addplot[
						color=blue, dashed,
						mark=pentagon*,
						mark size=2pt,
						mark options={solid},
						line width=0.75pt,
						] table[x=modes,y=E8979]{data/Asy-mesh-errors.txt};
						
						\legend{
							{\text{ECSW error $\varepsilon_{r}$ of $u$}},
							{\text{POD error $\varepsilon_{r}$ of $u$}}
						}
						\node[align=center] at (50,0.2) {(b)};
					\end{axis}
				\end{tikzpicture}
			\end{minipage}
			\caption{(a) Absolute errors for different numbers of linear nonlocal damage modes $l \in [5,155]$ using a mesh discretization of 2912 in the case with and without the field decomposition method. (b) Error comparison for different numbers of linear displacement modes $k \in [2, 165]$ for a mesh discretization of 8979 elements, with $l=150$ linear and $q_l=5$ nonlinear modes for the nonlocal damage field, $q_k=5$ nonlinear modes for the displacement field, $m=2$ linear and $q_m=1$ nonlinear modes for the temperature field.}
			\label{fig13}
		\end{figure*}
		To evaluate the accuracy of the reduced-order simulations, both the absolute error and the relative error are computed. As illustrated in \hyperref[fig13]{Fig.13 (a)}, the number of linear displacement modes is chosen as $k=15$, while the number of nonlinear displacement modes is set to $q_k=5$. For the temperature field, both the number of linear and the nonlinear temperature modes are selected as $m=2$ and $q_m=1$, respectively. Moreover, the number of linear nonlocal damage modes varies within the range $l\in[5,155]$, and the number of nonlinear damage modes is fixed to $q_l=5$. The results shown in \hyperref[fig13]{Fig.13 (a)} highlight that the field decomposition technique outperforms the non-decomposed method with respect to the error in the nonlocal damage variable. This advantage becomes increasingly significant for $k$ ranging from 2 to 165, as illustrated in \hyperref[fig13]{Fig.13 (b)}, which compares the POD-based and ECSW-based nonlinear manifold methods. As can be observed and could be expected, the error of the ECSW-based nonlinear manifold approach is higher than the error of the POD-based nonlinear manifold approach. This aligns with the trend observed in comparisons between POD- and hyperreduction-based methods that make use of linear projection techniques by \citet{radermacher2016pod}. The reason for this lies in the hyperreduction technique's selection of only a few effective elements aimed at significantly accelerating simulations. However, this naturally leads to an impact on the accuracy \cite{jain2019some,rutzmoser2018model}. The goal in engineering problems is usually to find a suitable compromise between computational efficiency and accuracy, see \hyperref[fig14]{Fig.14 (a)}. Although the computational cost can be significantly reduced by evaluating only a small subset of elements, this efficiency inherently comes at the expense of accuracy, due to the difficulty of accurately representing complex constitutive behavior with such a limited sampling.
		
		\subsection{Influence of the number of linear displacement modes in the nonlinear manifold approach}
		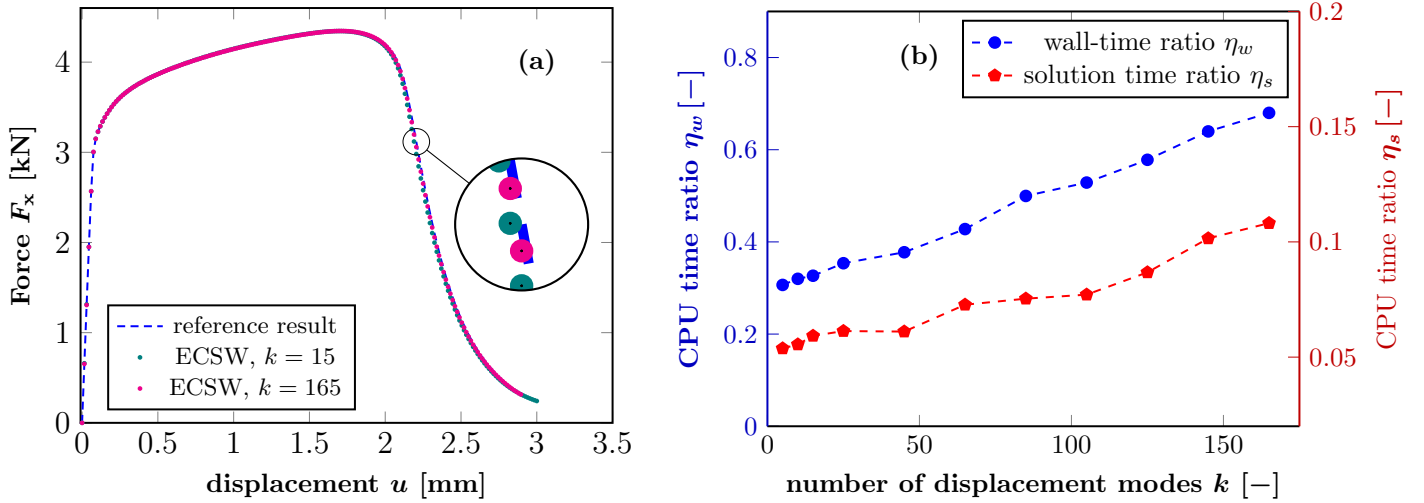
\begin{figure*}[!ht]
			\begin{minipage}[b]{0.5\linewidth}
				\centering
				\begin{tikzpicture}[spy using outlines={circle, magnification=5, connect spies}]
					\begin{axis}[
						name=plotA,
						clip=true,
						xmin=-0.01,xmax=3.5,
						ymin=0,ymax=4.6,
						scale only axis,
						log ticks with fixed point,
						ticklabel style={font=\large},
						xlabel={displacement $\bm{u}$ $\textbf{[mm]}$}, 
						ylabel={Force $\bm{F_{\rm{x}}}$ \textbf{[kN]}},
						line width=0.75pt,
						font=\bfseries,
						legend pos=south west,
						legend style={fill=none, font=\small, at={(0.05, 0.03)}},
						scatter/classes={%
							a={mark=*,draw=teal},
							b={mark=*,draw=magenta},
							c={mark=*,draw=black},
							d={mark=*,draw=teal}
						}
						]
						
						\addplot[
						color=blue, densely dashed,
						] table[x=u8979full,y=f8979full]{data/FU8979full.txt};
						
						\addplot[
						scatter, only marks,
						mark=*,mark options={solid},
						scatter src=explicit symbolic,
						mark size=0.5pt,
						] table[x=u8979mor,y=f8979mor,meta=label]{data/Asy8979_FU_ECSW_K15.txt};
						
						\addplot[
						scatter, only marks,
						mark=*,mark options={solid},
						scatter src=explicit symbolic,
						mark size=0.5pt,
						] table[x=u8979mor,y=f8979mor,meta=label]{data/Asy8979_FU_ECSW_K165.txt};
						\legend{\rm{reference result}, \rm{ECSW, $k=15$}, \rm{ECSW, $k=165$}}
						\node[align=center] at (3,4) {(a)};
						
						\coordinate (spypoint) at (axis cs:2.205, 3.117);
						\coordinate (magnifyglass) at (axis cs:2.9, 2.2);
					\end{axis}
					\spy [black, size=1.75cm] on (spypoint) in node[fill=white] at (magnifyglass);
				\end{tikzpicture}
			\end{minipage}
			\begin{minipage}[b]{0.5\linewidth}
				\centering
				\begin{tikzpicture}
					\pgfplotsset{
						scale only axis,
						xmin=0, xmax=175,
						y axis style/.style={
							yticklabel style=#1,
							ylabel style=#1,
							y axis line style=#1,
							ytick style=#1
						}
					}
					
					\begin{axis}[
						line width=0.75pt,
						axis y line*=left,
						y axis style=blue!75!black,
						ymin=0, ymax=0.9,
						font=\bfseries,
						xlabel={number of displacement modes $\bm{k}$ $\bm{[-]}$},
						ylabel={CPU time ratio $\bm{\eta_w} \, \bm{[-]}$},
						]
						
						\addplot[
						color=blue, dashed,
						mark=*,
						mark size=2pt,
						mark options={solid},
						line width=0.75pt,
						] table[x=modes,y=Tstep]{data/Asy_8979_CPU_Time.txt}; \label{plot_one},
						\node[align=center] at (50,0.8) {(b)};
					\end{axis}
					
					\begin{axis}[
						line width=0.75pt,
						axis y line*=right,
						axis x line=none,
						ymin=0.02, ymax=0.2,
						y tick label style={/pgf/number format/fixed, /pgf/number format/precision=2},
						ylabel={CPU time ratio $\bm{\eta_s} \, \bm{[-]}$},
						y axis style=red!75!black
						]
						\addlegendimage{/pgfplots/refstyle=plot_one}\addlegendentry{{wall-time ratio ${\eta_w}$}}
						\addplot[
						color=red, dashed,
						mark=pentagon*,
						mark size=2.25pt,
						mark options={solid},
						line width=0.75pt,
						] table[x=modes,y=Tsolve]{data/Asy_8979_CPU_Time.txt};\addlegendentry{solution time ratio ${\eta_s}$}
					\end{axis}
				\end{tikzpicture}
			\end{minipage}
			\caption{(a) Force-displacement curves comparison between full-order and reduced-order simulations for a fixed discretization of 8979 elements with $k = \{15, 165\}$. (b) CPU time ratio comparison for different values of $k$ using ECSW and nonlinear manifold-equipped reduced-order simulations.}
			\label{fig14}
		\end{figure*}
		
		\hyperref[fig14]{Fig.14 (a)} presents a comparison of the force-displacement curves for different values of $k \in [15, \, 165]$. Using the nonlinear manifold ECSW method, the computed error in the nonlocal damage field reaches $4.51 \times 10^{-3}$. This level of accuracy, achieved with relatively low values for $k$, is difficult to achieve using a linear approximation approach, unless the number of modes is significantly increased, e.g., to $k \geq 40$ \cite{zhang2025multi}. In return, this would lead to a substantial rise in computational cost. These results demonstrate that, even with a relatively small number of modes, the nonlinear manifold ECSW can lead to good performance. \hyperref[fig14]{Fig.14 (b)} shows the change in the CPU time ratio concerning both the wall-time $\eta _w$ and solution time $\eta_s$, when the number of displacement modes $k$ increases for a fixed mesh discretization of 8979 elements.
		
		\newpage
		\subsection{Comparisons between linear and nonlinear manifold approaches}
		A comparison between the linear and quadratic approximation techniques is conducted to illustrate the computational barrier posed by the linear approach in damage-induced problems, which is attributed to the slow decay (see \hyperref[AppendixC]{Appendix C}) in the Kolmogorov $n$-width. This comparison assesses both the relative error and the associated computational cost. To allow for a consistent comparison with the multi-field decomposed linear manifold approach recently proposed by \citet{zhang2025multi}, the mode numbers are prescribed as $l=155$ for the nonlocal damage field and $m=5$ for the temperature field. The number of nonlinear modes is set to the same value as in \autoref{sec:RAD}.
		
		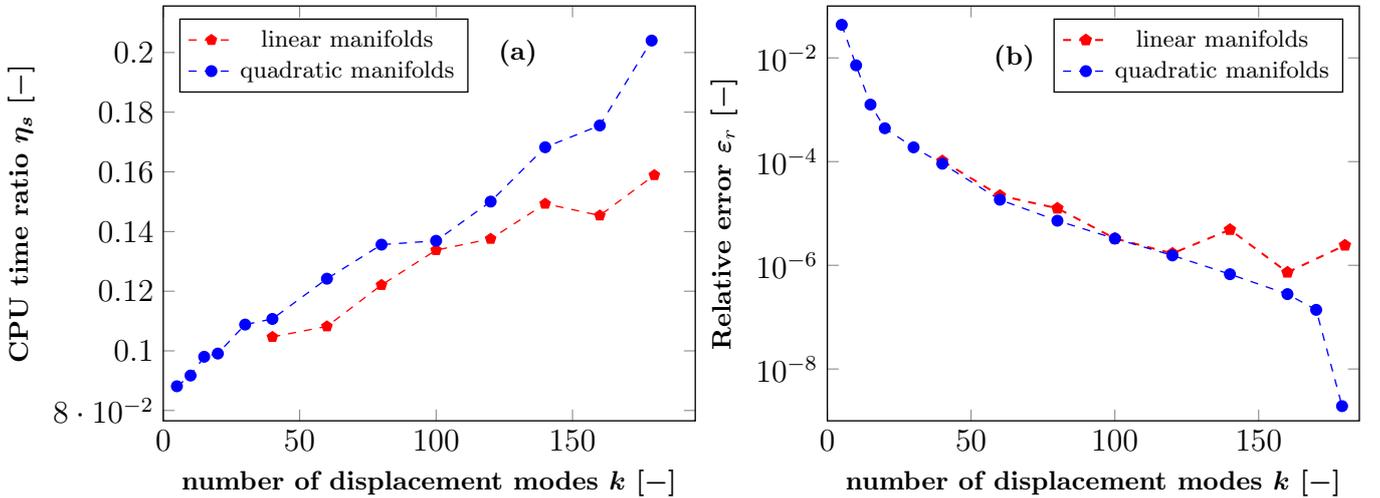
\begin{figure*}[!h]
			\begin{minipage}[b]{0.5\linewidth}
				\centering
				\begin{tikzpicture}
					\begin{axis}[
						clip=true,
						xmin=0,xmax=195,
						scale only axis,
						ticklabel style={font=\large},
						xlabel={number of displacement modes $\bm{k}$ $\bm{[-]}$}, 
						ylabel={CPU time ratio $\bm{\eta_s}$ $\bm{[-]}$},
						line width=0.5pt,
						font=\bfseries,
						legend pos=north west,
						legend style={fill=none, font=\small}
						]
						\addplot[
						color=red, dashed,
						mark=pentagon*,
						mark size=2.0pt,
						mark options={solid},
						] table[x=modes,y=CPUTime]{data/Asy-CPUTime-lin.txt};
						
						\addplot[
						color=blue, dashed,
						mark=*,
						mark size=2.0pt,
						mark options={solid},
						] table[x=modes,y=CPUTime]{data/Asy-CPUTime-nonlin.txt};
						
						\legend{
							{\rm{linear manifolds}},
							{\rm{quadratic manifolds}}
						}
						\node[align=center] at (130,0.2) {(a)};
					\end{axis}
				\end{tikzpicture}
			\end{minipage}
			\begin{minipage}[b]{0.5\linewidth}
				\centering
				\begin{tikzpicture}
					\begin{axis}[
						clip=true,
						xmin=0,xmax=185,
						ymin=1e-9,ymax=0.1,
						scale only axis,
						ticklabel style={font=\large},
						xlabel={number of displacement modes $\bm{k}$ $\bm{[-]}$}, 
						ylabel={Relative error $\bm{\varepsilon}_r$ $\bm{[-]}$},
						line width=0.5pt,
						font=\bfseries,
						legend pos=north east,
						legend style={fill=none, font=\small},
						ymode=log,
						log basis y={10},
						]
						\addplot[
						color=red, dashed,
						mark=pentagon*,
						mark size=2.0pt,
						mark options={solid},
						line width=0.75pt,
						] table[x=modes,y=uerror]{data/Asy-uerror-lin.txt};
						
						\addplot[
						color=blue, dashed,
						mark=*,
						mark size=2.0pt,
						mark options={solid},
						] table[x=modes,y=uerror]{data/Asy-uerror-nonlin.txt};
						
						\legend{
							{\rm{linear manifolds}},
							{\rm{quadratic manifolds}}
						}
						\node[align=center] at (65,0.01) {(b)};
					\end{axis}
				\end{tikzpicture}
			\end{minipage}
			\caption{Comparisons between linear and nonlinear manifold MOR approaches without using ECSW in terms of (a) the CPU time ratio $\eta _s$ in solving the equation system and (b) the relative error $\varepsilon_{r}$ for a mesh discretization of 2912 elements.}
			\label{fig15}
		\end{figure*}
		
		Typically, a fast reduced-order simulation is realized by choosing a low number of modes for a target equation system. However, in the context of damage-plasticity simulations under complex shear and tensile loadings, a linear projection often results in an ill-conditioned equation system, here observed when less than 40 displacement modes are used, as shown in \hyperref[fig15]{Fig.15}. This limitation is in agreement with the Kolmogorov barrier in approximation theory, whereby iteration is prematurely terminated due to non-convergence. In contrast to the linear projection, the relative error $\varepsilon_{r}$ associated with the nonlinear projection decreases smoothly and monotonically as the number of linear modes increases. As observed in \hyperref[fig15]{Fig.15 (b)}, the linear projection method also exhibits significant error fluctuations at higher numbers of modes (e.g., $k \geq 135$). In contrast, the quadratic manifold approach leads to stable results across a wide range of modes. This behavior reflects the capacity of nonlinear projections to approximate the strong nonlinearities present in thermo-mechanically coupled damage-plasticity simulations. It also indicates a significant reduction in the approximation barrier commonly encountered for linear approximations. Nevertheless, the nonlinear projection comes at the price of additional computational cost compared to the linear projection, as shown in \hyperref[fig15]{Fig.15 (a)}. This can be mitigated when integrating the ECSW technique into the quadratic manifold method. Since the dimension of the equation system will be small enough to neglect the computational cost of nonlinear operations. In conclusion, the key advantage of nonlinear projection lies in its capacity to mitigate the Kolmogorov $n$-width barrier in linear approximation, thereby enabling accurate and stable reduced-order simulations with a very small number of dominant modes.
		
		\section{Conclusion}
		\label{sec:Conclusion}
		
		In this study, a unified multi-perspective (multi-field and multi-state) quadratic manifold-based model order reduction framework has been proposed to mitigate the approximation barrier caused by the slow decay of the Kolmogorov $n$-width in thermo-mechanically coupled damage-plasticity simulations. To further enhance computational efficiency for solving the highly nonlinear equation systems, the framework integrated the ECSW hyperreduction method, which allows accurate results to be achieved using only a small subset of selected elements and modes.
		
		A key contribution of this work lies in the newly developed multi-field and multi-state decomposition strategy, which is grounded in the material’s physical states to guide the selection of mode numbers for each coupled field. Comparative analyses have demonstrated that, unlike the classical linear manifold approach, which often suffers from premature stagnation in reduced-order simulations, the proposed multi-perspective quadratic manifold approach ensured a smooth and monotonic decrease for the computed error as the number of modes increases, thereby effectively mitigating the Kolmogorov barrier.
		
		Another important characteristic of the proposed framework is its capability to effectively decouple material states and physical fields in complex damage-involved multiphysics problems. This decoupling has provided a clearer assessment of how each individual field contributes to and interacts within the overall multiphysics behavior, as shown through the normalized singular value distributions of the coupled solutions. In addition, a systematic study of the ECSW training tolerances has revealed that the number of selected elements played a critical role in balancing accuracy and computational efficiency for damage-involved multiphysics simulations.
		
		Despite the demonstrated effectiveness of the proposed framework in mitigating the Kolmogorov barrier, its dependence on a fixed discretization remains an inherent limitation of global Galerkin-projection-based MOR techniques. Furthermore, this study has focused exclusively on monotonically increasing loadings leading to structural failure. Extending the approach to address high-cycle fatigue-induced damage under multiphysics scenarios could open a broad avenue for future work, with significant potential for industrial applications, such as fatigue-life prediction and durability assessment.

		\section{Acknowledgements}
		\label{Acknowledgement}
		Q. Zhang, T. Brepols, and S. Reese acknowledge the financial support of subproject B05 within SFB/TRR 339 (project number: 453596084) by the German Research Foundation (DFG, Deutsche Forschungsgemeinschaft). J. Zhang acknowledges the financial support from the China Scholarship Council (CSC). S. Ritzert, S. Reese, and T. Brepols acknowledge the financial support of the subproject A01 within SFB/TRR 280 (project number: 417002380) by the German Research Foundation (DFG, Deutsche Forschungsgemeinschaft).
		
		\section{Conflict of interest}
		The authors declare that they have no known competing financial interests or personal relationships that could have appeared to influence the work reported in this paper.
		
		\section{Contributions by the authors}
		\textbf{Qinghua Zhang}: Writing – original draft, Writing – review \& editing, Conceptualization, Methodology, Theoretical description, Implementation, Data Curation, Formal analysis, Investigation, Visualization.
		\textbf{Stephan Ritzert}: Theoretical description, Writing – review \& editing.
		\textbf{Jian Zhang}: Writing – review \& editing.
		\textbf{Jannick Kehls}: Review \& editing. 
		\textbf{Stefanie Reese}: Supervision, Funding acquisition.
		\textbf{Tim Brepols}: Supervision, Funding acquisition, Writing – review \& editing.
		
		\appendix 
		
		\section{Thermodynamic consistency and \textit{two-surface} damage-plasticity}
		\label{AppendixA}
		Here, consistent constitutive relations are derived using an extended form of the Clausius-Duhem inequality:
		\begin{equation}
			{\bf{S}} \, {:} \, \frac{1}{2} \, \dot {\bm{C}} + \underbrace {{a_i} \, \dot {\bar D} + {\bm{b}_i} \cdot \nabla \bar D}_{\rm{micromorphic \, extension}} - \dot \psi  - \eta \, \dot \theta  - \frac{1}{\theta } \, \bm{q} \cdot \nabla \theta  \ge 0 .
			\label{eqA1}
		\end{equation}
		
		\noindent Inserting the time derivative of the Helmholtz free energy density function $\dot{\psi}$ (see \hyperref[eq4]{Eq.\eqref{eq4}}) into inequality \eqref{eqA1}, leads after some rearrangements to: 
		\begin{equation}
			\begin{gathered}
				{\bf{S}} \, {\text{:}} \, \frac{1}{2} \, {\dot{\bm {C}}} + \left( {{a_i} - \frac{{\partial {\psi _{\bar d}}}}{{\partial \bar D}}} \right)\dot {\bar D} + \left( {{{\bm{b}}_i} - \frac{{\partial {\psi _{\bar d}}}}{{\partial \nabla \bar D}}} \right) \cdot \nabla \dot {\bar D}   - {f_d}(D)\left[ {\frac{{\partial {\psi _e}}}{{\partial {{\bm{C}}_e}}} \, {\text{:}} \, {{{\dot{\bm {C}}}}_e} + \frac{{\partial {\psi _p}}}{{\partial {{\bm{C}}_p}}} \, {\text{:}} \, {{{\dot{\bm {C}}}}_p} + \frac{{\partial {\psi _p}}}{{\partial {\xi _p}}} \, {{\dot \xi }_p}} \right] \hfill \\
				- \left[ {\frac{{\partial {f_d}(D)}}{{\partial D}}\left( {{\psi _e} + {\psi _p}} \right) + \frac{{\partial {\psi _{\bar d}}}}{{\partial D}}} \right]\dot {D} - \frac{{\partial {\psi _d}}}{{\partial {\xi _d}}} \, {{\dot \xi }_d}  - \left( {\frac{{\partial \psi }}{{\partial \theta }} + \eta } \right)\dot \theta - \frac{1}{\theta } \, {\bm{q}} \cdot \nabla \theta  \geqslant 0. \hfill \\ 
			\end{gathered} 
			\label{eqA2}
		\end{equation}
		
		\noindent By assuming a zero dissipation for the micromorphic (nonlocal) damage variable $\bar{D}$ and its gradient $\nabla \bar{D}$, the second Piola-Kirchhoff stress ${\bf{S}}$, the micromorphic forces ($a_i$, $\bm{b}_i$), the entropy $\eta$, and heat flux $\bm{q}$ are obtained:
		\begin{equation}
			{{\bf{S}} = 2 \, {f_d}(D) \, \frac{1}{{{\vartheta ^2}}} \, {\bm{F}}_p^{ - 1} \, \frac{{\partial {\psi _e}}}{{\partial {{\bm{C}}_e}}} \, {\bm{F}}_p^{ - {\text{T}}}}, \quad
			{a_i} = \frac{{\partial {\psi _{\bar d}}}}{{\partial \bar D}}, \quad \bm{b}_{i} = \frac{{\partial {\psi _{\bar d}}}}{{\partial \nabla \bar D}}, \quad
			\eta  =  - \frac{{\partial \psi }}{{\partial \theta }}, \quad 
			\bm{q} =  - {\bm{K}_{\theta}} \, \nabla \theta.
			\label{eqA3}
		\end{equation}
		\noindent For brevity, the Mandel stress tensor $\bm{M}$ and the back stress tensor $\bm{\chi}$ in the intermediate configuration are introduced as:
		\begin{equation}
			\bm{M} :={2 \, f_{d}(D) \, {\frac{{\partial {\psi _e}}}{{\partial {{\bm{C}}_e}}}}  \,  \bm{C}_{e}  }, \quad
			\bm{\chi} :={2 \, f_{d}(D) \, {\bm{F}}_p^{\text{T}} \, \frac{{\partial {\psi _p}}}{{\partial {{\bm{C}}_p}}} \, {{\bm{F}}_p}}.
			\label{eqA4}
		\end{equation}
		\noindent The conjugated driving forces to damage ${Y}$, isotropic hardening $q_p$, and damage hardening $q_d$ are defined, respectively, as:
		\begin{equation}
			{Y}: =  - \left( {\frac{{\partial {f_d}(D)}}{{\partial D}} \, \left( {{\psi _e} + {\psi _p}} \right) + \frac{{\partial {\psi _{\bar d}}}}{{\partial D}}} \right), \quad
			{q_p} := {f_d}(D) \, \frac{{\partial {\psi _p}}}{{\partial {\xi _p}}}, \quad
			{q_d} := \frac{{\partial {\psi _d}}}{{\partial {\xi _d}}}.
			\label{eqA5}
		\end{equation}
		\noindent With equations \eqref{eqA3}, \eqref{eqA4}, and \eqref{eqA5}, the remaining part of inequality \eqref{eqA1} is reformulated as:
		\begin{equation}
			\left( {{\bm{M}} - {\bm{\chi }}} \right):{{\bm{D}}_p} - {q_p} \, {{\dot \xi }_p} - {q_d} \, {{\dot \xi }_d} + Y \, \dot D \geqslant 0,
			\label{eqA6}
		\end{equation}
		\noindent where $\bm{D}_{p}=\frac{1}{2} \, \bm{F}^{{\rm{-T}}}_{p} \, \dot {\bm{C}}_{p} \, \bm{F}^{{\rm{-1}}}_{p}$ represents the symmetric part of the plastic velocity gradient. The positive semi-definite conductivity tensor $\bm{K}_{\theta}$ in \hyperref[eqA3]{Eq.\eqref{eqA3}} is taken from \citet{dittmann2020phase}:
		\begin{equation}
			\bm{K}_{\theta} = \left( f_{d}(D) \, K_{0} + \left( 1-f_{d}(D) \right) \, K_{c} \right) \, \bm{C}^{-1}.
			\label{eqA7}
		\end{equation} 
		\noindent $K_0$ and $K_c$ denote the heat conduction parameters of the undamaged state ($f_{d}(D)=1$) and fully broken state ($f_{d}(D)=0$), respectively. Following \citet{dittmann2020phase}, as the temperature increases, the plastic material parameters are degraded by a thermal softening function $f_{\theta} (\theta)$ with a softening parameter $\omega$:
		\begin{equation}
			f_{\theta}(\theta) =  1 - \omega \, \left(\theta-\theta _{0}\right).
			\label{eqA8}
		\end{equation} 
		
		Next, the plastic yield function $\Phi_p$ is employed using a von Mises-type criterion. Regarding damage, the damage loading function $\Phi _d$ is expressed as:
		\begin{equation}
			{\Phi _p} = \sqrt {\frac{3}{2}} \, \| {{\bm{\tilde M}'} - {\bm{\tilde \chi }'}} \| - \left( {{\sigma _{y0}} + {{\tilde q}_p}} \right), \quad
			{\Phi _d} = Y - \left( {{Y_0} + {q_d}} \right).
			\label{eqA9}
		\end{equation} 
		\noindent Here, $\sigma _{y0}$ denotes the initial yield stress. The operator ${(  \bullet )^\prime } = (  \bullet  ) - \tfrac{1}{3}{\text{tr}} (  \bullet ) \, \bm{I}$ in \hyperref[eqA9]{Eq.\eqref{eqA9}} is the deviatoric part of a second-order tensor. $( {\tilde  \bullet } ) = \tfrac{(\bullet)}{{{f_d}\left( D \right)}}$ denotes an effective quantity. $Y_0$ represents the initial damage threshold.
		
		\section{Derivation of the internal energy}
		\label{AppendixB}
		The Legendre transformation of the internal energy and its time derivative are defined as:
		\begin{equation}
			e_{\text{int}} = \psi + \theta \, \eta, \quad \dot{e}_{\text{int}} = \dot{\psi} + \dot{\theta} \, \eta + \theta \, \dot{\eta}.
			\label{BA1}
		\end{equation}
		By inserting the entropy $\eta=-\frac{\partial \psi}{\partial \theta}$ into Eq.\eqref{BA1}, one obtains
		\begin{equation}
			e_{\text{int}} = \psi + \theta \, \eta = \psi - \theta \, \frac{\partial \psi}{\partial \theta}.
			\label{BA2}
		\end{equation}
		After taking the partial derivative of Eq.\eqref{BA2} with respect to the temperature $\theta$ and  integrating the obtained equation from the reference temperature $\theta_{0}$ to the current  temperature $\theta$, the equation becomes
		\begin{equation}
			\frac{\partial {e_{\text{int}}}}{\partial \theta} = \underbrace{- \theta \, \frac{\partial ^2 \psi}{\partial \theta ^2} }_{c} \quad \Rightarrow \quad {e_{\text{int}}} =  \int_{\theta_{0}}^{\theta} c \, d \theta + e_0,
			\label{BA3}
		\end{equation} 
		where $e_0$ denotes the internal energy at temperature $\theta_{0}$. Since the heat capacity $c$ is assumed to be constant, the following result is obtained:
		\begin{equation}
			e_{\text{int}} = c \, (\theta - \theta_{0}) + e_0, \quad \frac{d {e_{\text{int}}}}{dt} = \frac{d}{dt} \, \left( c \, (\theta - \theta_{0}) + e_0 \right) \quad \Rightarrow \quad {\dot{e}_{\text{int}} } = c \, \dot \theta.
			\label{BA4}
		\end{equation}
		
		\section{}
		\label{AppendixC}
		
\begin{figure*}[!ht]
			\centering
			\begin{minipage}[b]{0.5\linewidth}
				\begin{tikzpicture}
					\begin{axis}[name=plotA,
						xmin=-5,xmax=186,
						ymin=1e-17,ymax=10,
						scale only axis,
						ticklabel style={font=\large},
						xlabel={number of modes $\bm{[-]}$}, 
						ylabel={Normalized singular values $\bm{[-]}$},
						line width=0.5pt,
						ymode=log,
						log basis y={10},
						axis line style={line width=0.5pt},
						font=\bfseries,
						legend pos=north east,
						legend style={fill=none, font=\small}
						]
						\addplot[
						line width=1.0pt,
						color=green, densely dashed,
						] table[x=modes,y=U]{data/Asy-modes-SVD.txt};
						
						\addplot[
						line width=1.0pt,
						color=red, densely dashed,
						] table[x=modes,y=D]{data/Asy-modes-SVD.txt};
						
						\addplot[
						line width=1.0pt,
						color=blue, densely dashed,
						] table[x=modes,y=T]{data/Asy-modes-SVD.txt};
						
						\addplot[
						line width=1.0pt,
						color=violet, densely dashed,
						] table[x=modes,y=full]{data/Asy-modes-SVD.txt};
						
						\legend{
							{\text{$\bm{u}$ field   }},
							{\text{$\bar{D}$ field   }},
							{\text{$\theta$ field   }},
							{\rm{full field}}
						}
						
					\end{axis}
				\end{tikzpicture}
			\end{minipage}
			\label{figC16}
			\caption{Distribution of normalized singular values of the asymmetrically notched specimen. The individual normalized singular values denote the displacement field $\bm{u}$, nonlocal damage field $\bar{D}$, and temperature field $\theta$, see details in \cite{zhang2025multi}.}
		\end{figure*}
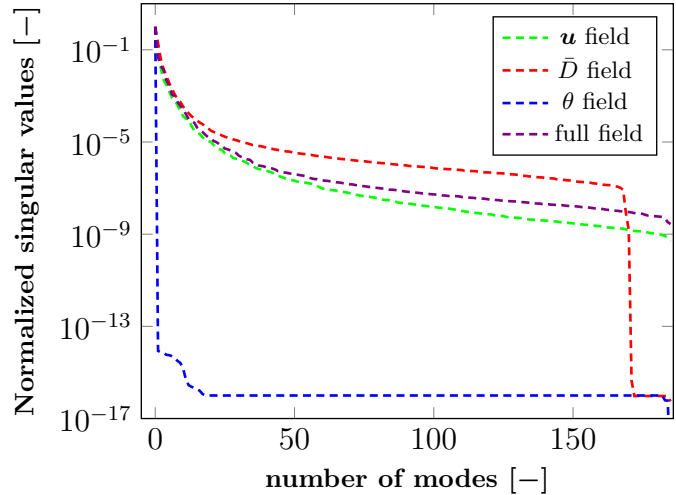

		Here, a uniaxial tensile test conducted on an asymmetrically notched specimen is used to illustrate the slow decay of the singular values in the presence of damage simulations. As shown in \hyperref[figC16]{Fig.C.16}, \hyperref[fig4]{Fig.4 (a)}, and \hyperref[fig9]{Fig.9}, the displacement and damage fields resulting from this test show complex localized features, leading to a slow decay in the distribution of singular values. These characteristics give rise to the Kolmogorov $n$-width approximation barrier when a linear projection-based model reduction approach is employed.
		
		\newpage	
		\section*{References}
		\bibliography{Article}
	\end{document}